\documentclass[aps,prx,superscriptaddress,amsmath,amsfonts,amssymb,showpacs,floatfix,reprint,nofootinbib,longbibliography]{revtex4-2}

\usepackage[utf8]{inputenc} 
\usepackage[T1]{fontenc}
\usepackage{mathtools}
\usepackage{newtxtext,newtxmath}
\usepackage{amssymb}
\usepackage{amsmath}
\usepackage{microtype}

\usepackage{graphicx}
\usepackage[dvipsnames,table]{xcolor}
\usepackage{natbib}
\usepackage{hyperref}

\usepackage{physics} 
\usepackage{bm}
\usepackage{bbm}
\usepackage[version=4]{mhchem}
\usepackage[percent]{overpic}
\usepackage[capitalise]{cleveref}

\usepackage{nicematrix}
\usepackage{longtable}
\usepackage{array}
\usepackage{dcolumn}

\usepackage{comment}
\usepackage{datetime}

\usepackage[mathcal]{eucal}

\definecolor{cblue}{RGB}{55,126,184}
\definecolor{ogreen}{RGB}{238,255,204}
\hypersetup{colorlinks=true,linkcolor=cblue,citecolor=cblue,urlcolor=cblue}

\DeclareMathAlphabet{\mathpzc}{OT1}{pzc}{m}{it}

\newcommand{\bq}{\bm{q}}

\newcommand{\integers}{\mathbbm{Z}}

\newcommand{\Jmat}{\mathcal{J}}
\newcommand{\Amat}{\mathcal{A}}

\newcommand{\diff}{\mathrm{d}}
\newcommand{\boundary}{\partial}

\newcommand{\uvec}[1]{\hat{\bm{#1}}}

\newcommand{\pe}{p_{\mathrm e}}
\newcommand{\pmag}{p_{\mathrm m}}

\newcommand{\vtiny}{\vcenter{\hbox{\includegraphics[height=0.15in]{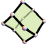}}}}
\newcommand{\memin}{\vcenter{\hbox{\includegraphics[height=0.31in]{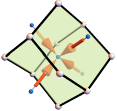}}}}
\newcommand{\memout}{\vcenter{\hbox{\includegraphics[height=0.31in]{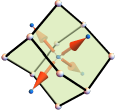}}}}

\begin{document}
\title{2-Form U(1) Spin Liquids: A Classical Model and Quantum Aspects}

\author{Kristian Tyn Kai Chung}
\altaffiliation[ktchung@pks.mpg.de, ktchung@uwaterloo.ca]{}
\affiliation{Department of Physics and Astronomy, University of Waterloo, Ontario, N2L 3G1, Canada}
\affiliation{Max Planck Institute for the Physics of Complex Systems, N\"othnitzer Strasse 38, 01187 Dresden, Germany}

\author{Michel J. P. Gingras}
\altaffiliation[]{gingras@uwaterloo.ca}
\affiliation{Department of Physics and Astronomy, University of Waterloo, Ontario, N2L 3G1, Canada}
\affiliation{Canadian Institute for Advanced Research, MaRS Centre, West Tower 661 University Ave., Suite 505, Toronto, ON, M5G 1M1, Canada}

\begin{abstract}
We introduce a novel geometrically frustrated classical Ising model, dubbed the ``spin vorticity model'', whose ground state manifold is a novel classical spin liquid: a 2-form Coulomb phase. We study the thermodynamics of this model both analytically and numerically, exposing the presence of algebraically decaying correlations and an extensive ground state entropy, and give a comprehensive account of its ground state properties and excitations. Each classical ground state may be decomposed into collections of closed 2-dimensional membranes, supporting fractionalized string excitations attached to the edges of open membranes. At finite temperature, the model can then be described as a gas of closed strings in a background of fluctuating membranes. The emergent gauge structure of this spin liquid is naturally placed in the language of 2-form electrodynamics, which describes 1-dimensional charged strings coupled to a rank-2 anti-symmetric gauge field. After establishing the classical spin vorticity model, we consider perturbing it with quantum exchange interactions, deriving an effective ``membrane exchange'' model of the quantum dynamics, analogous to ring exchange in quantum spin ice. We demonstrate the existence of a Rokhsar-Kivelson point where the quantum ground state is an equal-weight superposition of all classical ground state configurations, i.e. a quantum spin liquid. The quantum aspects of this spin liquid are exposed by mapping the membrane exchange model to a strongly-coupled frustrated 2-form U(1) lattice gauge theory. We further demonstrate how to quantize the string excitations by coupling a 1-form string field to the 2-form U(1) gauge field, thus mapping a quantum spin model to a 2-form U(1) gauge-Higgs model. We discuss the stability of the gapless deconfined phase of this gauge theory and the possibility of realizing a novel class of quantum phases of matter: 2-form U(1) quantum spin liquids.
\end{abstract}

\date{\today}
\maketitle

\section{Introduction}

A central aim of modern condensed matter physics is the search for and classification of novel phases of matter which challenge traditional notions of order and symmetry breaking~\cite{wenQuantumFieldTheory2004,mcgreevyGeneralizedSymmetriesCondensed2023}.
Some of the most experimentally promising routes to realizing such ``beyond-Landau'' possibilities are spin liquids---magnetic phases in which long-range order is evaded due to strong frustration caused by competing spin-spin interactions. 
At the classical level, spin liquids are described by a massively degenerate ground state manifold with an emergent gauge structure and fractionalized excitations~\cite{henleyCoulombPhaseFrustrated2010,castelnovoSpinIceFractionalization2012,rehnMaxwellElectromagnetismEmergent2016}. 
At the quantum level, the ground state of a quantum spin liquid (QSL) is a highly entangled superposition of such classical ground states~\cite{milaQuantumSpinLiquids2000,zhouQuantumSpinLiquid2017,broholmQuantumSpinLiquids2020,savaryQuantumSpinLiquids2016,yanClassificationClassicalSpin2024a}.
Theoretically, QSLs are characterized by emergent gauge fields, which mediate the interactions of deconfined fractionalized excitations~\cite{savaryQuantumSpinLiquids2016,broholmQuantumSpinLiquids2020,zhouQuantumSpinLiquid2017}. 
These can be roughly divided into those with a discrete gauge group, which are usually gapped, and those whose gauge group is U(1), which are generically gapless~\cite{savaryQuantumSpinLiquids2016}. 
In the latter case, the gapless excitations may be identified as emergent  or ``artificial'' photons~\cite{savaryQuantumSpinLiquids2016,hermelePyrochlorePhotonsU12004}.

While in two dimensions we are furnished with a number of models which admit exact or near-exact solutions, such as the toric code~\cite{kitaevFaulttolerantQuantumComputation2003} or Kitaev model~\cite{kitaevAnyonsExactlySolved2006}, the situation for three-dimensional U(1) QSLs is significantly less ideal, with a paucity of cases that one can conclusively identify as quantum spin liquids starting from a realistic microscopic spin model. 
Notwithstanding this shortcoming, significant effort has gone to classifying mean field ground states based on parton ans\"{a}tze of quasiparticles interacting with a gapless photon~\cite{wangTimeReversalSymmetricU12016,huangInterplayNonsymmorphicSymmetry2017,zouSymmetryEnrichedU12018,heringCharacterizationQuantumSpin2019,liuSymmetricU1Z22021,chernTheoreticalStudyQuantum2021}. 
While these classification results are powerful, it remains desirable to have microscopic three-dimensional spin models from which one can directly derive the emergent gauge theory without resorting to either a quasiparticle ansatz or parton mean field approximation.

\begin{figure*}[ht]
    \centering
    \begin{overpic}[width=0.33\textwidth]{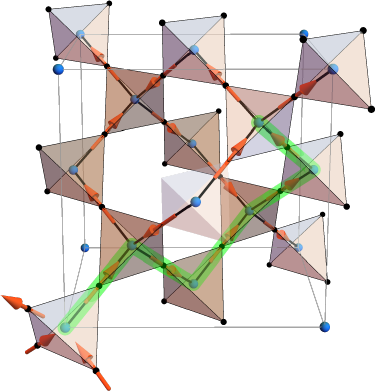}
        \put(47,0){(a)}
        \put(146,0){(b)}
        \put(244,0){(c)}
    \end{overpic}
    \hfill
    \begin{overpic}[width=0.33\textwidth]{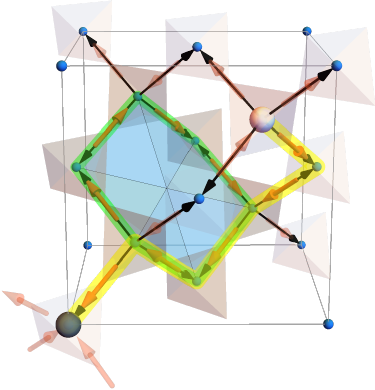}
        \put(60,55){1}
        \put(55,74){4}
        \put(76,65){2}
        \put(72,78){3}
        \put(28.3,75){$A$}
        \put(15.5,90){$B$}        
    \end{overpic}
    \hfill
    \includegraphics[width=0.33\textwidth]{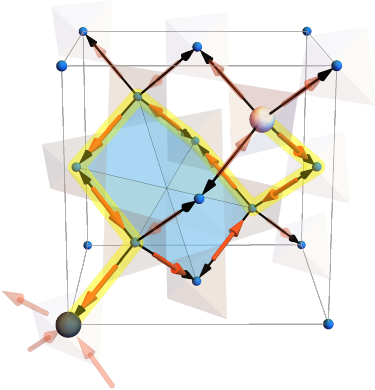}
    \caption{
    (a) The pyrochlore lattice of corner sharing tetrahedra, along with its parent diamond lattice. The diamond lattice is illustrated with light blue vertices at tetrahedron centers connected by thick black bonds. Each Ising spin (red arrow) sits at a pyrochlore site (black dots) at the midpoint of a diamond bond. 
    The two states of the Ising variable $\sigma_i^z=\pm 1$ correspond to the two orientations of a classical spin-1/2 vector aligned along each bond, $\bm{S}_i = (\hbar/2) \sigma_i^z \hat{\bm{z}}_i$. 
    Here, we show a portion of a 2-in-2-out ground state of the NNSI model, satisfying $Q_t=0$ on each tetrahedron. 
    An open chain of five head-to-tail spins is highlighted (green).
    (b) Black arrowheads at the ends of each diamond bond in panels (b) and (c) indicate our convention for the quantization axes $\hat{\bm{z}}_i$, which point from the diamond $A$ to diamond $B$ sublattice. Flipping the open (green) chain of spins in (a) creates a pair of opposite charges (\cref{eq:Q_t}) with $Q_t = +1$ (white) and $Q_t=-1$ (black) at the endpoints, connected by an electric flux string (yellow). The position of this string is only uniquely defined relative to the ground state in (a). 
    One plaquette of the diamond lattice, a buckled hexagon, is shaded (blue). The six spins on the boundary of this plaquette form a closed chain (green). 
    (c) Flipping the closed (green) chain of head-to-tail spins around the plaquette in (b) costs zero energy, and may be seen as moving the flux string (yellow) by effectively ``reconfiguring the vacuum''. 
    The minimal six-spin loop flips are precisely those generated in the ring exchange model of quantum spin ice~\cite{hermelePyrochlorePhotonsU12004}. 
    }
    \label{fig:pyro_diamond}
\end{figure*}

In this endeavor, the pyrochlore lattice of corner-sharing tetrahedra (\cref{fig:pyro_diamond}(a)) serves as the premier platform for the study of frustrated magnetism and spin liquidity in three dimensions, with a rich interplay of theoretical and experimental developments~\cite{gardnerMagneticPyrochloreOxides2010,hallasExperimentalInsightsGroundState2018,rauFrustratedQuantumRareEarth2019}.
In particular, it hosts a U(1) QSL called quantum spin ice (QSI)~\cite{gingrasQuantumSpinIce2014}, one of very few 3D QSLs which can be mapped at the microscopic level to a lattice gauge theory.
The QSI phase arises by perturbing the classical nearest-neighbor spin ice (NNSI) model~\cite{bramwellFrustrationIsingtypeSpin1998,moessnerReliefGenerationFrustration1998,bramwellHistorySpinIce2020} with spin-flipping operators such as a transverse field or, more physically, exchange interactions of the transverse spin components~\cite{gingrasQuantumSpinIce2014,hermelePyrochlorePhotonsU12004,molavianDynamicallyInducedFrustration2007,onodaQuantumMeltingSpin2010,savaryCoulombicQuantumLiquids2012}.
NNSI is an exemplar model of classical spin liquidity, whose strongly-correlated behavior arises from the geometrically frustrated corner-sharing geometry of the pyrochlore lattice~\cite{henleyPowerlawSpinCorrelations2005}.
The classical Ising Hamiltonian imposes a local microscopic constraint on every tetrahedron, such that the low-temperature phase is described by a coarse-grained vector field $\bm{E}$ satisfying a divergence-free condition, $\cramped{\nabla\cdot\bm{E} = 0}$~\cite{henleyPowerlawSpinCorrelations2005,gingrasQuantumSpinIce2014,henleyCoulombPhaseFrustrated2010,conlonAbsentPinchPoints2010,moessnerSpinIceCoulomb2021,castelnovoMagneticMonopolesSpin2008,castelnovoSpinIceFractionalization2012}.
These local constraints leave a highly-degenerate ground state manifold, in which spins collectively align in a head-to-tail fashion to form a network of closed strings~\cite{castelnovoSpinIceFractionalization2012,castelnovoMagneticMonopolesSpin2008,moessnerSpinIceCoulomb2021}, referred to as a Coulomb phase~\cite{henleyCoulombPhaseFrustrated2010,moessnerSpinIceCoulomb2021} or string condensate~\cite{levinStringnetCondensationPhysical2005}. 
This phase manifests itself in momentum-resolved correlation functions as characteristic ``pinch point'' singularities~\cite{isakovDipolarSpinCorrelations2004,henleyPowerlawSpinCorrelations2005}. 
Point-like quasiparticle excitations (spinons) then appear at the ends of \emph{open} strings of spins, acting as local ``electric charges''~$Q$ sourcing the divergence of $\bm{E}$ via an emergent Gauss law, $\cramped{\nabla\cdot\bm{E} = Q}$~\cite{castelnovoMagneticMonopolesSpin2008,jaubertSignatureMagneticMonopole2009,morrisDiracStringsMagnetic2009,gingrasObservingMonopolesMagnetic2009}.
Upon quantization, the quantum fluctuations of these strings become a collective gapless photon excitation described by an emergent U(1) gauge field~\cite{hermelePyrochlorePhotonsU12004,bentonSeeingLightExperimental2012,gingrasQuantumSpinIce2014}. 
Thus quantum spin ice realizes emergent quantum electrodynamics (QED), with electric charges (spinons), magnetic monopoles (visons)~\cite{hermelePyrochlorePhotonsU12004}, a gapless photon~\cite{hermelePyrochlorePhotonsU12004} and an associated fine structure constant~\cite{paceEmergentFineStructure2021}.\footnote{
    In this paper we adopt the convention, most natural when formulating the lattice gauge theory, that the strings formed by the spins are electric field lines, the spinons are electric charges, and the visons are magnetic charges. This is opposite to the convention used in much of the spin ice literature, where the spinons are identified as ``magnetic monopoles'', the strings are referred to as ``Dirac strings'', and the visons are electric charges. The two descriptions are dual~\cite{gingrasQuantumSpinIce2014,hermelePyrochlorePhotonsU12004}.
}

In addition to spin ice, the pyrochlore lattice has been found to host additional spin liquid phases by considering anisotropic spin-spin interactions in the full symmetry-allowed nearest-neighbor pseudospin Hamiltonian~\cite{rauFrustratedQuantumRareEarth2019}. 
Within the classical phase diagram of this model~\cite{yanTheoryMultiplephaseCompetition2017} one may find, in addition to spin ice, the Heisenberg antiferromagnet~\cite{henleyPowerlawSpinCorrelations2005,isakovDipolarSpinCorrelations2004,conlonAbsentPinchPoints2010} and so-called pseudo-Heisenberg antiferromagnet~\cite{taillefumierCompetingSpinLiquids2017}, both described by three decoupled Coulomb phases. 
In addition to these, a variety of further spin liquids have recently been identified, stabilized by nearest-neighbor anisotropic spin-spin interactions, which are described by emergent rank-2 \textit{tensor} gauge fields~\cite{bentonSpinliquidPinchlineSingularities2016,yanRank2U1Spin2020,lozano-gomezCompetingGaugeFields2024,lozano-gomezAtlasClassicalPyrochlore2024,chungMappingPhaseDiagram2024}. 
Of particular interest is the so-called rank-2 U(1) spin liquid~\cite{yanRank2U1Spin2020}, described by \textit{symmetric} rank-2 tensor gauge fields, which are fundamental to theories of fractons~\cite{pretkoFractonPhasesMatter2020}---excitations that can only move along subdimensional spaces~\cite{pretkoGeneralizedElectromagnetismSubdimensional2017,pretkoSubdimensionalParticleStructure2017,pretkoFractonGaugePrinciple2018}---and have close connections to the theories of elasticity~\cite{pretkoFractonElasticityDuality2018} and linearized gravity~\cite{pretkoEmergentGravityFractons2017,bertoliniGaugingFractonsLinearized2023}.
For all of these classical spin liquids arising from nearest-neighbor spin models on the pyrochlore lattice, the essential physics can be understood as a consequence of \emph{local constraints} enforced on each tetrahedron, which endow the system with a very large and highly correlated ground state manifold.
In principle, quantum fluctuations may turn  these classical spin liquids into exotic QSLs, as suggested by some numerical results~\cite{niggemannQuantumEffectsUnconventional2023,lozano-gomezCompetingGaugeFields2024}. 
However, to study these hypothetical QSLs in a controlled manner, one would generally hope to construct a map from the microscopic spin model to an appropriate lattice gauge theory, as has been done in the case of  quantum spin ice~\cite{hermelePyrochlorePhotonsU12004,savaryCoulombicQuantumLiquids2012,gingrasQuantumSpinIce2014}.
Unfortunately such a construction starting from the microscopic level has thus far not been achieved for these tensor spin liquid  models on the pyrochlore lattice, and it remains desirable to find exotic tensor spin liquids for which such a controlled analysis can be carried out.

In each of the pyrochlore spin liquids studied thus far, violations of the local constraints may be viewed as charges localized at the center of a tetrahedron.
Even in the ``exotic'' tensor spin liquids, the excitations are zero-dimensional charged quasiparticles coupled to gauge fields by emergent Gauss laws, i.e. the charge density acts as the source of an appropriate divergence of the gauge field~\cite{yanClassificationClassicalSpin2024a}.  
This prompts one to ask whether there are spin liquid phases falling beyond the framework of point-like quasiparticles coupled by a Gauss law to the divergence of a gauge field. 
Upon brief thought, taking inspiration from Maxwell electromagnetism, we are presented with another possibility: replacing the Gauss law with an emergent Ampere-like law. 
This idea might seem peculiar at first glance---the Gauss law is a constraint enforced by gauge symmetry, while the Ampere's law is a dynamical consequence of the equations of motion.
However, from a field theory point of view, this does not prevent us from asking a simple question: can a three-dimensional classical spin liquid exist described by an emergent coarse-grained vector field ${\bm E}$ with a suppressed \emph{curl}, ${\nabla\times\bm{E}=0}$, and if so, what happens when one adds quantum fluctuations to it?
Such a spin liquid would naturally have \emph{vortex} excitations---string-like ``currents'' acting as sources of the \emph{vorticity} of the emergent field, rather than point-like charges.

Motivated by this simple question, and guided by the intuition leveraged from the physics of NNSI, we introduce in this paper (\cref{sec:spin_vorticity_model}) a classical Ising model on the pyrochlore lattice, which is explicitly constructed to enforce a ``zero-curl'' constraint on every \emph{hexagonal plaquette} of the lattice at zero temperature, rather than enforcing a zero-divergence constraint on every tetrahedron.
We dub this the ``spin vorticity model'' since its fundamental excitations are violations of a zero-curl constraint, i.e. sources of vorticity.
We study this classical model both analytically and numerically to verify that it hosts a classical spin liquid ground state, characterized by the presence of flat bands in the Hamiltonian interaction matrix, an extensive zero-temperature entropy, and singularities in the spin-spin correlation functions.
Interestingly, Monte Carlo simulations find that this model exhibits an unusual weak symmetry-breaking transition at finite temperature, at which a very small but extensive fraction of the system appears to develop long range order. 
Nevertheless, we find that the ground state entropy remains extensive, remarkably close to that of NNSI~\cite{singhCorrectionsPaulingResidual2012}, and with correlation functions showing the distinct pinch point features of a spin liquid~\cite{henleyPowerlawSpinCorrelations2005,isakovDipolarSpinCorrelations2004}.

We elucidate the classical topological order~\cite{castelnovoSpinIceFractionalization2012} (\cref{sec:2-form-classical}) of this spin liquid and find that the ground state manifold of the spin vorticity model can be described as a \emph{condensate of closed membranes}. 
Its fractionalized excitations are 1-dimensional string objects attached to the edges of open membranes.
We thus demonstrate the existence of a novel type of spin liquid that is stabilized by further-neighbor interactions and which has \emph{no point-like quasiparticle excitations}, but instead has extended string objects as minimal excitations.
We then consider the emergent gauge structure of the spin vorticity model, constructing an electrodynamics analogy and associated U(1) gauge field description. 
Doing so exemplifies this spin liquid's novelty: whereas spin ice realizes an emergent Maxwell electrodynamics~\cite{henleyPowerlawSpinCorrelations2005,henleyCoulombPhaseFrustrated2010,castelnovoSpinIceFractionalization2012,moessnerSpinIceCoulomb2021,gingrasQuantumSpinIce2014}, the spin vorticity model realizes the generalization of Maxwell theory to a theory of charged strings, called \emph{2-form electrodynamics}~\cite{henneauxPFormElectrodynamics1986}.
Briefly, a differential $p$-form is a rank-$p$ \emph{antisymmetric} tensor field, an object which can be naturally integrated over $p$-dimensional surfaces (thus, roughly, a $p$-form is a ``$p$-dimensional density'').
In electrodynamics, the prototypical U(1) gauge theory, the vector potential is a 1-form whose line integral along the 1-dimensional worldline of a point charge encodes the rotation of the internal U(1) phase of the charge in the presence of the gauge field. 
In $p$-form electrodynamics~\cite{henneauxPFormElectrodynamics1986}, a U(1) $p$-form gauge field encodes the change in internal phase carried by a $(p-1)$-dimensional charged object as it traces out a $p$-dimensional ``worldsheet'' in spacetime. 
Spin liquids are most often characterized by emergent 1-form gauge fields which encode the statistics and interactions of their fractionalized 0-dimensional charged quasiparticles~\cite{savaryQuantumSpinLiquids2016}.
In contrast, the membrane condensate of the spin vorticity model realizes 2-form U(1) electrodynamics, where charged string excitations couple to an emergent 2-form U(1) gauge field. 

With a clear view of the 2-form gauge structure of the classical spin vorticity model in hand, we begin to lay the groundwork for the realization of a hypothetical novel phase of quantum matter: a \emph{2-form U(1) QSL} (\cref{sec:QSV}).
By perturbing the classical model with quantum (non-Ising) exchange interactions, we derive at using degenerate perturbation theory a \emph{membrane exchange} model which describes quantum tunnelling between the classical ground states, a direct analog of the ring exchange model of QSI~\cite{hermelePyrochlorePhotonsU12004}. 
We demonstrate that this model maps directly to a (frustrated) 2-form U(1) lattice gauge theory. 
The proposed 2-form U(1) QSL then corresponds to the deconfined phase of this gauge theory. 
Our primary aim in this work is to elucidate the qualitative gauge-theoretic features of this phase, while direct verification of its stability, or lack thereof,  will require significant future work.
Toward that end, we demonstrate how to embed the spin Hilbert space within the larger Hilbert space of a Higgs-like model by augmenting the system with a 1-form ``string field'' which is slaved to the 2-form gauge field by a generalized Gauss law when restricting to the physical spin Hilbert space.
This procedure then allows, in principle, for a gauge mean field theory treatment~\cite{savaryCoulombicQuantumLiquids2012,leeGenericQuantumSpin2012,savarySpinLiquidRegimes2013,savaryQuantumCoherenceQuantum2021}. 
We discuss the stability of this putative 2-form U(1) QSL, whose fate is ultimately dependent on the effects of the same sorts of magnetic instantons that destabilize 1-form U(1) QSLs in two spatial dimensions~\cite{polyakovQuarkConfinementTopology1977}. 
We conjecture that a 2-form U(1) QSL could be stabilized in three spatial dimensions analogous to the two-dimensional Dirac U(1) spin liquid~\cite{hermeleStabilityU1Spin2004,hermeleAlgebraicSpinLiquid2005}, whose stability (despite instantons) is strongly supported by renormalization~\cite{songUnifyingDescriptionCompeting2019}, symmetry~\cite{songSpinonBandTopology2020}, and numerical studies~\cite{wietekQuantumElectrodynamics2+12024,kiesePinchpointsHalfmoonsStars2023}. 
In summary, we establish the existence of a 2-form classical spin liquid and lay the ground work for the detailed study of a novel phase of quantum magnetic matter: 2-form U(1) QSLs. 
We briefly discuss in \cref{sec:experimental} the prospects of experimentally realizing 2-form spin liquids such as theoretically proposed in this work.
We end the paper with a discussion and general conclusions in Section \cref{sec:discussion}.

\section{Preliminaries}
\label{sec:NNSI}
To set the stage for the spin vorticiy model introduced below, we begin by briefly reviewing the physics of NNSI and its emergent gauge structure, which will serve as a crucial point of contrast throughout the rest of this paper. 
We then discuss its coarse-grained description to motivate an alternative mechanism for spin liquidity.

In NNSI, $N_{\text{spin}}$ Ising spins are arranged on the sites of the pyrochlore lattice of corner-sharing tetrahedra, shown in \cref{fig:pyro_diamond}(a). 
The centers of the tetrahedra (light blue spheres) form a bipartite diamond lattice, and each pyrochlore site (small black sphere) is located at the midpoint of a bond of this diamond lattice (thick black lines).
The Ising spins (red arrows) are constrained to point in one of the two directions along the diamond bond, corresponding to the eigenstates of the spin-1/2 operator $\cramped{S^z_i \equiv \bm{S}_i\cdot\hat{\bm{z}}_i \equiv (\hbar/2)\sigma_i^z}$, where $\hat{\bm{z}}_i$ is the local quantization (cubic $[111]$ crystalline) axis at site $i$ and $\sigma_i^z$ is the Pauli-$z$ operator with eigenvalues $\pm 1$~\cite{rauFrustratedQuantumRareEarth2019}.
We take the convention that the quantization axes point from the $A$ to $B$ diamond sublattices, represented in \cref{fig:pyro_diamond}(b,c) as black arrowheads at the ends of the diamond bonds (\cref{apx:conventions}).

\subsection{The Gauge Structure of NNSI}
\label{sec:gauge_structure_NNSI}

The NNSI Hamiltonian reads
\begin{equation}
    H_{\text{NNSI}} = J\sum_{\langle ij\rangle} \sigma_i^z \sigma_j^z, 
    \label{eq:H_NNSI_spin}
\end{equation}
where the sum is over nearest-neighbor sites of the pyrochlore lattice and $J>0$ sets the (antiferromagnetic) exchange energy scale. The essential physics of this model is exposed by defining a local lattice divergence~\cite{hermelePyrochlorePhotonsU12004,henleyCoulombPhaseFrustrated2010,lhotelFragmentationFrustratedMagnets2020} at each tetrahedron center (i.e. diamond vertex)~$t$,
\begin{equation}
    Q_t \equiv \frac{1}{2}(-1)^t (\sigma_1^z + \sigma_2^z + \sigma_3^z + \sigma_4^z)_t,
    \label{eq:Q_t}
\end{equation}
where $(-1)^t$ is positive (negative) on the $A$ ($B)$ diamond sublattices so that the divergence is positive, $\cramped{Q_t>0}$, when the majority of spins point ``out'' from the vertex $t$, such as illustrated in \cref{fig:pyro_diamond}(b) on the top right tetrahedron carrying labels 1,2,3,4. 
By construction, $Q_t$ takes values $0, \pm 1, \pm 2$, where the $+1$ ($-1$) configuration is illustrated in \cref{fig:pyro_diamond}(b,c) on the tetrahedron carrying a white (black) sphere.
In terms of $Q_t$, \cref{eq:H_NNSI_spin} takes the form
\begin{equation}
    H_{\text{NNSI}} = E_0 + 2J \sum_t Q_t^2,
    \label{eq:H_NNSI_charge}
\end{equation}
where the sum is over all tetrahedra (diamond vertices) $t$, and $E_0 = -J N_{\text{spin}}$ is the ground state energy. 
Ground states must satisfy the local zero-divergence constraint $Q_t = 0$ everywhere (two spins ``in'' and two ``out'' of each tetrahedron, the ``ice rules''~\cite{bramwellSpinIceState2001}). 
\Cref{fig:pyro_diamond}(a) shows a portion of such a 2-in-2-out ground state spin configuration, with spins denoted by red arrows.
In these ground states, spins may be viewed as aligning head-to-tail to form \emph{closed strings}~\cite{castelnovoSpinIceFractionalization2012,moessnerSpinIceCoulomb2021}.  
Flipping such a closed string of spins costs zero energy because it does not change $Q_t$ anywhere.
The number of such degenerate ground states is exponentially large in the system size and, consequently, NNSI has an extensive zero temperature entropy that is approximately equal to $N_{\text{spin}} S_{\text{P}}$,  where  $S_{\text{P}} = (k_{\text{B}}/2) \ln(3/2)$ is the  Pauling estimate for the residual zero-temperature entropy (per proton) 
for water ice---hence the name spin ice ~\cite{ramirezZeropointEntropySpin1999,melkoMonteCarloStudies2004,moessnerSpinIceCoulomb2021,singhCorrectionsPaulingResidual2012,bramwellSpinIceState2001}.

Minimal energy excitations are created by flipping a single spin,
which produces a pair of defective tetrahedra with opposite divergence $Q_t = \pm 1$.
More generally, flipping an open chain of head-to-tail spins, such as the five spins highlighted in green in \cref{fig:pyro_diamond}(a), creates a pair of defective tetrahedra located at the two ends of the chain, as shown in \cref{fig:pyro_diamond}(b).  
If we take the view that \cref{eq:Q_t} is a Gauss law on the lattice~\cite{hermelePyrochlorePhotonsU12004,savaryCoulombicQuantumLiquids2012,henleyCoulombPhaseFrustrated2010} then we may view these excitations as fictional 0-dimensional point charges which sit on the vertices of the diamond lattice, at the centers of the tetrahedra (large black and white spheres 
in Fig.~\ref{fig:pyro_diamond}(b,c)),
which act as sources and sinks of the divergence of the spin configuration.
These two charges appear to be connected by an \emph{open string} of head-to-tail spins, shown in yellow in \cref{fig:pyro_diamond}(b). 
The spin ice literature commonly describes these excitations as ``monopole-antimonopole'' pairs connected by ``Dirac strings'', since they appear as local sources of the  magnetization density~\cite{castelnovoMagneticMonopolesSpin2008}. 
We shall utilize the dual perspective, more natural for the gauge theory description, viewing them as electric charge pairs connected by a 1-dimensional string of electric flux~\cite{hermelePyrochlorePhotonsU12004,gingrasQuantumSpinIce2014}.

Crucially, the positions of these closed and open electric strings are not physically observable since the decomposition of each spin configuration into a collection of strings is not unique~\cite{castelnovoSpinIceFractionalization2012}.
The open (yellow) string shown in \cref{fig:pyro_diamond}(b) is only defined relative to the ground state spin configuration in \cref{fig:pyro_diamond}(a), and there are many possible strings one can identify connecting a given pair of charges. 
Having chosen one such string, it can be deformed by flipping an ``attached'' closed loop of spins~\cite{castelnovoSpinIceFractionalization2012,moessnerSpinIceCoulomb2021}. 
This is demonstrated in \cref{fig:pyro_diamond}(c), where the closed (green) string of six spins in \cref{fig:pyro_diamond}(b) has been flipped, thereby deforming the open (yellow) string connecting the two charges.
In a sense, the string between two charges is delocalized throughout a fluctuating vacuum of background ground state configurations, analogous to how the field of a point charge spreads over all space.\footnote{In the dual magnetic perspective, this is also equivalent to how the position of a Dirac string connected to a magnetic monopole is moved by a gauge transformation and is thus unobservable~\cite{castelnovoSpinIceFractionalization2012,moessnerSpinIceCoulomb2021}. 
}
This unobservability of the individual string positions characterizes the emergent gauge structure of the NNSI spin liquid as a \emph{string condensate}~\cite{castelnovoSpinIceFractionalization2012,moessnerSpinIceCoulomb2021,levinStringnetCondensationPhysical2005}.

\subsection{A Different Kind of Spin Liquid}

We now proceed to propose a new type of spin liquid lying beyond the description of charged zero-dimensional (point-like) quasiparticles connected by one-dimensional electric strings. 
To motivate its construction, let us recall that the long-wavelength description of the NNSI spin liquid is a Coulomb phase~\cite{henleyCoulombPhaseFrustrated2010}, described by a coarse-grained vector field $\bm{E}$ governed by a free energy of the form~\cite{isakovDipolarSpinCorrelations2004,henleyPowerlawSpinCorrelations2005,conlonAbsentPinchPoints2010,henleyCoulombPhaseFrustrated2010,castelnovoSpinIceFractionalization2012,moessnerSpinIceCoulomb2021}
\begin{equation}
    F = \int \mathrm{d}^3r \; \big ( \kappa \vert \bm{E}\vert^2 + \xi^2 \vert \nabla\cdot\bm{E}\vert^2 \big ) ,
    \label{eq:S_NNSI}
\end{equation}
where  $\xi$ is a temperature-dependent lengthscale (with the density of spinon defects $\rho\propto \xi^{-3}$~\cite{henleyCoulombPhaseFrustrated2010}). The quantity $\kappa$ is a stiffness parameter, which arises entropically (rather than energetically) from the fact that the spin length is bounded and thus the field strength cannot grow arbitrarily 
large~\cite{henleyPowerlawSpinCorrelations2005,henleyCoulombPhaseFrustrated2010,conlonAbsentPinchPoints2010,castelnovoSpinIceFractionalization2012}.
The Coulomb phase is obtained at zero temperature, at which point $\xi \to \infty$ and the divergence of $\bm{E}$ is fully suppressed~\cite{henleyCoulombPhaseFrustrated2010,moessnerSpinIceCoulomb2021}. The correlation functions $\cramped{\langle E^\alpha(\bm{r})E^\beta(\bm{0})\rangle}$ then decay algebraically with the same  $\cramped{1/r^3}$ form as that of a physical dipole-dipole interaction~\cite{isakovDipolarSpinCorrelations2004,henleyPowerlawSpinCorrelations2005,conlonAbsentPinchPoints2010,henleyCoulombPhaseFrustrated2010,moessnerSpinIceCoulomb2021}.

The vector field $\bm{E}$ can be decomposed into three components, known as the Helmholtz decomposition, with
\begin{equation}
    \bm{E} = \nabla \phi + \nabla\times\bm{A} + \bm{E}_0,
    \label{eq:Helmholtz}
\end{equation}
where $\phi$ is a scalar function, $\bm{A}$ is another vector field, and $\bm{E}_0$ satisfies Laplace's equation. The three terms in \cref{eq:Helmholtz} may be referred to as the irrotational (zero curl), rotational (zero divergence), and harmonic (both curl-free and divergence-free) components of $\bm{E}$, respectively.
In the Coulomb phase, $\bm{E}$ has vanishing irrotational component ($\cramped{\nabla\phi = 0}$), while the fluctuations of the rotational and harmonic degrees of freedom, each satisfying $\cramped{\nabla\cdot\bm{E} = 0}$, are controlled by the Gaussian stiffness term in \cref{eq:S_NNSI},~$\cramped{\kappa \vert \bm{E}\vert^2}$. 
One may heuristically associate the irrotational (divergence-full) component of $\bm{E}$ as sourced by the $\cramped{Q_t\neq 0}$ defects via a Gauss law, while the Gaussian-distributed rotational and harmonic component correspond to the degenerate ice manifold of closed-string configurations~\cite{lhotelFragmentationFrustratedMagnets2020}, with the harmonic component associated to strings that wind around the periodic boundaries~\cite{bramwellHarmonicPhasePolar2017}.

At this juncture, with \cref{eq:S_NNSI,eq:Helmholtz} in mind, one may naturally wonder whether a microscopic lattice spin model analogous to NNSI exists for which the coarse-grained description takes the converse form to \cref{eq:S_NNSI},
\begin{equation}
   \tilde{F} \sim \int \mathrm{d}^3r 
   \;
   \big (
   \tilde{\kappa} \vert \bm{E}\vert^2 + \tilde{\xi}^2 \vert \nabla\times\bm{E}\vert^2
   \big ).
    \label{eq:S_vort}
\end{equation}
If such a microscopic spin lattice model were to exist, it would na\"{i}vely have a degenerate ground state manifold consisting of all configurations with zero \emph{rotational} component of the coarse-grained field $\bm{E}$. 
The excitations would be sources of the curl of $\bm{E}$ via an emergent Ampere-like law which, in Maxwell electrodynamics, would be a current-carrying wire---a \emph{string-like} object---rather than the point-charge sources of the divergence of $\bm{E}$ in spin ice.

As we will see in much greater detail in \cref{sec:2-form-classical,sec:QSV}, such a coarse-grained theory can be naturally mathematically formulated in terms of an anti-symmetric rank-2 tensor field, called a 2-form, as alluded to in the Introduction.
We can see that the coarse-grained theory in \cref{eq:S_vort} can be expressed in terms of a tensor field by writing the curl in components as $(\nabla \times \bm{E})_i = \epsilon_{ijk}\partial_j E_k$, where $\epsilon_{ijk}$ is the fully anti-symmetric Levi-Civita symbol. 
The zero-curl condition enforced at zero temperature can then be expressed as a Gauss law $\partial_j \tilde{E}_{ij}=0$, where $\tilde{E}_{ij} = \epsilon_{ijk}E_k$ is a rank-2 anti-symmetric tensor, i.e. a 2-form. 
The charge density of this theory, the violation of the zero-curl condition, carries a vector index, defined by the Gauss law $\partial_j \tilde{E}_{ij} \equiv \mathscr{j}_i$. 
It automatically satisfies $\partial_i \mathscr{j}_i = 0$ since $\tilde{E}$ is anti-symmetric, meaning that the ``field lines'' of the vector field $\bm{\mathscr{j}}$ do not end, i.e. the excitations are \emph{closed} strings. 
Intuitively, they may be viewed as vortex strings which source the vorticity of $\bm{E}$.

\begin{figure}
    \centering
    \includegraphics[width=0.9\columnwidth]{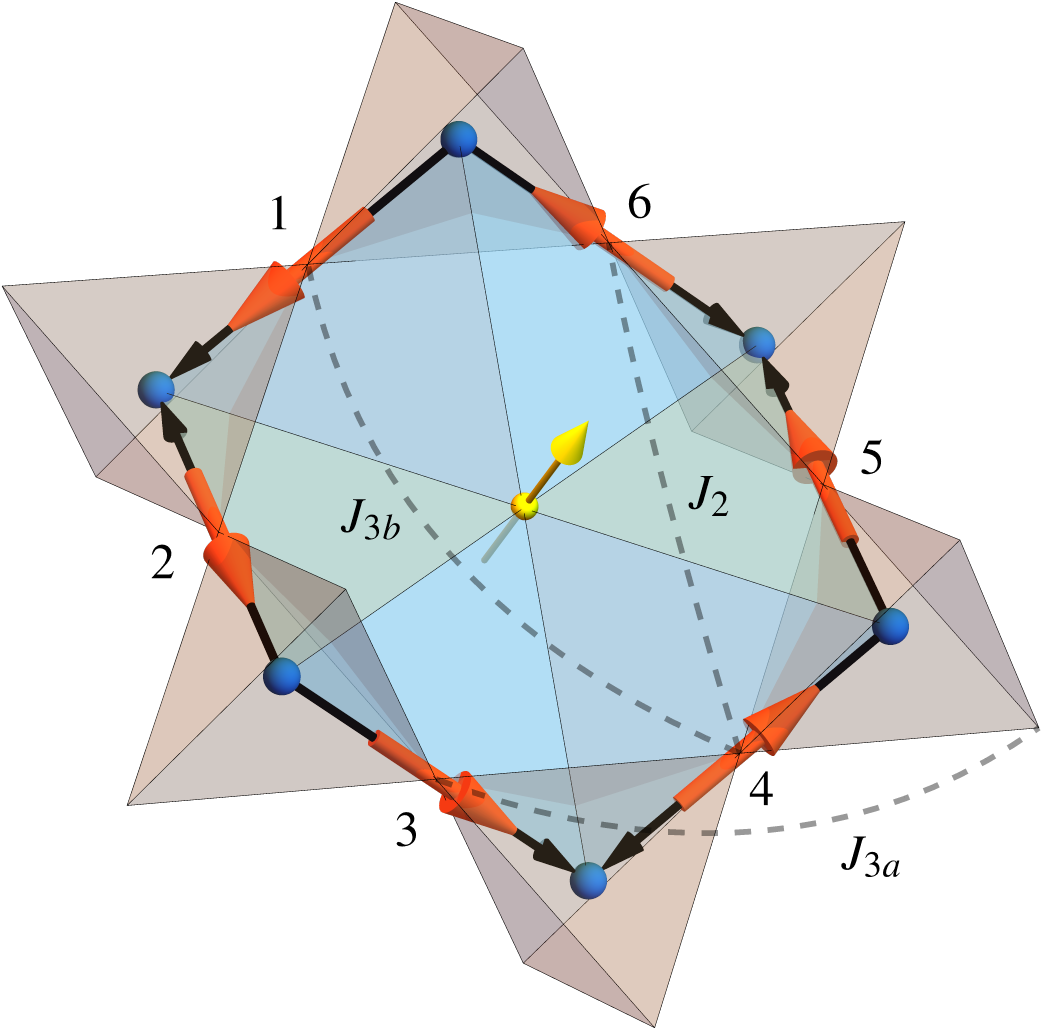}
    \caption{A single hexagonal plaquette of the diamond lattice and corresponding hexagonal arrangement of tetrahedra in the pyrochlore lattice. 
    Again, as in Fig.~\ref{fig:pyro_diamond}(b), the $A$ to $B$ orientation of the diamond bonds is indicated with black arrowheads.
    A spin configuration (six red arrows) with vorticity $\omega_p=+3$ (\cref{eq:omega_p}) with respect to the orientation indicated by the yellow arrow is shown.
    Second neighbors are indicated by $J_2$ and the two types of third neighbors are indicated by $J_{3a}$ and $J_{3b}$, where the $J_{3b}$ neighbors sit on opposite sides of the hexagonal plaquette. }
    \label{fig:hexagon}
\end{figure}

\section{Classical Spin Vorticity Model}\label{sec:spin_vorticity_model}

In this section, we put forward a family of microscopic spin models that realizes the physics suggested by the hypothetical coarse-grained free energy of \cref{eq:S_vort}. 
Guided by the construction of the microscopic NNSI model in \cref{eq:H_NNSI_spin,eq:Q_t,eq:H_NNSI_charge} which leads to the coarse-grained form \cref{eq:S_NNSI}, we consider reverse-engineering this process to obtain~\cref{eq:S_vort}.
As such, we aim to replace the lattice divergence $Q_t$ with an appropriate definition of lattice \emph{curl}, such that $\nabla\cdot\bm{E}$ in \cref{eq:S_NNSI} is replaced with $\nabla\times\bm{E}$ in \cref{eq:S_vort}.
To do so, we start from a line graph lattice, where spins sit on the edges of a parent lattice, for example pyrochlore spins sit on the edges of the diamond lattice, c.f. \cref{fig:pyro_diamond}.
We define the \emph{vorticity} $\omega_p$ as a signed sum of spins surrounding each plaquette $p$ of the parent lattice, defined to be maximal when the Ising spins are uniformly circulating around the edges of the plaquette.
We then define a \emph{spin vorticity model}, by replacing $Q_t$ with $\omega_p$ in \cref{eq:H_NNSI_charge}, i.e.
\begin{equation}
    H_{\text{SV}} 
    =  
    \tilde{E}_0 
     + 
    J\sum_p \omega_p^2  ,
    \label{eq:H_vort_wp}
\end{equation}
where the sum is now over all plaquettes~$p$.
We note that, just as spin ice models can be defined on a variety of lattices, spin vorticity models can as well.

\subsection{The Pyrochlore Spin Vorticity Model}

In this paper we will study in detail one example of a spin vorticity model, on the pyrochlore lattice.
This will allow us to juxtapose its properties with that of the well-known physics of spin ice~\cite{castelnovoSpinIceFractionalization2012,moessnerSpinIceCoulomb2021}.
The vorticity $\omega_p$ is defined on each hexagonal plaquette $p$ of the diamond lattice (buckled blue hexagon in  \cref{fig:hexagon}) as the circulation of the spins around the plaquette boundary,
\begin{equation}
    \omega_p = \frac{1}{2}(-1)^p(\sigma_1^z - \sigma_2^z + \sigma_3^z - \sigma_4^z + \sigma_5^z - \sigma_6^z)_p.
    \label{eq:omega_p}
\end{equation}
Here, the spins are indexed sequentially around the perimeter of the plaquette as shown in \cref{fig:hexagon}. 
The alternating signs account for the alternating orientations of the quantization axes $\hat{\bm{z}}_i$ (small black arrowheads), so that $\omega_p$ is maximum when the six spins (red arrows) are oriented head-to-tail around the hexagon, as shown. The sign $(-1)^p$ depends on the chosen orientation of the plaquette (yellow plaquette normal arrow), so that $\omega_p$ is positive when the circulation has the same sense as the plaquette orientation by the right hand rule (analogous to the sign $(-1)^t$ in \cref{eq:Q_t}).\footnote{For concreteness and to eliminate any ambiguity, we can fix all signs by choosing a unit normal  vector $\hat{\bm{n}}_p$ for every plaquette $p$ (i.e. the yellow vector in \cref{fig:hexagon}), with which the vorticity is defined as
    \[
        \omega_p = \sum_{i \in \boundary p} \mathrm{sign}[\hat{\bm{n}}_p \cdot (\hat{\bm{r}}_{i,p}\times \hat{\bm{z}}_i)] \sigma_i^z,
    \]
    where the sum is over all spins on the boundary of plaquette $p$ (denoted $\boundary p$). 
    Here, $\hat{\bm{r}}_{i,p}$ is the unit vector from the plaquette center to the spin position, and $\hat{\bm{z}}_i$ is the local site-dependent quantization axis (black arrowheads in \cref{fig:hexagon}).
}
By construction, $\omega_p$ takes values $0,\pm 1, \pm 2, \pm 3$, where the $+3$ configuration is illustrated in \cref{fig:hexagon}.

Plugging \cref{eq:omega_p} into \cref{eq:H_vort_wp} and expanding $\omega_p^2$, we obtain bilinear spin-spin interactions between all spins around each hexagon, i.e. first, second, and third-type-$b$ neighbors indicated in \cref{fig:hexagon}~\cite{conlonAbsentPinchPoints2010,heneliusRefrustrationCompetingOrders2016}. 
The classical ground state energy $\cramped{\tilde{E}_0 = -(3/2) J N_{\text{spin}}}$ has been added to cancel all single-site terms $\cramped{(\sigma_i^z)^2 = 1}$. 
The spin vorticity Hamiltonian can then be written as 
\begin{equation}
    H_{\text{SV}} = 
    -J\sum_{\mathclap{\langle ij \rangle}} \sigma_i^z\sigma_j^z 
    + 
    \frac{J}{2}\sum_{
    \langle ij \rangle_{\,\mathrlap{{2}}}
    } \sigma_i^z\sigma_j^z
    - 
    \frac{J}{2} \sum_{
    \langle ij \rangle_{\,\mathrlap{3b}}}\sigma_i^z\sigma_j^z.
    \label{eq:H_vort_spin}
\end{equation}
The reason the nearest-neighbor exchange energy is twice that of the further-neighbors is that every nearest-neighbor pair of spins belongs to two plaquettes (\cref{apx:SVM_Hamiltonian}). 
By construction, ground states of this Hamiltonian (with $\cramped{J>0}$) must satisfy the ``zero-curl'' condition $\cramped{\omega_p = 0}$ on every plaquette, just as NNSI ground states obey the zero-divergence condition $\cramped{Q_t = 0}$ at every diamond lattice vertex. 
Note that the nearest-neighbor exchange (first term in \cref{eq:H_vort_spin}) is \emph{ferromagnetic} with $J>0$, implying that this model is \emph{not} adiabatically connected to NNSI. 
To demonstrate that this model indeed hosts a stable spin liquid, we next turn to analyzing its thermodynamic properties, before studying its emergent gauge structure.

\begin{figure}
    \centering
    \begin{minipage}{\columnwidth}
        \hspace{7mm}
        \begin{overpic}[width=.32\textwidth]{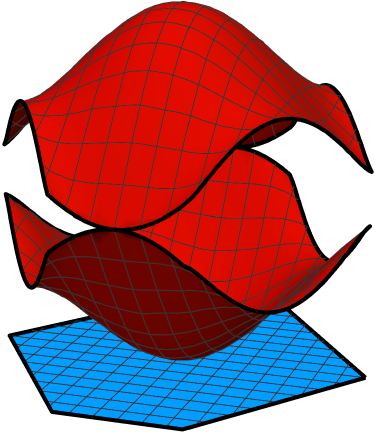}
            \put(22,-6){$X$}
            \put(40,-8){$U$}
            \put(60,-3){$K$}
            \put(80,3){$L$}
            \put(-22,90){(a)}
            \put(-22,-24){(c)}
            \put(-22,-220){(e)}
            \put(114,90){(b)}
            \put(114,-24){(d)}
            \put(114,-220){(f)}
        \end{overpic}
        \hfill
        \begin{overpic}[width=.36\textwidth]{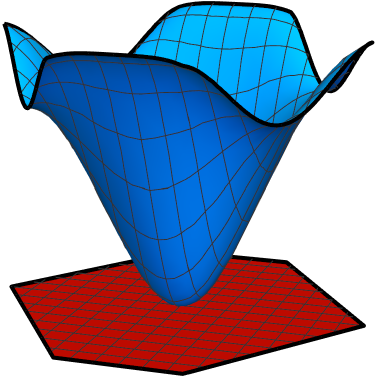}
            \put(26,-6){$X$}
            \put(46,-8){$U$}
            \put(68,-2){$K$}
            \put(91,4){$L$}
        \end{overpic}
        \hspace{2mm}
    \end{minipage}
    \\[2ex]
    \includegraphics[width=\columnwidth]{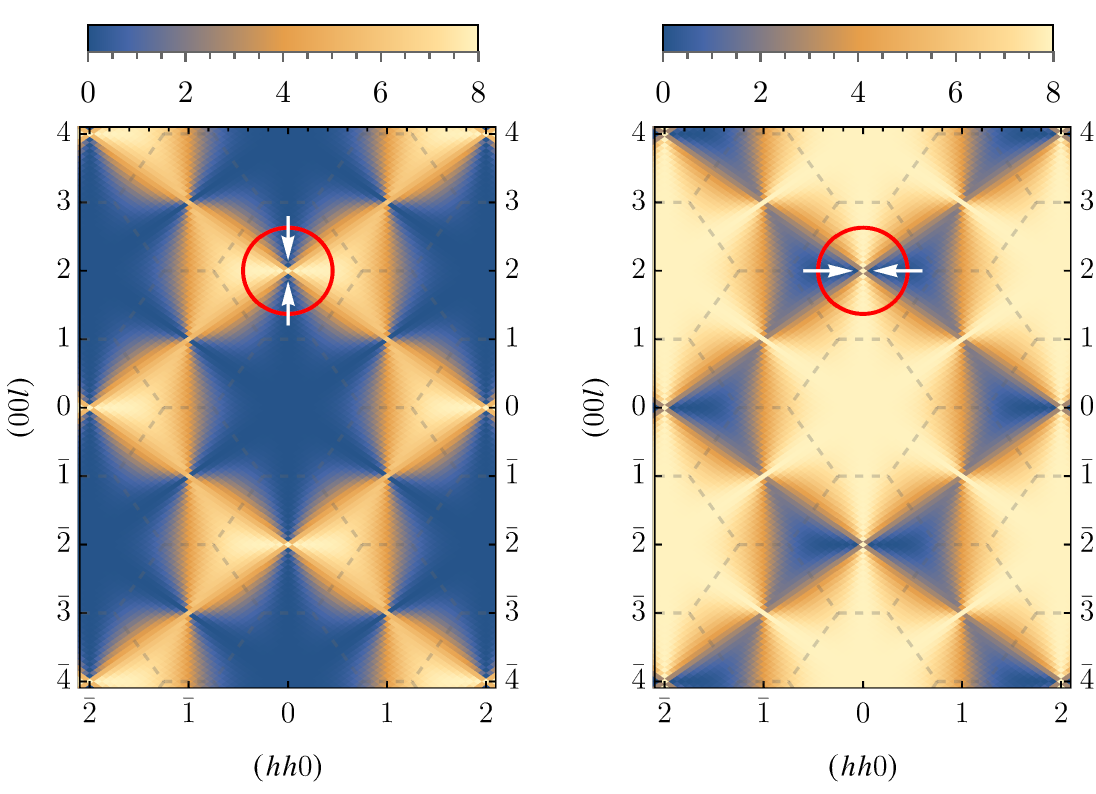}
    \\[1ex]
    \begin{minipage}{\columnwidth}
        \hspace{6mm}
        \begin{overpic}[width=.38\textwidth]{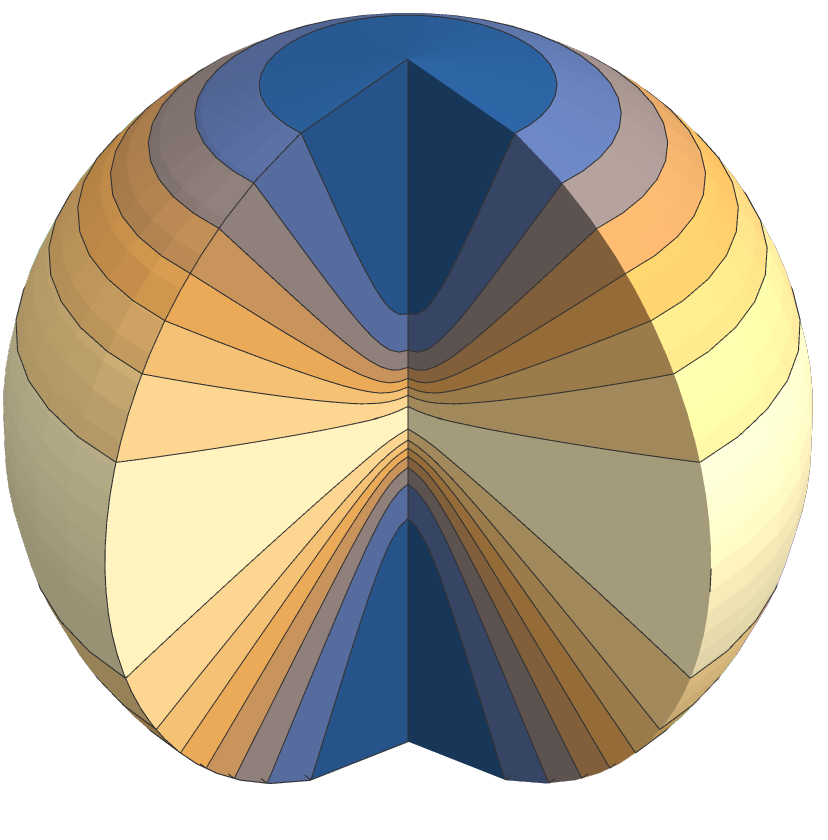}
        \end{overpic}
        \hfill
        \begin{overpic}[width=.38\textwidth]{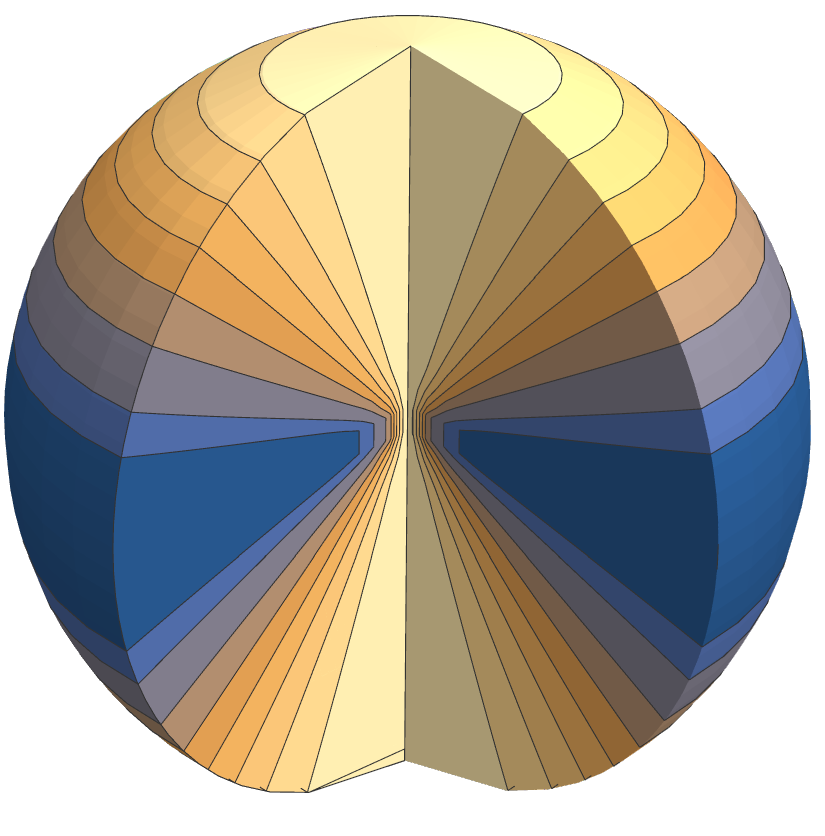}
        \end{overpic}   
        \hspace{.5mm}
    \end{minipage}
    \\[1ex]
    \caption{(a,b) The bands of the interaction matrix for (a) NNSI and (b) the spin vorticity model in the $(hhl)$ plane of reciprocal space. 
    The flat rotational bands of NNSI (blue) correspond to the dispersive bands of the spin vorticity model and the dispersive irrotational bands of NNSI (red) correspond to the flat bands of the spin vorticity model. 
    In both models the total bandwidth is $8J$.  
    (c,d) The spin structure factor in the $(hhl)$ plane, computed in the SCGA for (c) NNSI and (d) the spin vorticity model at $T/J = 0.01$, demonstrating that the correlations of these two models are ``reciprocal'', with high and low intensity regions swapped, corresponding to replacing a suppressed divergence in the NNSI with a suppressed curl in the spin vorticity model.
    (e,f) The pinch point at (002) for (e) NNSI and (f) the spin vorticity model, in the region indicted by the red circles in (c,d), showing that while the correlations in NNSI are pinched \emph{along} $\bq$ (along $(00l)$), the correlations in the spin vorticity model are pinched \emph{transverse to} $\bq$. }
    \label{fig:vort_bands_SCGA}
\end{figure}

\subsection{SCGA Band Analysis}
\label{sec:SCGA}

The self-consistent Gaussian approximation (SCGA) is a standard technique to study paramagnetic phases of spin models whose Hamiltonian takes the form (for Ising spins) $\cramped{H=(1/2)\sum_{ij} \sigma_i^z \Jmat_{ij} \sigma_j^z}$, a method that has proven successful as a starting point for the analysis of spin liquids~ \cite{garaninClassicalSpinLiquid1999,henleyPowerlawSpinCorrelations2005,isakovDipolarSpinCorrelations2004,conlonAbsentPinchPoints2010,lantagne-hurtubiseSpinIceThinFilms2018,plumbContinuumQuantumFluctuations2019,zhangDynamicalStructureFactor2019,chungProbingFlatBand2022}. 
Within the SCGA, the $\sigma_i^z$ are treated as continuous real variables and Lagrange multipliers $\lambda_i (\sigma_i^z)^2$ are added to the Hamiltonian, subject to the average spin length constraints $\cramped{\langle (\sigma_i^z)^2 \rangle = 1}$ (\cref{apx:SCGA}).
The resulting Gaussian theory is entirely controlled by the spectral decomposition of the interaction matrix $\Jmat$, whose eigenvalues form four bands in reciprocal space for pyrochlore Ising models. 
Most classical spin liquids can be associated to a set of flat bands at the bottom of the spectrum of the interaction matrix, with the zero-temperature correlation matrix, $
\cramped{\lim_{T \to 0}\langle \sigma_i^z \sigma_j^z\rangle}$, being proportional to the projector to the space spanned by the eigenvectors of the flat bands~\cite{henleyPowerlawSpinCorrelations2005,chungProbingFlatBand2022,yanClassificationClassicalSpin2024,yanClassificationClassicalSpin2024a} (\cref{apx:SCGA}). 
    
For NNSI, $\Jmat$ has two degenerate flat bands at the bottom of its spectrum~\cite{gingrasOriginSpiniceBehavior2001,isakovDipolarSpinCorrelations2004,isakovWhySpinIce2005,conlonAbsentPinchPoints2010,chungProbingFlatBand2022} with a three-fold band touching point at the zone center, shown in \cref{fig:vort_bands_SCGA}(a) in the high-symmetry $(hhl)$ plane of reciprocal space (\cref{apx:conventions}). 
Performing a long-wavelength expansion of the interaction matrix for NNSI, one obtains an effective theory taking the form of ~\cref{eq:S_NNSI}~\cite{conlonAbsentPinchPoints2010}, demonstrating that the gapless band dispersing quadratically upwards from the band touching point corresponds to the irrotational component of the coarse-grained field $\bm{E}$, while the two dispersionless flat bands correspond to its rotational components (\cref{apx:Long_wavelength_Expansion}).

\Cref{fig:vort_bands_SCGA}(b) shows the bands of the interaction matrix for the spin vorticity Hamiltonian, \cref{eq:H_vort_spin}.
As in NNSI, there are two degenerate flat bands throughout the zone, indicative of a potential spin liquid ground state, and two degenerate dispersive bands and a fourfold band touching point at the zone center. These results already presage a significant contrast between the spin vorticity model and NNSI. 
Performing a long-wavelength expansion of the spin vorticity model interaction matrix $\Jmat$ demonstrates that the two degenerate dispersing bands correspond to the two \emph{rotational} components of the coarse-grained field $\bm{E}$, while the flat bands correspond to the irrotational components (\cref{apx:Long_wavelength_Expansion}).
This confirms that the long-wavelength form of the spin vorticity Hamiltonian \cref{eq:H_vort_spin} has the form of the coarse-grained theory proposed in \cref{eq:S_vort}.
    
The relationship between the NNSI and the pyrochlore spin vorticity model may be understood by observing first that the interaction matrices for these two models commute. 
Therefore they share eigenvectors, and one can continuously interpolate between them, changing the shape of the bands (the eigenvalues) but preserving the distinction of rotational and irrotational bands. 
Doing so (\cref{fig:sm_interpolate} in \cref{apx:interpolation_to_NNSI}) demonstrates that the flat rotational bands of NNSI continuously deform to the dispersive bands of the spin vorticity model
(blue bands in \cref{fig:vort_bands_SCGA}(a,b)), while the dispersive irrotational bands of NNSI deform to the flat bands of the spin vorticity model (red bands). 
In other words, the rotational and irrotational modes exchange roles in the two models---the flat and dispersive bands of the spin vorticity model are irrotational and rotational, respectively. 
Thus, within the SCGA, the correlation matrices of the NNSI and spin vorticity models are projectors to complementary subspaces, and sum to the identity, except at $\bq=\bm{0}$, where the two models share three eigenvectors at the threefold band touching point.

This ``recipocality'' of the spin ice and spin vorticity models is strikingly exposed when considering the low-temperature SCGA spin-spin correlation functions, encoded in the spin structure factor, 
\begin{equation}
    S(\bq) = \frac{1}{L^3}\sum_{ij} \langle \sigma_i^z \sigma_j^z \rangle \,e^{-i\bq\cdot(\bm{r}_j - \bm{r}_i)},
    \label{eq:Sq}
\end{equation}
    where $\bm{r}_i$ is the position of spin $i$ in the lattice and $L^3$ is the number of primitive FCC unit cells, assuming periodic boundary conditions.
\Cref{fig:vort_bands_SCGA}(c) and (d) show $S(\bq)$ for NNSI and the spin vorticity model, respectively, computed using the SCGA. 
They appear ``inverted'' relative to each other: the pinch points in NNSI reflect the thermal extinction of the irrotational component of the coarse-grained field~\cite{henleyCoulombPhaseFrustrated2010,conlonAbsentPinchPoints2010}, while the inverted intensity of the pinches of the spin vorticity model reflects the thermal depopulation of the rotational component of the field. 
This is most clearly seen in the three-dimensional structure of the pinch points,  as shown in \cref{fig:vort_bands_SCGA}(e,f) for $\bq=(002)$. For NNSI, the pinch is \emph{along} $\bq$, i.e. the longitudinal (irrotational) mode is suppressed, while for the spin vorticity model the pinch is \emph{orthogonal} to $\bq$, i.e. the transverse (rotational) modes are suppressed.

\begin{figure*}
    \centering
    \raisebox{.11\height}{
        \begin{overpic}[width=.195\textwidth]{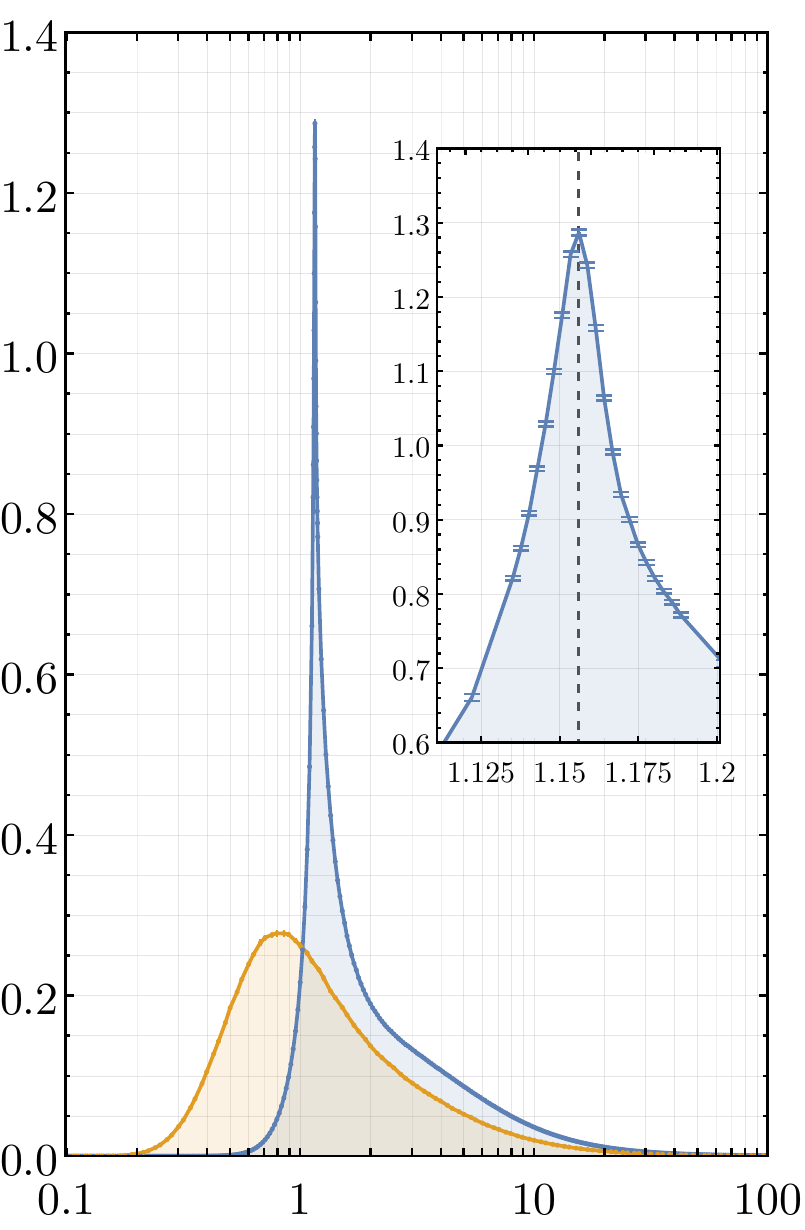}
            \put(6.5, 85){\footnotesize{Vorticity}}
            \put(  8, 25){\footnotesize{NNSI}}
            \put(44,31){\tiny{$T/J$}}
            \put( 28,-6){$T/J$}
            \put(  4,102){Specific Heat Per Spin}
            \put( 29.5,-17){(a)}
            \put(106.5,-17){(b)}
            \put(163.5,-17){(c)}
            \put(220.0,-17){(d)}
            \put(288.5,-17){(e)}
        \end{overpic}
        }
    \begin{overpic}[width=0.56\textwidth]{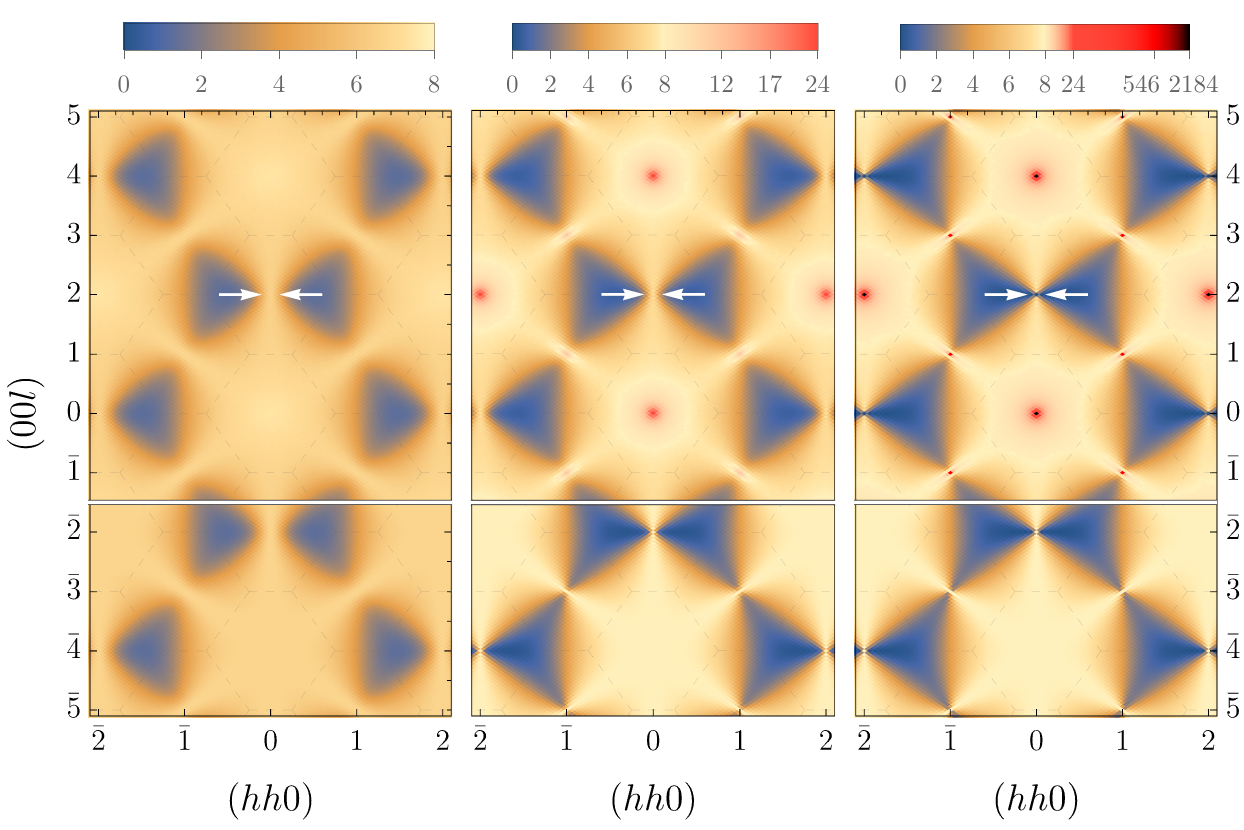}
        \put(16.0,55){\scriptsize{$T/J=2.0$}}
        \put(47.2,55){\scriptsize{$T/J=1.2$}}
        \put(78.6,55){\scriptsize{$T/J=0$}}
        \put(7.9,23.5){\footnotesize{SCGA}}
        \put(7.9,27.5){\footnotesize{MC}}
    \end{overpic}
    \hfill
    \raisebox{.11\height}{
        \begin{overpic}[width=.2\textwidth]{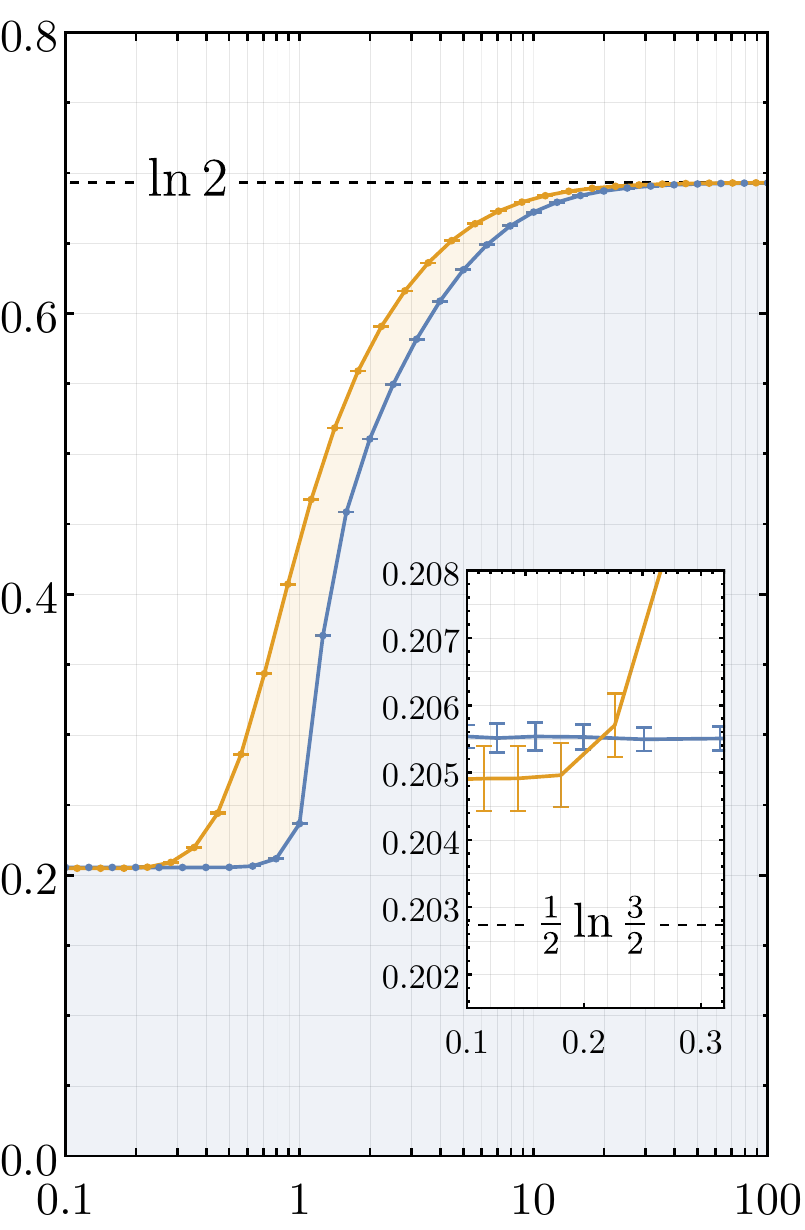}
            \put(11,102){Entropy per Spin}
            \put(35,68){\footnotesize{Vorticity}}
            \put(13,63){\footnotesize{NNSI}}
            \put(44,10){\tiny{$T/J$}}
            \put(28,-6){$T/J$}
        \end{overpic}
        }
    \\[1ex]
    \caption{(a) The specific heat per spin as a function of temperature, $c(T)$, for the spin vorticity (blue) and NNSI (yellow) models measured in Monte Carlo simulations with $\cramped{L=32}$, with all results averaged over $10^4$ sample configurations per temperature. The spin vorticity model shows a sharp anomaly near $\cramped{T_*/J\approx 1.15}$ while NNSI only has a smooth crossover associated to the thermal depopulation of the tetrahedra that violate the  two-in/two-out ice rule ($Q_t=0$).
    (b,c,d) The $\cramped{L=32}$ spin structure factor $S(\bq)$ for the spin vorticity model at (b) $\cramped{T/J=2.0}$, (c) $\cramped{T/J=1.2}$ and (d) $\cramped{T/J=0}$, showing Monte Carlo data in the top panels and SCGA data below at a temperature chosen to match the diffuse intensity. 
    The Monte Carlo correlations are consistent with the SCGA, but with the formation of weak Bragg peaks below $T_*$. 
    The intensity of the peaks in (d) at (000), (111), and (002) are in the ratio $\cramped{(4:1:0)}$, respectively, consistent with all-in-all-out (AIAO) symmetry breaking (\cref{apx:AIAO}). 
    The color scale is linear from 0 to 8 in (b), matching the corresponding intensity in \cref{fig:vort_bands_SCGA}(c,d), and logarithmic from 8 onward, to demonstrate that the Bragg peaks contain a negligible fraction of the total spectral weight.
    No interpolation or symmetrization has been applied to the data.
    Line cuts along high-symmetry paths are reported in~\cref{apx:LineCuts}.
    Note that at $T/J=0$, the pinch point intensity at (002) in the Monte Carlo simulation (top panel of \cref{fig:vorticity_monte_carlo}(d)) is zero to numerical precision, while it is not in the SCGA (bottom panel of \cref{fig:vorticity_monte_carlo}(d)). 
    This difference is due to the lack of topological sector fluctuations in the Monte Carlo simulations using only local updates (\cref{sec:top_sector_ergodicity}).
    (e) The entropy per spin (in units of $k_{\textrm B}$) for the spin vorticity and NNSI models as a function of temperature, obtained by integrating $c(T)/T$, demonstrating that the spin vorticity model has an extensive ground state entropy remarkably close to that of NNSI, 
    which is itself very close to 
    Pauling's estimate of the ground state entropy of water ice, $S_{\textrm{P}}=(k_{\text{B}}/2)\ln(3/2)\approx 0.203$ $k_{\text{B}}$~\cite{bramwellSpinIceState2001, singhCorrectionsPaulingResidual2012}.
    }
    \label{fig:vorticity_monte_carlo}
\end{figure*}

\subsection{Monte Carlo Analysis} 
\label{subsec:MonteCarlo}
The SCGA analysis suggests that the spin vorticity model may indeed host an extensively degenerate ground state and a novel spin liquid phase quite distinct from that of NNSI.  
While the SCGA is known to provide a good qualitative description of NNSI, 
it is only the first term in a large-$N$ expansion~\cite{garanin1ExpansionClassical1996}, where $N$ is the number of spin components. 
It remains to be demonstrated that the putative spin liquid phase implied by the flat bands of the interaction matrix survives in the $N=1$ Ising limit, as it does in NNSI~\cite{isakovDipolarSpinCorrelations2004}.
To do so, we perform classical Monte Carlo simulations of the spin vorticity Hamiltonian, \cref{eq:H_vort_spin}, with $4L^3$ spins on an $\cramped{L\times L\times L}$ FCC lattice with periodic boundaries. 
We also simulate NNSI for comparison using a highly efficient cluster algorithm introduced in Ref.~\cite{otsukaClusterAlgorithmMonte2014}.
Details of our Monte Carlo procedures are provided in \cref{apx:MonteCarlo_sampling}.

In order to equilibrate the system at low temperatures, we introduce a zero-energy cluster move for the spin vorticity model.
In NNSI, a single spin flip out of an ice-rules obeying ground state generates two tetrahedra with $\cramped{Q_t = \pm 1}$, and occurs with a probability exponentially suppressed by a $\exp(-4J/T)$ Boltzmann factor.
The ground state manifold of NNSI can instead be explored by flipping closed loops of spins discussed in \cref{sec:gauge_structure_NNSI}~\cite{melkoMonteCarloStudies2004,jaubertAnalysisFullyPacked2011}. 
The minimal such ``loop move'' consists of flipping six spins  around a hexagon as illustrated in \cref{fig:pyro_diamond}(b,c). 
For the spin vorticity model, an analogous minimal cluster move is to flip four spins in an all-in or all-out tetrahedron, i.e. one with $Q_t = \pm 2$. 
We refer to this as a \emph{star move}, since the four diamond bonds emanating from a single diamond vertex form a ``star''. 
Each hexagonal plaquette of the diamond lattice (\cref{fig:hexagon}) shares either two spins with a given star or none.
For each plaquette~$p$ touching a given flippable star, the two spins contributing to $\omega_p$ are consecutive in \cref{eq:omega_p} and satisfy $\cramped{\sigma_1^z = \sigma_2^z}$, i.e. both spins are in or both spins are out. 
They therefore contribute $\cramped{\sigma_1^z - \sigma_2^z = 0}$ in \cref{eq:omega_p}, so that flipping them both does not change $\omega_p$ for that plaquette. 
Flipping all four spins in a star therefore does not change $\omega_p$ on any plaquette, and thus costs zero energy.
We will discuss these zero energy cost star updates further in \cref{sec:2-form-classical}. 
We utilize both star moves and single spin flips in our Monte Carlo simulations of the spin vorticity model, with the primary results presented in \cref{fig:vorticity_monte_carlo}.

\subsection{Spin Liquid and Weak Symmetry Breaking}

\Cref{fig:vorticity_monte_carlo}(a) shows the specific heat per spin, $c(T)$, as a function of temperature $T$ for both NNSI and the spin vorticity model. In NNSI, the specific heat displays a broad Schottky-like peak, whose primary contribution is from the thermal depopulation of the minimal $\cramped{Q_t = \pm 1}$ excitations with energy cost of $2J$~\cite{ramirezZeropointEntropySpin1999,melkoMonteCarloStudies2004,moessnerSpinIceCoulomb2021,applegateVindicationYb2Ti2O7Model2012,linNonmonotonicResidualEntropy2014}, with a smaller contribution at higher temperatures coming from the depopulation of $\cramped{Q_t = \pm 2}$ excitations with energy $8J$.
In contrast, we observe a significant departure from this behavior in the spin vorticity model in the form of a sharp anomaly at $\cramped{T_{*}/J\sim 1.15}$, strongly suggestive of a finite-temperature phase transition that is not predicted by the SCGA. 
As we will see just below, however, this does not preclude a low-temperature classical spin liquid phase with extensive ground state entropy and power-law decaying correlations.

To investigate in more detail the potential spin liquidity and the nature of this specific heat anomaly, we track the evolution of the spin structure factor $S(\bq)$ (\cref{eq:Sq}) with temperature, shown in \cref{fig:vorticity_monte_carlo}(b-d) with Monte Carlo data in the top panels and the SCGA comparisons below. 
Starting with \cref{fig:vorticity_monte_carlo}(b) at $\cramped{T/J=2.0}$, well above $T_*$, the spin-spin correlations are highly consistent with those computed within the SCGA, showing the formation of pinch point features (indicated by the white arrows) broadened at finite temperature. 
Moving to lower temperature, \cref{fig:vorticity_monte_carlo}(c) shows $S(\bq)$ at $\cramped{T/J=1.2}$, just above~$T_*\sim 1.15 J$. 
The pinch points continue to sharpen, but we begin to see a deviation from the SCGA in the form of Lorentzian peaks (red) forming near the zone center ($\cramped{\bq=\bm{0}}$). 

Going below~$T_*$, we can continue to thermalize all the way to zero temperature using the zero-energy star moves,\footnote{This is so up to topological sector fluctuations which would require non-local updates to generate, discussed in detail in \cref{sec:gs_manifold}.
    } 
shown in \cref{fig:vorticity_monte_carlo}(d). 
Here we observe the formation of Bragg peaks characteristic of long-range order. 
This strongly suggests that the specific heat anomaly at~$T_*$ is a thermodynamic phase transition to a magnetically ordered phases, with the pattern of peak intensities consistent with an all-in-all-out (AIAO) symmetry breaking (\cref{apx:AIAO}). 
However, the Bragg peak intensity is \emph{extremely small}.
Fully saturated AIAO order would yield $\cramped{S(\bm{0}) = 16L^3}$ (setting $\cramped{\sigma_i^z =1}$ in \cref{eq:Sq}), about $\cramped{5.2\times 10^4}$ for the $\cramped{L=32}$ system used to generate \cref{fig:vorticity_monte_carlo}.
The peak intensity seen at $T=0$ of about $\cramped{2.2\times 10^3}$ in panel (d) is only about~$0.4\%$ of full saturation, indicating a weak entropic preference for a sublattice asymmetry.
We have checked the system size scaling of the Bragg peak intensity to verify that this remains the case in the thermodynamic limit (see \cref{apx:AIAO}), corresponding to an AIAO order parameter partially saturated at about 7\% of its maximum value.

This behavior should not be confused with the phenomenon of ``fragmentation'' in spin ice, whereby a rigid crystalline arrangement of $Q_t = \pm 1$ charges is formed below a critical temperature while an extensive number of degrees of freedom continue to fluctuate as a spin liquid~\cite{brooks-bartlettMagneticMomentFragmentationMonopole2014,lhotelFragmentationFrustratedMagnets2020}.
We can rule out the presence of such a rigid charge crystal in the spin vorticity model, which has magnetic Bragg peaks with $25\%$ of the intensity of full AIAO order~\cite{brooks-bartlettMagneticMomentFragmentationMonopole2014}.
Moreover, the diffuse ``tails'' beneath the Bragg peaks in \cref{fig:vorticity_monte_carlo}(d) retain a Lorentzian intensity profile with finite width all the way to $\cramped{T=0}$ (see \cref{apx:LineCuts} for line cuts of  $S(\bq)$).

Together, these observations imply that the phase transition at $T_*$ is not a transition to a ``rigid'' low-entropy phase.
Rather, the system retains predominant liquid-like correlations, macroscopic entropy, and dynamic fluctuations below $T_*$ all the way to zero temperature, coexisting with unsaturated magnetic order.
This conclusion is evident when considering the diffuse intensity away from the Bragg peaks, which contains the vast majority of the spectral weight and remains highly consistent with the SCGA down to $T=0$. 
In other words, while the disconnected correlation functions, $\langle\sigma_i^z \sigma_j^z \rangle$, decay to a non-zero constant at large distances, the connected correlation functions, $\langle\sigma_i^z \sigma_j^z \rangle - \langle \sigma_i^z \rangle \langle \sigma_j^z \rangle $, decay algebraically, with the connected structure factor containing the majority of the spectral weight.
In sum, the structure factor data implies that the low-temperature phase is a spin liquid which exhibits a weak spontaneous breaking of the sublattice inversion symmetry.

\begin{figure*}[ht]
    \centering
   \raisebox{-0.5\height}{
       \begin{overpic}[width=0.29\textwidth]{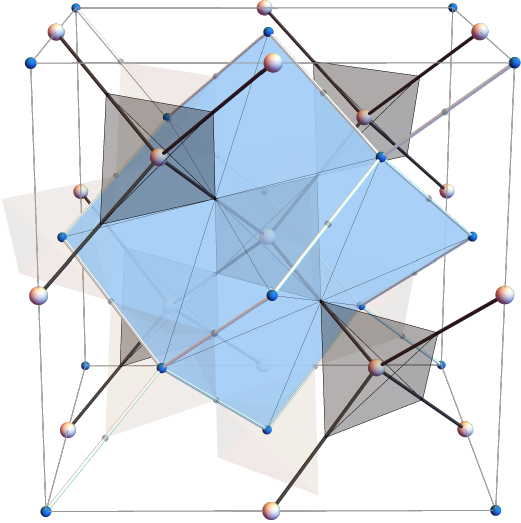}
            \put(048,-13){(a)}
            \put(156,-13){(b)}
            \put(273,-13){(c)}
       \end{overpic}
    }
    \hfill
   \raisebox{-0.5\height}{\begin{overpic}[width=0.29\textwidth]{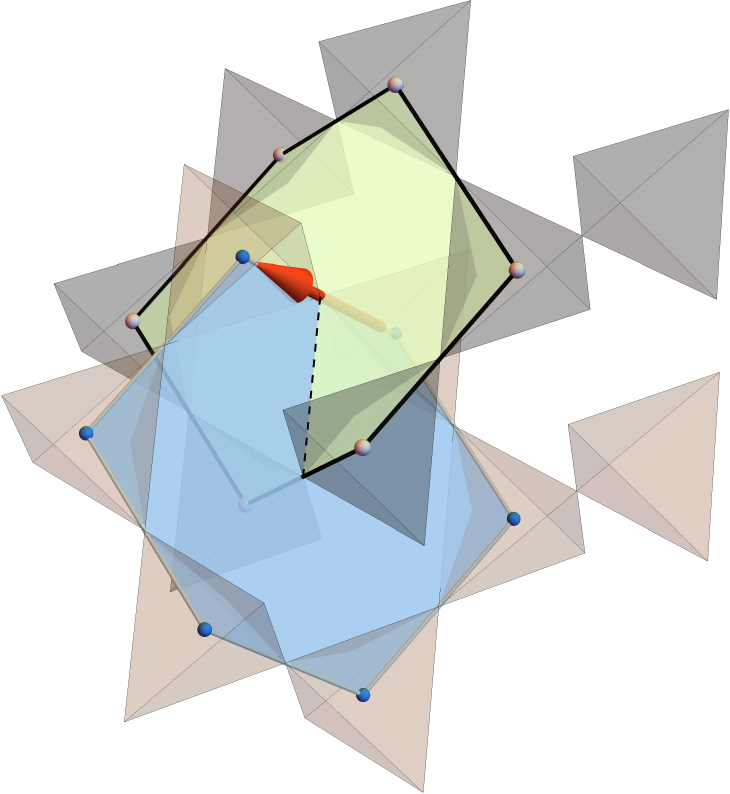}
        \put(82,40){$\mathcal{A}$}
        \put(82,72){$\mathcal{B}$}
   \end{overpic}}
    \hfill
    \raisebox{-0.5\height}{\includegraphics[width=0.37\textwidth]{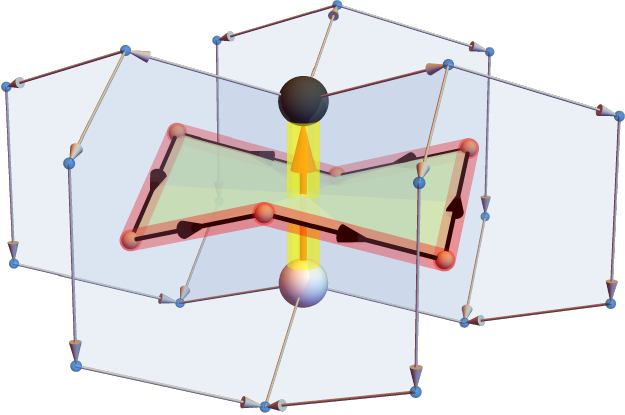}}
    \\[3ex]
    \caption{
        (a) A partial illustration of the relation between the direct and dual diamond lattices. The 
        \emph{direct} diamond lattice is shown with blue vertices (spheres), white edges, blue plaquettes, and a three-dimensional volume encapsulated by four blue plaquettes (two plaquettes in the back are not visible).
        This may be compared to the depiction in \cref{fig:pyro_diamond}(c), where the direct edges were colored black rather than white.
        The \emph{dual} diamond lattice, which was not illustrated in Fig.~\ref{fig:pyro_diamond}, is now shown with white dual vertices (spheres) and black dual edges. 
        Dual plaquettes are not shown.
        The dual diamond lattice has an associated dual pyrochlore lattice, illustrated by dark gray tetrahedra.
        Note that a dual diamond vertex sits exactly in the center of the blue direct lattice volume.
        (b) Illustration of the interpenetrating direct and dual diamond lattices, and their associated direct and dual pyrochlore lattices, labeled $\mathcal{A}$ (direct) and $\mathcal{B}$ (dual). Incidentally, the$\mathcal{A}$ and $\mathcal{B}$ pyrochlore lattices correspond to the $\mathcal{A}_2 \mathcal{B}_2\mathrm{O}_7$ oxide structure~\cite{gardnerMagneticPyrochloreOxides2010}.
        Each edge (plaquette) of the direct diamond lattice is dual to a plaquette (edge) of the dual diamond lattice.  
        Each physical Ising spin (red arrow) is associated to an edge of the direct diamond lattice, and may equivalently be viewed as associated to a plaquette (green) of the dual diamond lattice. 
        (c) The minimal excitation of the spin vorticity model is produced by flipping a single physical spin, which creates six defective plaquettes in the direct diamond lattice (light  blue hexagons) with non-zero vorticity $\omega_p$ (indicated by thin gray arrows). 
        The non-zero vorticity may be viewed as being sourced by a ``current'' loop (a string, red) circulating around the edges of a dual plaquette (green). 
        Thus, in the spin vorticity model, each spin is fractionalized into an oriented string charge in the dual lattice (red), attached to a flux membrane (green). 
        This contrasts with NNSI, where each spin is fractionalized into a pair of charges (black and white spheres) attached by a flux string (yellow).
        One may notice that these are the two fundamental representations of a dipole: a loop of current or a pair of charges.
    }
    \label{fig:diamond_dual_membranes}
\end{figure*}

\subsection{Pinch Points}

It remains to be ascertained whether the pinch points become singular at $\cramped{T=0}$, in \cref{fig:vorticity_monte_carlo}(d), reflecting the algebraic direct-space correlations of a Coulomb phase~\cite{henleyCoulombPhaseFrustrated2010}. 
There are two symmetry-inequivalent pinch points, at the $\{002\}$ and $\{111\}$ positions. 
The width of the pinch points at the  $\{111\}$ positions cannot be readily assessed as they are hidden beneath the Lorentzian tails accompanying the Bragg peaks (red) associated with AIAO order.
On the other hand, the pinch point at the $\{002\}$ position is not covered by a Bragg peak, since this peak is forbidden for AIAO order (\cref{apx:AIAO}). 
The shape of the (002) pinch point found in the Monte Carlo simulations, shown in the top panel of \cref{fig:vorticity_monte_carlo}(d), is highly consistent with the SCGA calculation, shown in the bottom panel. 
This is most clearly seen from line cut comparisons, provided in \cref{apx:LineCuts}, which are in excellent quantitative agreement. 
There is, however, one notable difference between the two sets of results: in the Monte Carlo data, the intensity of the structure factor precisely at $\cramped{\bq=(002)}$ is \emph{exactly} zero.
This can be seen by closely inspecting the top panel of \cref{fig:vorticity_monte_carlo}(d) at the location marked by the two white arrows.
We will return later to clarify the origin of this ``missing'' intensity in \cref{sec:top_sector_ergodicity} after developing an understanding of the emergent gauge structure of the ground state manifold, where we identify it with a lack of topological sector fluctuations at low temperature under the local dynamics of our simulations.

\subsection{Ground State Entropy}

To conclude our discussion of the thermodynamics of the spin vorticity model, we turn finally to estimating 
the ground state degeneracy of the spin vorticity model.
First, we can construct a set of ground states exponentially large in the system size utilizing the zero-energy star moves (flipping four spins in an all-in or all-out tetrahedron). 
Setting every $\sigma_i^z = +1$ (or $-1$) in \cref{eq:omega_p} demonstrates that the two AIAO configurations satisfy $\omega_p = 0$ on every plaquette and are ground states of the spin vorticity model.
Next, note that in these configurations \emph{every} star is flippable.
In particular, all $L^3$ stars on one diamond sublattice may be flipped independently to generate $\smash{2^{L^3}}$ ground states, giving a lower bound on the ground state entropy per spin of $\cramped{k_\text{B}\ln(2)/4\approx 0.173 k_\text{B}}$. 
\Cref{fig:vorticity_monte_carlo}(e) shows the entropy per spin measured in Monte Carlo as a function of temperature for both NNSI and the spin vorticity model, which is computed by integrating the $\cramped{L=32}$ specific heat data from \cref{fig:vorticity_monte_carlo}(a) (see \cref{apx:specific_heat_entropy} for numerical integration details).
We find that the zero-point entropy of the spin vorticity model is approximately $0.2055 k_{\text{B}}$ per spin.
We therefore conclude, based on the extensive ground state entropy, the diffuse paramagnetic correlations, and the pinched singularities in the structure factor, that the pyrochlore spin vorticity model, \cref{eq:H_vort_spin}, does indeed host a classical spin liquid ground state satisfying the local constraints $\omega_p = 0$.

There are two open questions regarding the pyrochlore spin vorticity model, namely the origin and character of the very weak AIAO symmetry breaking transition and the remarkable closeness of the ground state entropy to that of NNSI, which is well approximated by Pauling estimate's of the ground state entropy of water ice, $S_{\textrm{P}}=(k_{\text{B}}/2)\ln(3/2) \approx 0.203$ $k_{\text{B}}$~\cite{bramwellSpinIceState2001, singhCorrectionsPaulingResidual2012}.
For the latter, it seems likely that the transition is a type of order-by-disorder~\cite{henleyOrderingDueDisorder1989} driven by the fact that the AIAO configurations are maximally flippable with respect to the zero-energy star moves. 
For the former, the closeness of the entropy should be related to the fact that both NNSI and the pyrochlore spin vorticity model have two linearly-independent ground state constraints per unit cell, i.e. there are two tetrahedra cell and two linearly-independent hexagonal plaquettes per unit cell~\cite{bergmanBandTouchingRealspace2008}, and correspondingly both models have two flat bands (c.f \cref{fig:vort_bands_SCGA}).
We believe these are nonuniversal lattice-specific issues, which will be addressed in future work.
The central purpose of this paper is to explore the physics of the lattice-independent properties of spin liquidity in spin vorticity models \cref{eq:H_vort_wp}, which we now turn to.

\section{2-Form Classical Spin Liquid}
\label{sec:2-form-classical}

We now proceed to give a microscopic description of the ground state spin liquid, its fractionalized excitations, and the emergent gauge structure of the spin vorticity model.   
To set the stage, and recalling the discussion in \cref{sec:NNSI}, let us quickly summarize how these properties can be deduced in NNSI, following from the local constraint $Q_t=0$ imposed by \cref{eq:H_NNSI_charge} at zero temperature.
Starting from a ground state where the constraint holds, flipping a single spin creates the minimal excitation---a pair of equal and opposite charges on neighboring tetrahedra. 
In other words, each spin is fractionalized into a ``dumbbell'' of charges~\cite{castelnovoMagneticMonopolesSpin2008,castelnovoSpinIceFractionalization2012}.
Treating these dumbbells as ``building blocks'', we imagine inserting them into the lattice one-by-one to construct a ground state. 
In order to ensure that $Q_t =0$ is satisfied, they must be placed in a head-to-tail fashion until they form a closed loop, so that all the charges cancel pairwise. 
This implies that every NNSI ground state can be decomposed (non-uniquely) into a collection of closed strings and, furthermore, that excitations (un-cancelled charges) appear at the ends of open strings (c.f. Fig.~\ref{fig:pyro_diamond}(b,c)).
In this section, we shall parallel this approach to expose the fractionalization and gauge structure of the spin vorticity model.
In order to do so, we first need to introduce the dual diamond lattice, which will play a central role in the rest of this paper.

\subsection{Dual Diamond Lattice}

The physical spin degrees of freedom located on the sites of the pyrochlore lattice may equivalently be viewed as residing at the midpoints of the nearest-neighbor bonds of a diamond lattice.
We call this the \emph{direct} diamond lattice, illustrated in \cref{fig:pyro_diamond} with light blue vertices, thick black edges, and with a single plaquette (buckled hexagon) colored blue. 
Four of these plaquettes together form the surface of a three-dimensional volume of the direct diamond lattice, illustrated in \cref{fig:diamond_dual_membranes}(a), with the direct diamond lattice edges now colored white for visualization convenience.

The dual lattice (or dual cell structure~\cite{hatcherAlgebraicTopology2002}) construction replaces each $p$-dimensional element of the direct lattice (vertex, edge, plaquette, volume) with a $\cramped{(d-p)}$-dimensional element of the dual lattice, where $d$ is the dimension of space.
The dual lattice of the diamond lattice is another diamond lattice.\footnote{It is sometimes erroneously stated that the pyrochlore lattice is dual to the diamond lattice. Rather, the pyrochlore lattice is the ``line graph'' or ``medial lattice'' of the diamond lattice, and the diamond lattice is the ``parent lattice'' or ``premedial lattice'' of the pyrochlore lattice~\cite{henleyCoulombPhaseFrustrated2010}
    }.
This dual diamond lattice is illustrated in \cref{fig:diamond_dual_membranes}(a), with the dual vertices indicated by white spheres and the dual edges colored black.
Considering \cref{fig:diamond_dual_membranes}(b), note that each direct plaquette (a $\cramped{p=2}$-dimensional element, blue buckled hexagon) is pierced by a dual edge ($\cramped{p=1}$, black line). 
Each dual vertex ($\cramped{p=0}$, white sphere) sits at the center of a direct volume ($\cramped{p=3}$, encapsulated by four buckled hexagonal plaquettes). 

Associated to the dual diamond lattice is a dual pyrochlore lattice, shown in \cref{fig:diamond_dual_membranes}(a) with dark gray tetrahedra, where each dual diamond vertex (white sphere) sits in the center of a dual pyrochlore tetrahedron. 
We note that in the common \ce{$\mathcal{A}$2$\mathcal{B}$2O7}  
pyrochlore oxide compounds, the direct pyrochlore lattice sites are occupied by the $\mathcal{A}$ ions and the dual pyrochlore sites are occupied by the $\mathcal{B}$ ions~\cite{gardnerMagneticPyrochloreOxides2010}.
This is further illustrated in \cref{fig:diamond_dual_membranes}(b), which shows how the plaquettes of the dual diamond lattice (green) relate to the plaquettes of the direct diamond lattice (blue) and how the corresponding pyrochlore lattices (labeled $\mathcal{A}$ and $\mathcal{B}$) interpenetrate one another. 
Each edge of the dual diamond lattice (black line) pierces a plaquette of the direct diamond lattice (blue hexagon) and each edge of the direct lattice (gray line) pierces a dual plaquette (green hexagon), so that the plaquettes ``link'' each other, as shown. 
We can now make a key observation: each microscopic Ising spin (red arrow), located at a site of the direct ($\mathcal{A}$) pyrochlore lattice may be viewed \emph{either} as a 1-dimensional flux along a direct diamond lattice edge, as in NNSI, or as a \emph{2-dimensional flux} through a dual lattice plaquette.  
The latter perspective is essential to understand the emergent 2-form gauge structure of the spin vorticity model.

\begin{figure*}[ht]
    \centering
    \raisebox{-0.5\height}{
        \begin{overpic}[width=.3\textwidth]{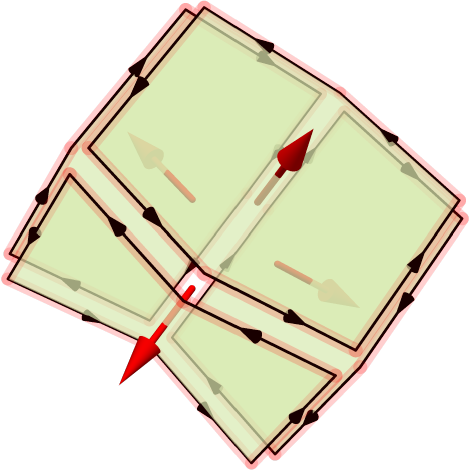}
            \put(041,-8){(a)}
            \put(160,-8){(b)}
            \put(277,-8){(c)}
        \end{overpic}
        }
    \hfill
    \raisebox{-0.5\height}{\includegraphics[width=.3\textwidth]{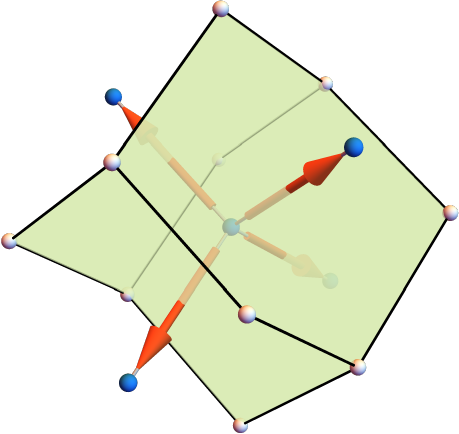}}
    \hfill
    \raisebox{-0.5\height}{\includegraphics[width=.3\textwidth]{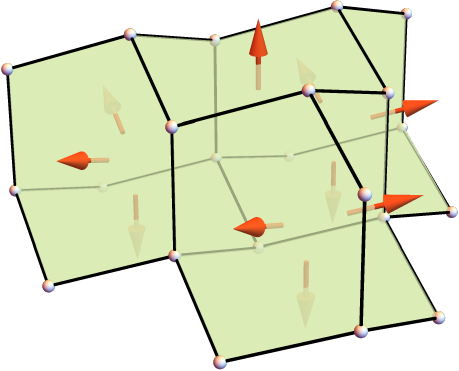}}
    \\[3ex]
    \caption{
        (a) Ground states are formed by arranging spins in such a way that all of the ``currents'' running around the dual plaquettes cancel. This is achieved when the spins form a closed, oriented surface. 
        The smallest such closed membrane is the one shown in (b), corresponding to a ``star'' of the direct diamond lattice. 
        A larger membrane with ten spins is shown in (c).  
        Flipping a collection of spins forming such a closed membrane costs zero energy, directly analogous to the flipping of closed loops in NNSI. 
    }
    \label{fig:small_membranes}
\end{figure*}

\subsection{String Fractionalization}

Starting from any ground state satisfying $\omega_p = 0$ on every direct plaquette (\cref{fig:hexagon}), consider flipping a single spin---the lowest-energy microscopic excitation.
After the flip, six direct plaquettes have non-zero vorticity, $\omega_p =\pm 1$, illustrated in \cref{fig:diamond_dual_membranes}(c) by the six lightly shaded blue hexagons, costing a total energy $6J$ (one $J$ for each hexagon). 
Thus a single spin flip is at first glance fractionalized into \emph{six} excitations. 
However, interpreting the vorticity $\omega_p$ as the lattice curl on direct plaquette $p$ (i.e. the circulation around the boundary of the plaquette), we may imagine that it is sourced by a ``current'' via an \emph{emergent Ampere's law}, where the current is ``flowing'' along the 1-dimensional dual edge piercing direct plaquette $p$.
Together, the six defective plaquettes carrying $\omega_p\neq 0$ after the single spin flip then appear to be sourced by a \emph{closed loop of current} running around the boundary of the green dual plaquette, indicated in \cref{fig:diamond_dual_membranes}(c) by a red string, with black arrowheads indicating the direction of ``flow''. 
Note that, as we will see in more detail in \cref{sec:2-form_electrodynamics}, these string objects are better thought of as string charges rather than currents, i.e. there is nothing flowing along the string. 
We utilize this ``current'' wording here only to draw analogy with the intuition of classical Maxwell electrodynamics.

We conclude that in the spin liquid of the spin vorticity model, spins are fractionalized into closed strings. 
These may be viewed as fictional currents ``flowing'' along the edges of the dual lattice, acting as sources of the spin vorticity via an emergent Ampere's law. 
The spins may be thus be viewed as fluxes through the 2-dimensional dual plaquettes, shown in green in \cref{fig:diamond_dual_membranes}(b,c), attached to oriented strings on their boundaries.
This contrasts with the fractionalization of NNSI, where each spin acts as a short string of electric flux along a direct diamond edge, attached to a pair (i.e. dumbbell) of opposite electric charges, which act as (fictional) sources of divergence of the electric field ${\bm E}$ representing the spin configurations~\cite{castelnovoMagneticMonopolesSpin2008} (see Introduction). 
These two perspectives are juxtaposed in \cref{fig:diamond_dual_membranes}(c):
The spin ice dumbbell is shown as a pair of point charges (black and white) attached to the ends of a 1-dimensional flux string (yellow) along a direct lattice edge; 
the equivalent fractionalized object in the spin vorticity model is an oriented string charge (red) attached to a two-dimensional flux membrane (green) on a dual plaquette.
These two pictures are naturally complementary---they correspond to the two classical representations of a dipole: either a pair of opposite charges, or a loop of current. 
In NNSI, each spin (dipole) is fractionalized in the former way since its Hamiltonian enforces a zero-divergence condition (a Gauss-like law), while in the spin vorticity model they fractionalize in the latter way since its Hamiltonian enforces a zero-curl condition (an Ampere-like law).

\subsection{Ground State Manifold: Topological Membrane Condensate}
\label{sec:gs_manifold}

In order to construct a ground state, we imagine placing the spins into the lattice one by one, each thought of according to \cref{fig:diamond_dual_membranes}(c) as a 2-dimensional flux through a dual plaquette with a 1-dimensional current (string charge) running around its boundary.
To form a ground state, these plaquettes must be placed so that all of the ``currents'' along their boundary edges cancel pairwise, meaning that there are no direct plaquettes with $\omega_p \neq 0$.
This is achieved when they are arranged to form a \emph{closed oriented surface} in the dual lattice (analogous to how the spins in spin ice must form closed oriented loops).
\Cref{fig:small_membranes}(a) shows four spins arranged ``all-out'' to form a closed membrane, where the string charges (black arrows) cancel pairwise. 
This implies that every ground state of the spin vorticity model is decomposable, non-uniquely, into a collection of closed, oriented membranes in the dual lattice. 
We conclude that the spin vorticity model spin liquid ground state may be described as a \emph{membrane condensate}, analogous to the string condensate of NNSI.

The smallest such closed membrane consists of four dual plaquettes (green) shown in \cref{fig:small_membranes}(b), with the four spins pointing all-out, and a larger membrane consisting of ten spins  shown in \cref{fig:small_membranes}(c). 
Flipping all of the spins forming one of these closed membranes in the dual lattice, no matter its size or shape, produces another configuration with the same energy, analogous to flipping closed loops of spins in spin ice~\cite{melkoLongRangeOrderLow2001,melkoMonteCarloStudies2004}.
This is because the circulation around any loop of direct-lattice edges piercing the membrane is unchanged by this collective spin flip---such a loop passes ``in'' and ``out'' of the membrane an equal number of times, and so ``flipping a membrane'' does not change $\omega_p$ anywhere (\cref{apx:string_membrane_relation}). 
Notice that the minimal membrane shown in \cref{fig:small_membranes}(b) is precisely the ``star'' of four spins surrounding a direct diamond lattice vertex, corresponding to an all-in or all-out tetrahedron of the pyrochlore lattice.
Thus the star moves, described in \cref{subsec:MonteCarlo} and used to generate the Monte Carlo results in \cref{fig:vorticity_monte_carlo}, are precisely the smallest possible ``membrane flipping'' Monte Carlo update.

Note that flipping any membrane that is the surface of a 3-dimensional volume is equivalent to applying a star move to every minimal membrane (i.e. every star or tetrahedron) contained inside the volume.
Therefore, using only the minimal star move updates in our Markov chain Monte Carlo is equivalent, after a sufficient number of such updates, to finding and flipping large membranes which bound a three-dimensional volume.  
On the other hand, there exist topologically non-trivial surfaces which wind around the periodic boundaries (or around any hole), which are not the boundary of any volume. 
Flipping such a non-contractible membrane is never equivalent to any sequence of local flips of topologically trivial membranes. 
The ground state manifold is therefore divided into topological sectors, which are connected by flipping non-contractible membranes.
This parallels the topological sectors of NNSI, which are connected by flipping non-contractible strings~\cite{jaubertAnalysisFullyPacked2011,jaubertTopologicalSectorFluctuationsCurieLaw2013}.
Notice that a star move (flipping an all-in to all-out tetrahedron or vice versa) does not change the net magnetization, from which it follows that flipping a contractible membrane conserves the net magnetization of the system, $\sum_i \sigma_i^z \hat{\bm{z}}_i$. 
On the other hand, flipping a flat surface of all-down spins winding around the periodic boundaries to all-up clearly changes the magnetization. 
We conclude that the topological sectors differ by their net magnetization (see \cref{apx:Topological_Sectors} for further details).

\begin{figure*}[ht]
    \centering
    \raisebox{-.5\height}{
        \begin{overpic}[width=0.28\textwidth]{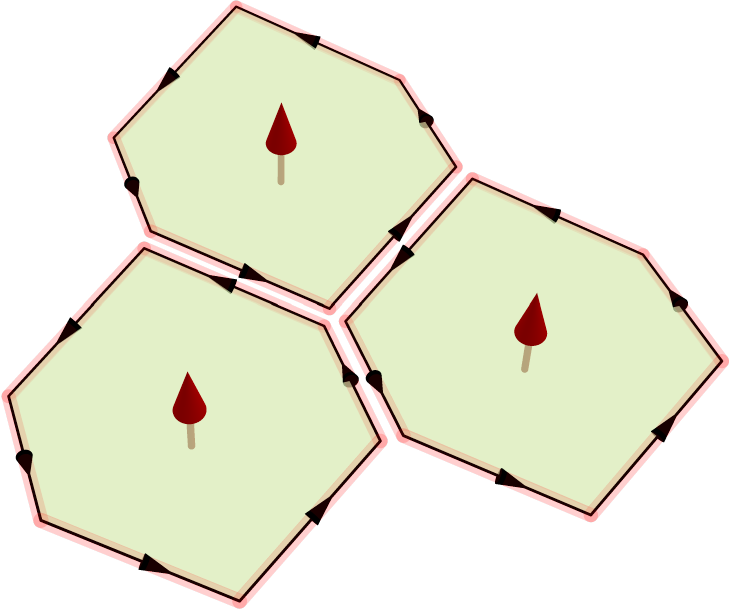}
            \put( 45,-25){(a)}
            \put(165,-25){(b)}
            \put(300,-25){(c)}
        \end{overpic}
        }
    \hfill
   \raisebox{-.5\height}{\includegraphics[width=0.36\textwidth]{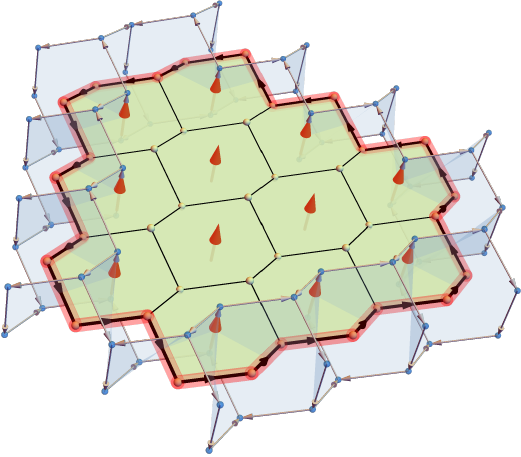} }
   \hfill
    \raisebox{-.5\height}{\includegraphics[width=0.30\textwidth]{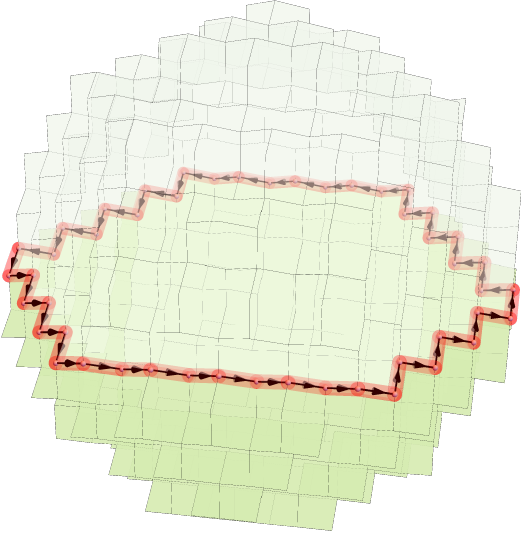}}
    \\[3ex]
    \caption{
        (a) Flipping a collection of spins (red arrows) which tile an open surface, the fractional string charges (red with black arrows, c.f. \cref{fig:diamond_dual_membranes}(c)) cancel in the interior of the surface, leaving a single string circulating around the boundary.
        (b) A large string excitation created by flipping twelve spins. 
        Each direct plaquette (blue) pierced by the string carries non-zero vorticity, $\omega_p \neq 0$. 
        The energy of the string is $J$ times the number of defect plaquettes, i.e. the length of the string. 
        In the direct lattice, there is a non-zero circulation of the spins around any closed contour that links the string (see \cref{apx:string_membrane_relation}).
        (c) The membrane attached to a given string is not observable, it is only well-defined relative to a fixed ground state configuration, just as is the case for the strings in spin ice (see \cref{fig:pyro_diamond}). 
        Here we show two possible membranes attached to the same string (spins are not shown for clarity). 
        Flipping one or the other differs by flipping the closed membrane formed by gluing them together. 
    }
    \label{fig:large_membranes_strings}
\end{figure*}

Why a ground state manifold satisfying a zero-curl condition should be described by closed-membrane configurations can be understood via the Helmholtz-Hodge decomposition, \cref{eq:Helmholtz}. 
Any vector field configuration with zero rotational component can be written as a gradient of a scalar function (neglecting boundaries), such that the vector field always points orthogonal to the  equipotential surfaces. 
This implies that the ground states of the spin vorticity model in the zero-magnetization sector may be expressed as an integer-valued function $\phi$ on the diamond lattice vertices, whose value $\phi_t$ increases or decreases by 1 across each direct diamond link, such that the spin on the edge connecting direct diamond lattice site $t$ to the one at $t'$ has value $\sigma^z_{tt'} = \phi_{t'}-\phi_t$. 
Thus each topologically-trivial ground state configuration can be decomposed into a collection of equipotential surfaces in the dual diamond lattice, along which the spins all point in the same direction, such as illustrated in \cref{fig:small_membranes}(b,c). 
This may be viewed as a 3D version of a height-representation~\cite{huseCoulombLiquidDimer2003,henleyCoulombPhaseFrustrated2010}, with $\phi$ the ``height'' field. 
Ground states in a non-trivial topological sector configuration (one with non-zero net magnetization) must be expressed as a combination of $\phi$ plus an extra piece, the harmonic component of the Helmholtz-Hodge decomposition, \cref{eq:Helmholtz}. 
This may be done by fixing the spins on a set of non-contractible membranes, which fixes a topological sector, effectively dividing the system into a set of contractible domains with spins fixed along their boundaries, within each of which the remaining spin configuration can be represented as the gradient of some $\phi$.  
The topological sector is determined by the winding number $w[C] := \sum_{i \in C}(-1)^i \sigma_i^z$ along non-contractible loops $C$ which link one of the three handles of the torus, where $(-1)^i$ alternates along the loop (see~\cref{apx:string_membrane_relation}). 
These winding numbers can only change in a ground state by flipping a non-contractible membrane of spins, but are ill-defined in the presence of excitations. 
The analogous winding numbers in spin ice are the net flux through three planes of the lattice~\cite{hermelePyrochlorePhotonsU12004,lantagne-hurtubiseSpinIceThinFilms2018}. 
See \cref{apx:string_membrane_relation} for further details.

In summary, the ground states of the spin vorticity model can be decomposed into a collection of closed oriented membranes, directly analogous to how ground states of spin ice can be decomposed into a collection of closed oriented strings~\cite{jaubertAnalysisFullyPacked2011}. 
Such a decomposition is not unique, thus characterizing this phase as a condensate of membranes.
Flipping one of these closed membranes costs zero energy because it does not change the vorticity $\omega_p$ on any direct plaquette, analogous to how flipping a closed loop in spin ice does not change the charge $Q_t$ on any tetrahedron~\cite{melkoLongRangeOrderLow2001,melkoMonteCarloStudies2004}.  
The ground state manifold is divided into topological sectors, differing by the system's net magnetization but connected by the flipping of non-contractible~membranes.

\subsection{Excitations: String Charges}
\label{sec:string_excitations}

The minimal excitation, created by flipping a single spin, is a small oriented string in the dual diamond lattice, shown in \cref{fig:diamond_dual_membranes}(c). 
More generally, excitations are created by flipping a collection of spins forming an \emph{open} surface, such as shown in \cref{fig:large_membranes_strings}(a) with three spins. 
The oriented string charges created in the interior of the surface formed by the three plaquettes cancel each other pairwise, while the uncancelled edges on the boundary of the surface carry a longer 1-dimensional string excitation, attached to a two-dimensional flux membrane in the interior of the surface.
\Cref{fig:large_membranes_strings}(b) shows a string excitation created by flipping twelve spins forming a flat open membrane. 
Each direct plaquette (blue) pierced by the string carries a non-zero vorticity, $\omega_p$. 
The net circulation around any loop in the direct lattice which links the string in the dual lattice is non-zero (\cref{apx:string_membrane_relation}).
There is no unique open membrane attached to a given string excitation, just as the closed membranes are not uniquely defined.
Having identified an open membrane attached to a string, it may be deformed by flipping a closed membrane that touches it.
This parallels how flipping closed loops in NNSI deforms the position of the string connecting a pair of charges, shown in \cref{fig:pyro_diamond}(b,c). 
\Cref{fig:large_membranes_strings}(c) depicts two different membranes attached to the same string.
Given a spin configuration, there are many possible membranes one can find, of arbitrary size and shape, connected to the same string excitation.

\subsubsection{String-String Interactions}

These extended string excitations cost energy proportional to their length, specifically $J$ times the number of direct plaquettes pierced by the string, which have $\omega_p = \pm 1$ in the dilute limit. 
For a plaquette pierced by two overlapping strings with the same local orientation, so that $\omega_p = \pm 2$, the energy cost on that plaquette is $4J$. 
This is higher than the $2J$ energy cost of two separate single-defect plaquettes, meaning there is an extra energy cost associated to parallel strings which overlap. 
Thus, strings with the same local orientation energetically repel each other, and may be considered to have the same charge locally.
On the other hand, strings with opposite local orientation may fuse, annihilating along their overlap and thus lowering the total energy by minimizing the total string length, so there is an energetic attraction between anti-parallel strings, and they may be considered to have opposite charges. 
One may view the strings as being made of individual segments (on the dual links), each of which carries a magnitude and direction defining the local charge.

This is quite distinct from NNSI, where the charge excitations are (\textit{i}) point-like, and (\textit{ii}) come in two species, either positive or negative. 
By contrast, the strings of the spin vorticity model do not come in two charge species; instead, they come with an orientation, and the relative charge of two local string segments is determined by their relative orientations. 
A pair of strings may be locally parallel and thus repelling in one place, but anti-parallel and thus attracting in another and, indeed, a single string can interact with itself. 
In addition to these local energetic contact interactions, we also expect an entropic interaction between the strings, analogous to the entropic $1/r$ Coulomb potential between charges in NNSI~\cite{castelnovoMagneticMonopolesSpin2008,castelnovoSpinIceFractionalization2012}. 
One can also consider tuning further-neighbor spin-spin interactions in the Hamiltonian of the form $\omega_p \omega_{p'}$ for nearby plaquettes $p$ and $p'$ to control the energetic interaction potential between strings, analogous to Refs.~\cite{rauSpinSlushExtended2016,mizoguchiMagneticClusteringHalfmoons2018} which consider terms of the form $Q_t Q_{t'}$ to control the attraction/repulsion of point charges.

\begin{figure}[t]
    \centering
    \vspace{5ex}
    \begin{overpic}[width=\columnwidth]{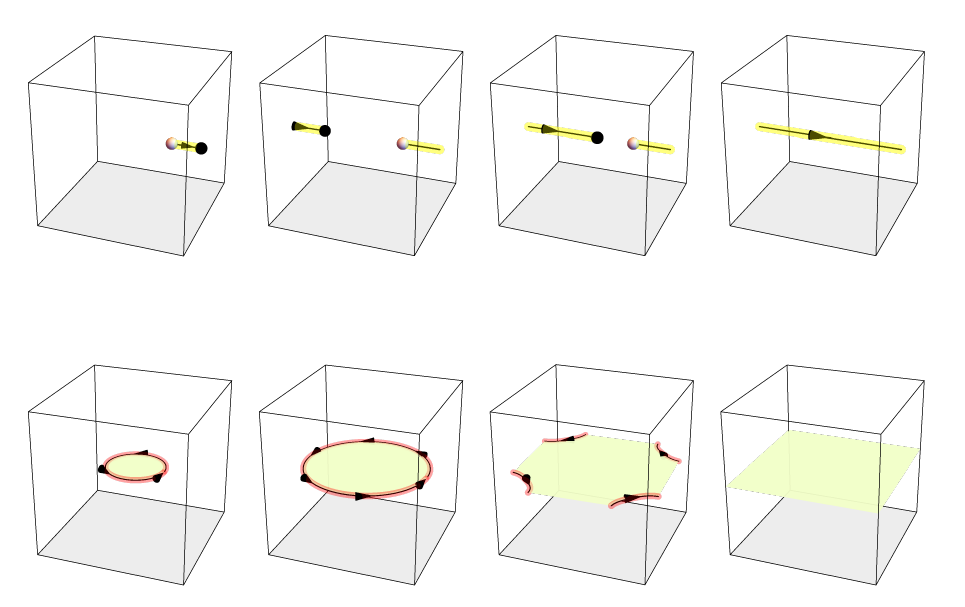}
        \put(00,60){(a)}
        \put(09,64){\underline{$E=4J$}}
        \put(33,64){\underline{$E=4J$}}
        \put(57,64){\underline{$E=4J$}}
        \put(81,64){\underline{$E=0$}}
        \put(00,25){(b)}
        \put(09,29){\underline{$E\ll L$}}
        \put(33,29){\underline{$E\sim L$}}
        \put(57,29){\underline{$E\ll L$}}
        \put(81,29){\underline{$E=0$}}
    \end{overpic}
    \caption{(a) In spin ice, the topological sector can fluctuate by creation of a pair of charges, running one around the periodic boundaries, and re-annihilating the pair, equivalent to flipping a non-contractible string, the only energy cost being the creation of the initial pair of charges. (b) In the spin vorticity model, the equivalent mechanism is to nucleate a single string, grow it to the size of the system, then re-contract it around the periodic boundaries until it annihilates itself, equivalent to flipping a non-contractible membrane. 
    Since strings cost energy proportional to their length, a string with length of the system size $L$ is forbidden in the thermodynamic limit.
    }
    \label{fig:topo_sector_flip}
\end{figure}

\subsubsection{Topological Sector Ergodicity}
\label{sec:top_sector_ergodicity}

The fact that the energy of a string scales with its length highlights a crucial qualitative distinction between the spin vorticity model and spin ice. 
In spin ice, the topological sector can be changed by nucleating a pair of charges out of a ground state, costing a total energy $4J$, moving one around the handle of the torus (which costs no energy), and annihilating them.
This process is illustrated in \cref{fig:topo_sector_flip}(a), which effectively inserts a string of flux around the periodic boundaries.
In the spin vorticity model on the other hand, the analogous operation is to nucleate a single string, grow it to the size of the torus until it wraps around the periodic boundaries, then re-contract it to annihilate itself, as illustrated in \cref{fig:topo_sector_flip}(b).
However, a string the size of the torus costs energy proportional to the linear size of the system, and is thus forbidden in the thermodynamic limit. 
This observation implies that the time to reach thermal equilibrium between the topological sectors with only local dynamics diverges at low temperature as the density of string defects and their average length falls below some threshold, such that on a finite timescale the topological sectors fall out of equilibrium. 
This loss of topological sector ergodicity naturally explains why the pinch point intensity of the structure factor measured in the Monte Carlo simulations, shown in \cref{fig:vorticity_monte_carlo}(d), is exactly zero at the $(002)$ point at $T=0$. 
Since our algorithm utilizes only local updates, the consequential lack of topological sector fluctuations at low temperatures means that the net magnetization effectively freezes, resulting in an extinction of the zone-center pinch point intensities (see \cref{apx:Topological_Sectors} for further details).

\subsection{2-Form Electrodynamics \texorpdfstring{\&}{and} Coulomb Spin Liquid}
\label{sec:2-form_electrodynamics}

Recall that we initially set out to swap the rotational-irrotational description of the Coulomb phase of spin ice by replacing the divergence in \cref{eq:S_NNSI} with a curl in \cref{eq:S_vort}.
What has been  achieved in the process is to increase the dimension of the microscopic charged particles and flux strings of NNSI by one, to charged strings and flux membranes, respectively.
Whereas NNSI realizes a Coulomb phase where charged particles couple to an electric field whose field lines are strings, the spin vorticity model realizes a generalized Coulomb phase, where charged strings couple to an electric field whose ``field lines'' are membranes.

The natural language to describe such ``generalized'' Coulomb phases is so-called ``$p$-form electrodynamics'', developed by Henneaux and Teitelboim~\cite{henneauxPFormElectrodynamics1986}.  
This is a straightforward generalization of Maxwell electrodynamics to describe $(p-1)$-dimensional charged objects interacting with a rank-$p$ antisymmetric tensor electric field, 
called a $p$-form.\footnote{This should not be confused with the rank-2 \emph{symmetric} tensor fields relevant to fracton physics~\cite{yanRank2U1Spin2020,pretkoSubdimensionalParticleStructure2017}.} 
In differential geometry, $p$-forms are the natural objects which can be sensibly integrated over a $p$-dimensional surface. 
A brief semi-technical but self-contained introduction to continuum $p$-form electrodynamics in the language of differential forms is provided in \cref{apx:2-form-gauge-theory} for the interested reader.

The case $p=1$ describes traditional Maxwell electrodynamics of point charges interacting with a 1-form (vector) electric field, governed by the Gauss law
\begin{equation}
    \partial_i E_i = \rho,
\end{equation}
where E is the 1-form electric field and $\rho$ is the scalar (0-form) charge density.
This is the coarse-grained description of the microscopic point charges attached to 1-dimensional electric field strings in spin ice.
The case $p=2$ describes string charges interacting with a 2-form electric field, governed by the Gauss law
\begin{equation}
    \partial_j E_{ij} = \mathscr{j}_i,
    \label{eq:2-form-gauss-law-continuum}
\end{equation}
where $E$ is now a 2-form, i.e. an anti-symmetric rank-2 tensor field, and $\mathscr{j}$ is a 1-form charge density, carrying a vector index.
It automatically satisfies $\partial_i \mathscr{j}_i = 0$ due to the anti-symmetry of $E$, meaning that the vector charges must form closed strings.
This is the appropriate coarse-grained description of the closed string excitations attached to 2-dimensional electric flux membranes in the spin vorticity model described thus far. 

\begin{figure}[t]
    \centering
    \includegraphics[width=.7\columnwidth]{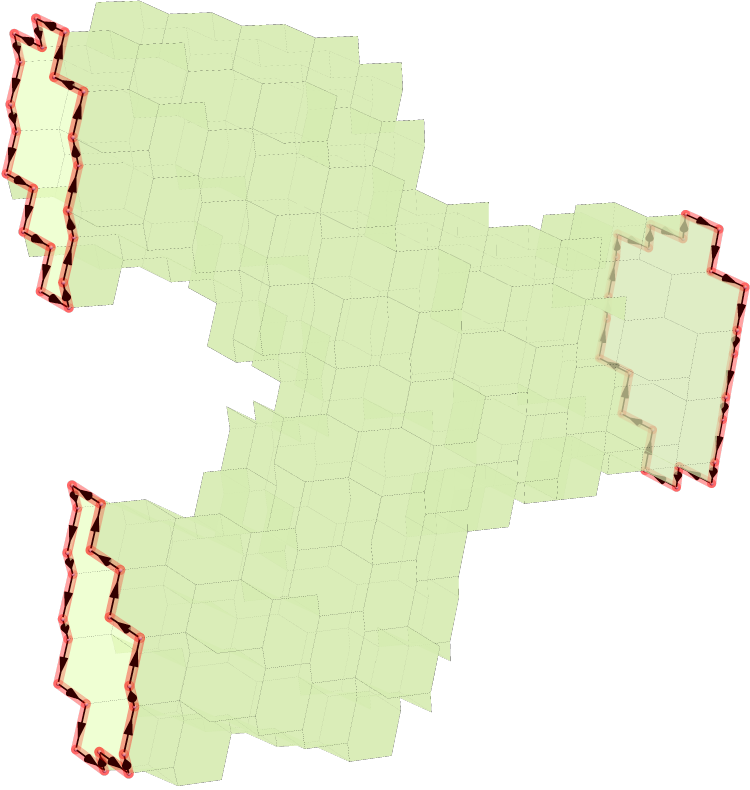}
    \caption{The classical spin liquid Coulomb phase of the spin vorticity model is described by 2-form electrodynamics, with charged string excitations interacting with a 2-form gauge field. This has properties not shared by the emergent 1-form Coulomb phase of NNSI, where point charges are always created in positive-negative pairs connected by a string of electric flux (\cref{fig:pyro_diamond}). Each string charge (red) of the spin vorticity model carries local charge but is globally charge-neutral, which is reflected in the fact that many strings may be connected by a single flux membrane (green). Here we show three such string charges. 
    }
    \label{fig:three-string-pants}
\end{figure}

The fact that the string charge density $\mathscr{j}$ carries a single vector index corresponds to the fact that the charge of a string depends on the local orientation as discussed in \cref{sec:string_excitations}.
In the spin vorticity model, the string charges may be viewed as consisting of charged segments (one per dual link), each of which carries a magnitude and direction, providing the microscopic analog the 1-form (vector) charge density $
\mathscr{j}$. 
A hallmark feature of $p$-form electrodynamics is that the charged excitations for $p>1$ have \emph{net-zero charge}.
In a sense, extended charged objects are their own ``anti-particles''~\cite{henneauxPFormElectrodynamics1986}. 
We can clearly see this in the spin vorticity model, because flipping a single spin creates only one string excitation, rather than a pair of opposite-charged excitations.
Correspondingly, an isolated string excitation can shrink and vanish into the vacuum without having to annihilate with another string.
This behavior contrasts with the ordinary (1-form) Coulomb physics of spin ice, where point charges can only ever be created in positive-negative pairs at the ends of an open string, i.e connected by a line of electric flux.
We have already seen that the minimal excitations of the spin vorticity 2-form spin liquid are not pairs of charges but rather a single string charge loop in \cref{fig:diamond_dual_membranes}(c).
\Cref{fig:large_membranes_strings}(b) shows a larger contiguous string charge, again attached to a single membrane.
Indeed, since a membrane may have an arbitrary number of holes, a single electric membrane may connect an arbitrary number of charged strings together, such as the three strings connected by a single membrane shown in \cref{fig:three-string-pants}.

We conclude that the spin vorticity model realizes a \emph{classical 2-form spin liquid}, i.e. an emergent 2-form Coulomb phase. 
To the best of our knowledge, this is the first example of a lattice spin model realizing such a phase.
As we emphasized at the very beginning of this section, analogous spin vorticity models can be defined on other line graph lattices in three dimensions~\cite{henleyCoulombPhaseFrustrated2010}, just as ice models are not unique to the pyrochlore lattice, and the 2-form spin liquid physics we have described in this section can be directly translated to various other geometries where the generic Hamiltonian \cref{eq:H_vort_wp} can be correspondingly defined.

\section{2-Form U(1) Quantum Spin Liquids}
\label{sec:QSV}

Up to this point we have considered the classical spin vorticity model defined by the Hamiltonian of~\cref{eq:H_vort_spin}, and demonstrated that it realizes an emergent 2-form Coulomb phase, i.e. a 2-form classical spin liquid.
In the language of 2-form electrodynamics~\cite{henneauxPFormElectrodynamics1986}, the membranes are 2-form electric fluxes, which are attached by the Gauss law to fractionalized charged string excitations.
This classical model is interesting in its own right, and could potentially be realized at finite temperatures in a materials setting, but more theoretically intriguing is the possibility that it may serve as a basis for a novel class of quantum phases: \emph{2-form U(1) quantum spin liquids}.
As such, we now study the effects of perturbing the classical spin vorticity model with weak quantum fluctuations which can drive tunneling between the classical ground states.

As a minimal definition of what such a liquid is, we search for a quantum model whose ground state is a coherent superposition of the classical ground states of the classical spin vorticity model. 
In this section, we demonstrate in broad strokes how to construct such a ``quantum spin vorticity'' (QSV) model.
We will show that the resulting model perturbatively maps to a 2-form U(1) lattice gauge theory which can be tuned with a four-spin interaction to a point where the ground state is a massive superposition of all classical ground states, serving as a proof of concept for a 2-form U(1) QSL.
Building on this, we demonstrate how the string excitations may then be treated as a 1-form Higgs field within the lattice gauge theory framework.
We end with a discussion of the stability of 2-form U(1) QSL phases, placing these phases within the context of well-known U(1) superfluid and spin liquid phases. 
In the remainder of this section, we switch to quantum spin-1/2 operators $S^z_i = (\hbar/2)\sigma^z_i$, along with the local transverse spin components $S_i^x$ and $S_i^y$ defined with the standard symmetry convention used for the pyrochlore lattice~\cite{rauFrustratedQuantumRareEarth2019} (\cref{apx:conventions}), and henceforth work in units with $\hbar = 1$.

\subsection{Quantum Membrane Exchange Dynamics}
\label{sec:membrane_exchange}

We consider now perturbing the classical pyrochlore spin vorticity model 
$H_{\textrm {SV}}$ in \cref{eq:H_vort_spin} with terms generating quantum dynamics (spin flips).
While a simple transverse field would suffice for perturbative arguments, more realistic terms are symmetry-allowed nearest-neighbor spin-spin interactions. 
For the pyrochlore lattice, these have the form $\cramped{S_i^{\pm}S_j^{\mp}}$, $\cramped{S_i^z S_j^{\pm}}$, or $\cramped{S_i^{\pm}S_j^{\pm}}$, where $\cramped{S^{\pm} = S^x \pm i S^y}$~\cite{rauFrustratedQuantumRareEarth2019}.
Treating these terms within degenerate perturbation theory generates operators acting on up to $2n$ spins at order $n$. 
We are then interested in the lowest order operators with matrix elements connecting one degenerate classical ground state to another, assuming that a large gap to the excited states is preserved. 
When perturbing the NNSI model, the $\cramped{S_i^{\pm}S_j^{\mp}}$ term generates at third order the ring exchange model of quantum spin ice (QSI), which flips maximal-circulation loops of six spins (such as illustrated in \cref{fig:hexagon}). This generates quantum dynamics within the classical ground state manifold and drives the system into a quantum spin liquid state~\cite{hermelePyrochlorePhotonsU12004,gingrasQuantumSpinIce2014,katoNumericalEvidenceQuantum2015,huangDynamicsTopologicalExcitations2018}.

\begin{figure}[t]
    \centering
    \vspace{2ex}
    \begin{overpic}[width=\columnwidth]{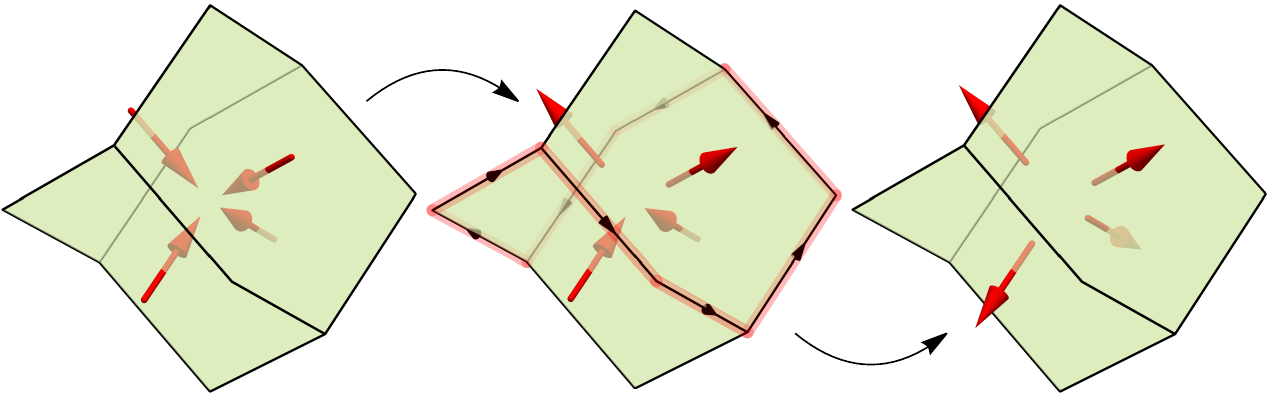}
        \put(31,29){$S_i^{\pm} S_j^{\pm}$}
        \put(64,-2){$S_k^{\pm} S_l^{\pm}$}
    \end{overpic}
    \\[2ex]
    \caption{Applying $S_i^\pm S_j^{\pm}$ from \cref{eq:H_QSV} twice to a classical ground state creates an intermediate excited state with a string costing energy $8J$, then annihilates it to generate another classical ground state with four spins flipped, thus generating membrane exchange. 
    }
    \label{fig:membrane_exchange}
\end{figure}

We consider now perturbing the classical pyrochlore spin vorticity model, \cref{eq:H_vort_spin}, with the three above-mentioned $S_i^\alpha S_j^\beta$ transverse terms. 
The desired leading-order effect is generated by the $\cramped{S_i^\pm S_j^\pm}$ term, so we consider a minimal ``quantum spin vorticity model'' (QSV) with Hamiltonian given by\footnote{
    From a materials perspective, this term is physically relevant to the dipolar-octupolar doublet of certain Kramers pyrochlore rare earth insulators~\cite{rauFrustratedQuantumRareEarth2019}. 
    For the more general pseudospin-1/2 doublets, additional complex phase factors should be included in the $S_i^\pm S_j^\pm$ term~\cite{rauFrustratedQuantumRareEarth2019}.
    Furthermore, the $\cramped{S_i^z S_j^\pm}$ interaction (only allowed for Kramers doublets) generates a non-trivial $S_i^z S_j^z$ operator at second order in perturbation theory ($J_{3a}$ in \cref{fig:hexagon})~\cite{savaryQuantumCoherenceQuantum2021}. 
    In this work, we omit such materials-specific details to focus more specifically on the quantum gauge theory description. 
}
\begin{equation}
    H_{\text{QSV}} = H_{\text{SV}} + J_{\pm\pm}\sum_{\langle ij \rangle} (S_i^+S_j^+ + S_i^- S_j^-).
    \label{eq:H_QSV}
\end{equation}
At second order in degenerate perturbation theory, one obtains quantum tunneling between two classical ground states, via the flipping of four spins shown in \cref{fig:membrane_exchange}. This process defines an effective membrane exchange (ME) Hamiltonian given by
\begin{align}
    H_{\text{ME}} &= J_{\text{mem}} \sum_t (S_1^- S_2^- S_3^- S_4^- + S_1^+ S_2^+ S_3^+ S_4^+)_t\,,
    \nonumber
    \\
    &\equiv J_{\text{mem}} \sum_{\vtiny} \left(\,\,\ket{\memin}\bra{\memout} + \ket{\memout}\bra{\memin}\,\,\right),
    \label{eq:H-mem-exch} 
\end{align}
where $\cramped{J_{\text{mem}} \propto J_{\pm\pm}^2/J > 0}$, and the numbering in the first line is over the four spins in a tetrahedron $t$ (see \cref{fig:pyro_diamond}(b)). 
In the second line, we pictorially represent the four-spin operators as a flip of the minimal membrane consisting of four dual plaquettes (shown in \cref{fig:small_membranes}(b)) from an all-out to an all-in configuration or vice versa.
In other words, this operator generates quantum ``star move'' dynamics.
We refer to the effective Hamiltonian in \cref{eq:H-mem-exch} as a \emph{membrane exchange} (ME) model, in  direct analogy to the ring exchange model of QSI~\cite{hermelePyrochlorePhotonsU12004,bentonSeeingLightExperimental2012,gingrasQuantumSpinIce2014}.

The membrane exchange model provides an intuitive picture for the quantum dynamics of the proposed 2-form quantum spin liquid phase, but we can go further.
By adding to the Hamiltonian a generalization of the Rokhsar-Kivelson (RK) term~\cite{hermelePyrochlorePhotonsU12004,rokhsarSuperconductivityQuantumHardCore1988,gingrasQuantumSpinIce2014,moessnerQuantumDimerModels2011}, which acts as a chemical potential for flippable membranes, we can tune the model to an exactly-solvable point. 
The RK term includes interactions of four spins on each tetrahedron, given by
\begin{align}
    H_{\text{RK}} &= V \sum_{t}\left[\sum_{\eta = \pm 1} \prod_{i \in t} \left(\frac{1}{2} + \eta S_i^z\right)\right]\,,
    \nonumber 
    \\[2ex]
    &\equiv V \sum_{\vtiny}  \left(\,\,\ket{\memin}\bra{\memin} + \ket{\memout}\bra{\memout}\,\,\right),
    \label{eq:H_RK}
\end{align}
which counts the number of flippable stars (i.e. minimal closed membranes), with $V>0$. 
The RK operator is diagonal in the $S^z$ basis, with positive entries $V$ times the number of flippable stars, while the membrane exchange term \cref{eq:H-mem-exch} is off-diagonal with positive matrix elements $J_{\text{mem}}$.
The sign of $J_{\text{mem}}$ can be changed via a unitary transformation, by applying a $\pi$ rotation about the local $z$-axis to every spins that belong to one of the four FCC sublattices of the pyrochlore lattice (\cref{apx:conventions}), which changes $S_i^\pm \to - S_i^\pm$ in \cref{eq:H-mem-exch} for exactly one spin on every tetrahedron.
This makes all of the off-diagonal matrix elements of the Hamiltonian non-positive (``stoquastic''~\cite{bravyiComplexityStoquasticLocal2007}).
Then, tuning $V \to J_{\text{mem}}$, the Hamiltonian can be written as a sum of projectors
\begin{equation}
    H_{\text{RK}-\text{ME}} = V \sum_{\vtiny} \left(\,
    \ket{\memin}-\ket{\memout}\,\right)\left(\,
    \bra{\memin}-\bra{\memout}\,\right).
\end{equation}
Since this Hamiltonian is a projector, with eigenvalues $0$ or $V$, any state it annihilates is a ground state. The ground state is then an equal-weight superposition of every classical closed-membrane ground state~\cite{rokhsarSuperconductivityQuantumHardCore1988,hermelePyrochlorePhotonsU12004,moessnerQuantumDimerModels2011},
\begin{equation}
    \ket{0}_{\text{RK}} = \mathcal{P}\,\prod_i \,\frac{1}{\sqrt{2}}\left(\ket{S_i^z = +\tfrac{1}{2}} +  \ket{S_i^z = -\tfrac{1}{2}}\right),
    \label{eq:RK_wavefunction}
\end{equation}
where $\mathcal{P}$ is the projector to the classical ground state manifold of closed-membrane configurations (those which satisfy the zero-vorticity constraint), and the state it acts on is the equal-weight superposition of every classical Ising configuration.~\footnote{
    Technically, at the RK point there is one quantum ground state for each topological sector, given by an equal weight superposition of all classical ground states in the same sector~\cite{rokhsarSuperconductivityQuantumHardCore1988}. \cref{eq:RK_wavefunction} is then an equal-weight superposition of all of these ground states.
    }
This demonstrates the existence of a fine-tuned point whose quantum ground state may be described as a 2-form QSL, i.e. a massive coherent superposition of the classical 2-form spin liquid ground states. 
The RK point itself is fine-tuned (in the ring exchange model of QSI, it corresponds to the boundary between the QSL and an ordered phase~\cite{hermelePyrochlorePhotonsU12004,bentonSeeingLightExperimental2012,shannonQuantumIceQuantum2012,gingrasQuantumSpinIce2014}), but we take its existence in the membrane exchange model as a proof of concept that a Hamiltonian with a 2-form QSL ground state exists. 
The issue of the stability of the putative QSL ground state \emph{away} from the fined-tuned RK point is discussed further in \cref{sec:instanton}.

In the next section we discuss how to map the membrane exchange model to a 2-form lattice gauge theory.
Before moving on, however, the existence of an RK point naturally begs the question of whether 2-form spin liquids, either quantum or classical, can be found in dimer models with local constraints~\cite{rokhsarSuperconductivityQuantumHardCore1988,moessnerShortrangedResonatingValence2001,moessnerThreedimensionalResonatingvalencebondLiquids2003,huseCoulombLiquidDimer2003,moessnerQuantumDimerModels2011,gingrasQuantumSpinIce2014}.
For example, spin ice can be mapped to a dimer model by mapping the two Ising spin states, $\sigma_i^z=\pm 1$, to the presence or absence of a dimer on the corresponding diamond lattice edge, such that the constraint $Q_t = 0$ translates to having two dimers touching at every diamond vertex. Then every spin ice ground state corresponds to fully-packed closed-loop coverings of the diamond lattice~\cite{gingrasQuantumSpinIce2014}, with similar loop-gas descriptions widely utilized for studying Coulomb phases~\cite{moessnerQuantumDimerModels2011,henleyCoulombPhaseFrustrated2010}.  
Na\"{i}vely applying this map to the spin vorticity model, the constraint $\omega_p=0$ translates to a non-local constraint on dimers sitting on the edges of the direct lattice hexagonal plaquettes. 
A more natural description would involve replacing 1-dimensional dimers by 2-dimensional hard plaquettes~\cite{pankovResonatingSingletValence2007} on the dual lattice, which then touch at each dual edge. 
One may then seek  to find a basis for which the constraint $\cramped{\omega_p = 0}$ translates to the constraint that exactly three hard plaquettes touch at each dual edge. 
However, such a basis would correspond to choosing an orientation of the edges of the direct diamond lattice with maximal circulation around every loop, which is not possible (\cref{apx:dimers-hard-hexagons}).
Further work is required to determine how to realize 2-form spin liquids in either dimer or hard-plaquette models.

\subsection{2-Form Lattice Gauge Theory}
\label{sec:2-form-lattice-gauge-theory}

We now demonstrate how to map the membrane-exchange Hamiltonian, \cref{eq:H-mem-exch}, to a lattice gauge theory in a way that parallels the analogous mapping of the ring exchange model of QSI~\cite{hermelePyrochlorePhotonsU12004}.
The key difference is that the membrane exchange model maps to a 2-form U(1) gauge theory, while the ring exchange model maps to a conventional 1-form gauge theory.
This means that for the membrane exchange model, the electric field $E$ and canonically conjugate compact U(1) gauge potential $A$ are 2-form operators on the \emph{plaquettes} of the dual lattice, whereas for the ring exchange model of QSI~\cite{hermelePyrochlorePhotonsU12004}, they are 1-form operators on \emph{links} of the direct lattice. 
For the interested reader, a brief overview of continuum differential forms and $p$-form U(1) gauge theory is provided in \cref{apx:2-form-gauge-theory}.

While 2-form gauge theory may seem esoteric, its study dates back to seminal work by Kalb and Ramond~\cite{kalbClassicalDirectInterstring1974} in string theory to describe string-string interactions.
Two-form gauge fields also arise in superfluids by a duality transformation, where they describe the interaction of vortex strings~\cite{lundUnifiedApproachStrings1976,kaoQuantumNucleationVortex1995,quevedoPhasesAntisymmetricTensor1997,nastaseParticleVortexDuality2019,fischerTheoryVortexQuantum1998,fischerMotionQuantizedVortices1999,savitDualityFieldTheory1980,davisGlobalStringsSuperfluid1989,orlandExtrinsicCurvatureDependence1994,vilenkinCosmicStringsOther1994}.
They have also been used to describe the radiation of cosmic strings resulting from symmetry breaking in the early universe~\cite{battyeGlobalStringRadiation1994,quevedoCondensationPBranesGeneralized1997,vilenkinCosmicStringsOther1994}. 
More recently, the study of higher-form gauge fields have been gaining significant attention in the context of ``higher-form symmetries''---symmetries acting on extended $p$-dimensional objects~\cite{gaiottoGeneralizedGlobalSymmetries2015,lakeHigherformSymmetriesSpontaneous2018,mcgreevyGeneralizedSymmetriesCondensed2023}.
Put simply, the presence of a $k$-form symmetry implies the conservation of flux through $(d-k)$-dimensional surfaces, where $d$ is the dimension of space. 
For example, in ordinary Maxwell gauge theory without charged matter there is a 1-form symmetry corresponding to conservation of electric flux through closed surfaces~\cite{mcgreevyGeneralizedSymmetriesCondensed2023}. 
Higher-form symmetries are part of a broader notion of ``generalized symmetries'', a broad term encompassing ordinary, higher-form, subsystem, non-invertible, categorical, and more types of symmetries~\cite{gaiottoGeneralizedGlobalSymmetries2015,mcgreevyGeneralizedSymmetriesCondensed2023}.
Such symmetries are rather common as emergent symmetries at low energies~\cite{paceEmergentGeneralizedSymmetries2024} or at phase transitions~\cite{bhardwajLatticeModelsPhases2024, chatterjeeQuantumPhasesTransitions2024}.
Indeed, higher-form gauge fields appear in characterizing various types of topological order~\cite{kapustinHigherSymmetryGapped2017,bullivantTopologicalPhasesHigher2017,huxfordExcitationsHigherlatticeGauge2023a,huxfordExcitationsHigherlatticeGauge2023,huxfordExcitationsHigherlatticeGauge2024}, for example 2-form gauge fields arise in BF theories and their generalizations~\cite{horowitzQuantumFieldTheoretic1990,hanssonSuperconductorsAreTopologically2004,choTopologicalBFField2011, kapustinTopologicalFieldTheory2014,kapustinHigherSymmetryGapped2017}, the topological response of symmetry protected topological orders (SPT's) protected by higher-form symmetries~\cite{yoshidaTopologicalPhasesGeneralized2016,verresenHiggsCondensatesAre2022,thorngrenHiggsCondensatesAre2023,paceTopologicalAspectsBrane2024}, higher-form generalizations of Berry phase topological invariants~\cite{palumboTensorBerryConnections2019,kapustinHigherdimensionalGeneralizationsBerry2020,hsinBerryPhaseQuantum2020}, and in the theory of multipole topological phases~\cite{dubinkinHigherformGaugeSymmetries2020} and fracton gauge theories~\cite{hironoSymmetryPrincipleGauge2024}.

The formalization of the generalized symmetries framework has recently led to an explosion of work, with one major advance being the significant extension of Landau's symmetry classification of phases to include many exotic phases such as topological orders~\cite{gaiottoGeneralizedGlobalSymmetries2015,mcgreevyGeneralizedSymmetriesCondensed2023,bhardwajIllustratingCategoricalLandau2024}.
In particular, deconfined phases of gauge theories, and by extension quantum spin liquids, can be understood as spontaneously breaking the higher-form electric symmetry of the gauge field, i.e. the ground state is a sort of coherent state of closed electric field configurations~\cite{mcgreevyGeneralizedSymmetriesCondensed2023}.
In this language, the ground state manifold of NNSI, consisting of closed string configurations, has an emergent 1-form symmetry, while the ground state manifold of the spin vorticity model, consisting of closed membrane configurations, has an emergent 2-form symmetry.
In the quantum theory, the QSL phase corresponds to spontaneously breaking this symmetry.\footnote{
    These 1- and 2-form symmetries are explicitly broken at finite energies by the presence of sources and sinks of the electric fields (point charges in spin ice, string charges in the vorticity model) but, since the symmetry-breaking excitations are gapped, these symmetries are emergent at energies below the charge gap~\cite{mcgreevyGeneralizedSymmetriesCondensed2023,paceEmergentGeneralizedSymmetries2024}.
}

As we will demonstrate in detail in the remainder of this section, the classical spin vorticity model plus quantum perturbations can be mapped to a 2-form U(1) lattice gauge theory~\cite{pearsonPhaseStructureAntisymmetric1982,reyHiggsMechanismKalbRamond1989}.
To proceed, we require a notation to refer to the oriented edges and plaquettes of the direct and dual lattices.
We use $\ell$ to denote an oriented link and $p$ to denote an oriented plaquette, both of the direct diamond lattice. 
We will use a tilde, $\tilde{\ell}$ and $\tilde{p}$, to indicate the oriented links and plaquettes of the dual diamond lattice.
We use a minus sign to denote reversing the orientations, e.g. $-\ell$ denotes the reversed orientation of link $\ell$. 
Lastly, we use the notation $\boundary$ to denote the $(k-1)$-dimensional \emph{boundary} of a $k$-dimensional cell. 
For example, the boundary of a 2-dimensional oriented direct plaquette~$p$, denoted $\cramped{\boundary p}$, consists of six direct links $\ell$, each oriented to form a clockwise circulation by the right-hand rule relative to the plaquette orientation, such as illustrated by the red arrows in \cref{fig:hexagon}. 
This allows us to write the vorticity, \cref{eq:omega_p}, in the compact form\footnote{This notation is unambiguous, the sign $(-1)^p$ in \cref{eq:omega_p} is determined by the chosen orientation of $p$, while the signs in front of the $\sigma_i^z$ are determined by the orientation of each $\ell\in\boundary p$ relative to the chosen quantization axes of the spins.
}
\begin{equation}
    \omega_p = \sum_{\mathclap{\ell \in \boundary p}} S_{\ell}^z,
    \label{eq:omega_p_boundary}
\end{equation}
where $\cramped{\ell\in\boundary p}$ is to be read ``oriented links $\ell$ on the boundary of oriented plaquette $p$'', and we identify $S_{-\ell}^z \equiv -S_{\ell}^z$. 
\Cref{eq:omega_p_boundary} is a more intrinsic, basis-independent way of writing $\omega_p$, whereas \cref{eq:omega_p} was written after choosing the $A$ to $B$ quantization axis convention. 
\Cref{eq:omega_p_boundary} may be read intuitively as a discrete (lattice) analog of the line integral
\begin{equation}
\omega_p \sim \oint_{\boundary p} \bm{S}\cdot\mathrm{d}\bm{\ell}.
\end{equation}

To map the membrane exchange Hamiltonian, \cref{eq:H-mem-exch}, to a lattice gauge theory, we associate each spin to an oriented plaquette $\tilde{p}$ of the dual diamond lattice, identifying the $S^z$ component as the 2-form electric field and the corresponding raising and lowering operators $S^\pm$ to the 2-form gauge potential, via the mapping\footnote{As an intermediate step, one may first map to quantum rotors~\cite{hermelePyrochlorePhotonsU12004,bentonSeeingLightExperimental2012}.
    }
\begin{equation}
    S^z_{\tilde{p}} \to E_{\tilde{p}}, 
    \quad S^{\pm}_{\tilde{p}} \to e^{\pm i A_{\tilde{p}}},
    \label{eq:spin_to_gauge_map}
\end{equation}
where the electric field $E$ and gauge potential $A$ satisfy the standard canonical commutation relation $\cramped{[A_{\tilde p},E_{\tilde p}] = i}$.
We associate the reversal of the plaquette orientation ($\cramped{\tilde{p} \to - \tilde{p}}$) to the following transformation of spin and corresponding gauge field operators,
\begin{alignat}{3}
    S_{-\tilde{p}}^z &\equiv -S_{\tilde p}^z 
    \quad &\Rightarrow  \quad
    E_{-\tilde p} &\equiv -E_{\tilde p}, 
    \nonumber \\
    S^{\pm}_{-\tilde{p}}&\equiv S_{\tilde{p}}^{\mp} 
    \quad &\Rightarrow \quad 
    A_{-\tilde p} &\equiv -A_{\tilde p}.
    \label{eq:inversion_sign}
\end{alignat}
The compact U(1) vector potential operators $A_{\tilde p}$ have eigenvalues in $\cramped{[0,2\pi)}$. 
In ordinary U(1) lattice gauge theory the electric field operators would then have integer eigenvalues, but the physical spin states correspond to $\cramped{E_{\tilde p} = \pm 1/2}$, so the electric field must take half-odd-integer values, $\cramped{E_{\tilde p}\in \integers + 1/2}$, meaning that the gauge field is \emph{frustrated}~\cite{hermelePyrochlorePhotonsU12004}, or ``odd''~\cite{gingrasQuantumSpinIce2014}.\footnote{The ``oddness'' of the gauge field, i.e. the half-odd-integer spectrum of the electric field operators, is equivalent in a first-quantized description to considering \emph{anti-periodic} wave functions of the compact gauge potential $A_{\ell} \in [0,2\pi)$ (equivalent to the angular coordinate of a quantum rotor)~\cite{kwasigrochSemiclassicalApproachQuantum2017}.
    }
The physical spin states are recovered by going to the strongly-coupled limit where the $E_{\tilde p}$ take their minimum eigenvalues.

Applying the map \cref{eq:spin_to_gauge_map} to the membrane exchange Hamiltonian, \cref{eq:H-mem-exch}, we obtain a compact 2-form lattice gauge theory Hamiltonian,
\begin{equation}
    H_{2\text{-LGT}} = \frac{U}{2} \sum_{\tilde{p}} E_{\tilde{p}}^2 - K \sum_{\vtiny} \cos\left[B(\vtiny)\right],
    \label{eq:H-2-form}
\end{equation}
where the second term comes directly from \cref{eq:H-mem-exch}, with $\cramped{K \sim J_{\text{mem}}}$, while the added first term controls the magnitude of the electric field. Here, $B$ is the generalized magnetic field, a \emph{3-form} associated to each 3-dimensional dual cell, given by
\begin{equation}
    B({\vtiny})\,\, = \sum_{\mathclap{\tilde{p}\, \in\, \boundary\,\vtiny}} A_{\tilde{p}}.
    \label{eq:3-form-B}
\end{equation}
Here, the sum is taken over the four oriented dual plaquettes $\tilde{p}$ making up the boundary ($\boundary$) of the oriented 3-dimensional dual cell $\vtiny$. 
The original spin-1/2 membrane exchange model corresponds to the strong-coupling limit, $U\to\infty$, of this compact 2-form gauge theory, which forces the frustrated electric field to take the physically allowed values $E_{\tilde p}=\pm 1/2$.

\begin{figure}
    \centering
    \begin{overpic}[width=\columnwidth]{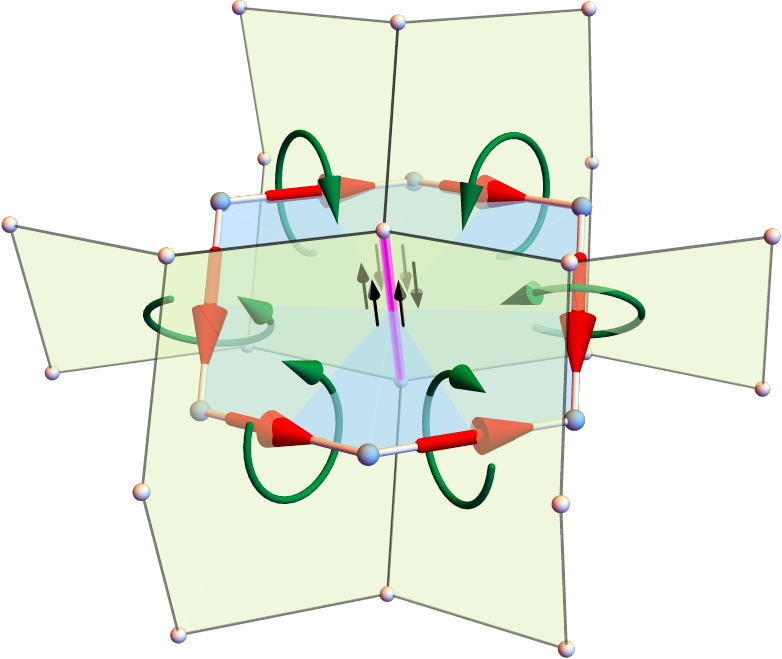}
        \put(55,46){$\tilde{\ell}$}
    \end{overpic}
    \caption{An illustration of the 2-form Gauss law, \cref{eq:2-form-gauss-law-continuum} in the continuum or \cref{eq:1-form-gauss-spins} on the lattice. We represent the 2-form electric field configurations by drawing a circulating arrow (green) on each dual plaquette (light green). In the direct lattice, these translate to the spin configuration (red arrows), where the circulation of the green arrows is given by right hand rule relative to the spins. The 2-form Gauss law says that at each dual edge $\tilde{\ell}$ (pink) the 2-form electric field on the six dual plaquettes (light green) which touch it (i.e. form its coboundary, $\boundary^\dagger \tilde{\ell}$) add to zero. Here the configuration shows that three of the green arrows circulate in one sense relative to $\tilde{\ell}$ while the other three circulate in the opposite sense. Each plaquette's contribution is indicated by a small black arrow next to the central link, three of which go ``up'' and three ``down''. 
    We think of this as a generalized (higher-form) zero-divergence condition. Correspondingly, the six spins (red arrows) have zero vorticity around the direct plaquette (blue) pierced by $\tilde{\ell}$, with three going clockwise and three counter-clockwise.
    Non-zero vorticity excitations violate the generalized zero-divergence constraint, corresponding to a net imbalance of the six 2-form electric fields at the central link, i.e. a non-zero string charge $\mathscr{j}_{\tilde{\ell}}$ on the pink dual edge sourcing the electric field via the Gauss law \cref{eq:1-form-gauss-spins}.
    }
    \label{fig:two-form-gauss-law}
\end{figure}

The Hamiltonian \cref{eq:H-2-form} is invariant under a gauge transformation of the 2-form gauge potential, 
\begin{equation}
    A_{\tilde{p}}\to A_{\tilde{p}} - \sum_{\mathclap{\cramped{\tilde{\ell} \in \boundary \tilde{p}}}} \theta_{\tilde{\ell}}\,,
    \label{eq:1-form-gauge}
\end{equation}
where the sum is over the six dual links $\tilde{\ell}$ on the boundary of the dual plaquette $\tilde{p}$ with appropriate orientation (six black arrows surrounding the green plaquette in \cref{fig:diamond_dual_membranes}(c)), 
and the $\theta_{\tilde{\ell}}\in[0,2\pi)$ are arbitrary compact 1-form variables.\footnote{
    \cref{eq:3-form-B} may be viewed as the discrete lattice equivalent of the integral equality
    \[
        \iiint_{\vtiny} B = \oiint_{\boundary\,\vtiny}A,
    \]
    where $\boundary$ denotes the boundary of the 3-dimensional volume, $A$ is a 2-form and $B$ is a 3-form. By Gauss' theorem, we can then identify $B \simeq \nabla\cdot A$, which is the 2-form version (in three spatial dimensions) of the more familiar $B=\nabla\times A$ when $A$ is a 1-form and $B$ is a 2-form.
    In the gauge transformation \cref{eq:1-form-gauge}, the sum may be viewed as the discrete analogue of the integral expression
    \begin{equation*}
        \iint_{\tilde p} A \to \iint_{\tilde p} A - \oint_{\boundary \tilde p} \theta.
    \end{equation*}
    By Stoke's theorem, this transformation then corresponds to $A \to A + \nabla\times \theta$. 
    This is the 2-form generalization of the usual transformation for a 1-form vector potential, $A \to A - \nabla\lambda$. 
    See \cref{apx:2-form-gauge-theory} for further discussion. 
    }
The gauge invariance of the Hamiltonian enforces that physical states are created by gauge-invariant operators. 
Since $A_{\tilde p}$ generates the creation and annihilation of the electric field, gauge-invariant electric field eigenstates are therefore generated by ``Wilson surface'' operators of the form
\begin{equation}
    W[\tilde{\Sigma}] = \exp(i\sum_{\tilde{p}\in \tilde{\Sigma}} A_{\tilde p}),
    \label{eq:wilson_surface}
\end{equation}
where $\tilde{\Sigma}$ is a \emph{closed} surface in the dual lattice, such as those depicted in \cref{fig:small_membranes}(b,c).
The Hilbert space thus contains only closed electric membrane configurations.
Equivalently, this gauge symmetry enforces a generalized Gauss law---the net 2-form electric flux emanating from each dual link $\tilde\ell$ is zero.
This can be stated as
\begin{equation}
    \sum_{\tilde{p}\in \boundary^\dagger \tilde{\ell}}E_{\tilde p} = 0,
    \label{eq:2-form-gauss}
\end{equation}
where the sum is over all plaquettes in the ``coboundary'' ($\boundary^\dagger$) of dual link $\tilde{\ell}$, which is defined by $\boundary^\dagger \tilde{\ell} = \cramped{\{\tilde{p} \vert \tilde{\ell}\in\boundary\tilde{p}\}}$, to be read ``the coboundary of oriented dual link $\tilde{\ell}$ is the set of all oriented dual plaquettes $\tilde{p}$ which contain $+\tilde{\ell}$ on their oriented boundary''.
This is illustrated in \cref{fig:two-form-gauss-law}, where we depict the 2-form electric field configuration by a set of circulating arrows (green) on the face of each dual plaquette (light green hexagons). These six dual plaquettes form the coboundary of the central dual edge $\tilde{\ell}$ (pink). 
Three of them circulate in one direction relative to $\tilde{\ell}$ and three in the opposite direction, yielding zero net sum in \cref{eq:2-form-gauss}. 
This is indicated by the six small black arrows next to $\tilde{\ell}$, three of which go ``up'' and three ``down'', denoting the contribution to the net flux along $\tilde{\ell}$ coming from each of the six dual plaquettes.
The 2-form electric field configuration (circulating green arrows), corresponds by the right-hand rule to the spin configuration (red arrows) on the direct lattice edges. 
The spins have zero net vorticity around the direct plaquette (blue) pierced by $\tilde{\ell}$, such that the 2-form Gauss law, \cref{eq:2-form-gauss}, is precisely equivalent to the constraint that the vorticity, \cref{eq:omega_p_boundary}, is zero.

The 2-form U(1) quantum spin liquid corresponds to the deconfined phase of the 2-form compact U(1) gauge theory defined by the Hamiltonian~\cref{eq:H-2-form}, in which there is a long-wavelength gapless excitation that generalizes the photon of the 1-form compact U(1) gauge theory~\cite{hermelePyrochlorePhotonsU12004,bentonSeeingLightExperimental2012}.
At the mean field level of the unfrustrated lattice gauge theory, this occurs at small $U/K$~\cite{reyHiggsMechanismKalbRamond1989} which, in line with the arguments put forward in the QSI case~\cite{hermelePyrochlorePhotonsU12004}, may survive in the large-$U/K$ limit in the frustrated model~\cite{hermelePyrochlorePhotonsU12004}.

\subsection{2-Form Gauge-Higgs Model}
\label{sec:2-form-gauge-Higgs-model}

Thus far, we have effectively discussed the zero-charge sector of the theory consisting of only closed membrane configurations, without consideration of open-membrane excitations.
We now proceed to incorporate the string charge excitations into the lattice gauge theory framework.
We adapt the method introduced by Ref.~\cite{savaryQuantumCoherenceQuantum2021} for QSI.
There, the excitations are the electric charges of the gauge theory.
These can be explicitly treated in the lattice gauge theory formalism by embedding the spin Hilbert space in an auxiliary Hilbert space, supplemented with additional bosonic degrees of freedom located on the vertices of the diamond lattice which represent the charge. 
In other words, the $Q_t$ in \cref{eq:Q_t} are treated as independent degrees of freedom.
Projecting back to the physical subspace where \cref{eq:Q_t} is satisfied~\cite{savaryQuantumCoherenceQuantum2021}, now interpreted as a Gauss law coupling the charges to the gauge field, one effectively maps the QSI model to a gauge-Higgs model~\cite{fradkinPhaseDiagramsLattice1979,savitDualityFieldTheory1980}.

Since the analogue of the point-like spinons of QSI in the spin vorticity model are strings (\cref{sec:string_excitations}), we must take a modified approach to make these string excitations manifest. 
We augment the spin Hilbert space with a bosonic 1-form field taking integer eigenvalues, denoted $\mathscr{j}_{\tilde\ell}$, on the oriented dual links $\tilde{\ell}$, with canonically conjugate compact operators $\mathscr{a}_{\tilde \ell}\in[0,2\pi)$ satisfying $\cramped{[\mathscr{a}_{\tilde \ell},\mathscr{j}_{\tilde \ell}]=i}$. 
The integer eigenvalues of $\mathscr{j}_{\tilde \ell}$ count the number of strings on each dual link, or equivalently the vorticity $\omega_p$ on the corresponding direct plaquette pierced by $\tilde{\ell}$. 
Note that we can think of $\mathscr{j}$ as a (unfrustrated) 1-form electric field and $\mathscr{a}$ as its corresponding 1-form U(1) gauge potential.

This 1-form ``string field'' is then ``slaved'' to the original spins by treating the strings as sources of the spin vorticity, \cref{eq:omega_p_boundary}, or equivalently as sources of the 2-form electric field, via the constraint
\begin{equation}
    \sum_{\mathclap{\ell\in\boundary p}} S^z_{\ell} = \mathscr{j}_{\tilde\ell}
    \quad
    \Leftrightarrow
    \quad   
    \sum_{\tilde{p} \in \boundary^\dagger \tilde{\ell}}
    E_{\tilde p} 
    = 
    \mathscr{j}_{\tilde \ell} 
    \,,
    \label{eq:1-form-gauss-spins}
\end{equation}
where $\tilde{\ell}$ is the link dual to plaquette $p$.
The first formulation states that the vorticity (circulation of $S^z$ around the boundary of direct plaquette $p$) is equal to the number of strings piercing the plaquette, i.e. it reads ``$\nabla\times\bm{S}^z = \bm{\mathscr{j}}$''. 
In this form, the constraint reads as an Ampere-like law, with a ``current'' acting as source of the circulation of a field. 
The second, equivalent, formulation is a generalized Gauss law, enforcing that the amount of 2-form electric flux emanating from the dual link $\tilde{\ell}$ is equal to the total string charge on that link, generalizing \cref{eq:2-form-gauss} in the presence of charged sources. 
We refer to this as a 2-form Gauss law, taking in the continuum the form $\partial_j E_{ij} = \mathscr{j}_i$, \cref{eq:2-form-gauss-law-continuum}.

We can now map the spin Hamiltonian into the enlarged Hilbert space of a 2-form gauge-Higgs model~\cite{reyHiggsMechanismKalbRamond1989,lozanoEmergentKalbRamondFields2019}.
The previous map \cref{eq:spin_to_gauge_map} does not account for the creation or annihilation of strings when flipping spins that form an open membrane, and so was only accurate for operators which only flip closed membranes.
Flipping a single spin in a ground state creates a string excitation attached to the edge of an electric flux membrane on a dual plaquette, as depicted in \cref{fig:diamond_dual_membranes}(c). 
Therefore \cref{eq:spin_to_gauge_map} must be modified to account for exciting the string field. 
This is achieved via the map\footnote{
    One can restrict the Hilbert space of the gauge-Higgs model to the physical spin states with eigenvalues $\pm 1/2$~\cite{savaryCoulombicQuantumLiquids2012,savaryQuantumCoherenceQuantum2021}, which reduces the gauge-Higgs model to a 2-form generalization of a quantum link model~\cite{chandrasekharanQuantumLinkModels1997}. See Ref.~\cite{savaryQuantumCoherenceQuantum2021} for further discussion of projecting from the unrestricted gauge-Higgs model back to the physical spin Hilbert space in the context of QSI. 
}
\begin{equation}
    S_{\tilde p}^z \to E_{\tilde p}^z, 
    \quad 
    S_{\tilde p}^{\pm} \to e^{\pm i A_{\tilde{p}}}\prod_{\tilde{\ell} \in \boundary \tilde{p}} e^{\pm i \mathscr{a}_{\tilde \ell}}.
    \label{eq:gauge-higgs-mapping}
\end{equation}
Referring to \cref{fig:diamond_dual_membranes}(c), 
the product of $e^{i\mathscr{a}_{\tilde\ell}}$ on the links surrounding the dual plaquette (green) creates a unit charged string on the plaquette boundary (red), while $e^{i A_{\tilde p}}$ inserts a unit electric flux through the plaquette.
By construction, these operators can never create states that violate the Gauss law. 
As such, they are invariant under the (higher-form) gauge transformation
\begin{equation}
    \mathscr{a}_{\tilde\ell}\to \mathscr{a}_{\tilde\ell} + \theta_{\tilde \ell}
    \quad 
    A_{\tilde p} \to A_{\tilde p} - \sum_{\tilde{\ell} \in \boundary \tilde{p}}\theta_{\tilde \ell}  ,
    \label{eq:1-form-gauge-transform}
\end{equation}
for an arbitrary compact 1-form $\theta_{\tilde \ell}$.
Products of spin flip operators generate gauge-invariant ``dressed Wilson surface'' operators of the form
\begin{equation}
    W[\tilde{\Sigma}] =
    \prod_{\tilde p \in \tilde{\Sigma}} S^+_{\tilde p}
    \equiv
    \exp(
        i\sum_{\tilde p \in \tilde{\Sigma}} A_{\tilde p} 
        \,+\, 
        i \sum_{\tilde{\ell}\in\boundary \tilde{\Sigma}} \mathscr{a}_{\tilde\ell}
        ),
    \label{eq:dressed_wilson_surface}
\end{equation}
where $\tilde{\Sigma}$ can be any \emph{open} surface in the dual lattice.\footnote{Equation \cref{eq:dressed_wilson_surface} is the discrete analog of the continuum expression
    \begin{equation*}
        \cramped{
            W[\tilde{\Sigma}]
            \sim
            \exp(
                i \iint_{\mathrlap{\tilde{\Sigma}}}\,\,\bm{A}\cdot\diff\bm{S} 
                + 
                i \oint_{\mathrlap{\boundary \tilde{\Sigma}}} \,\,\,\bm{\mathscr{a}}\cdot\diff\bm{\ell}
            )
        }.
\end{equation*}
}
Enforcing the Gauss law constraint \cref{eq:1-form-gauss-spins}  ``glues'' the 1-dimensional strings to the 2-form electric flux membranes---a string always comes attached to the edge of an open electric membrane and, conversely, electric membranes can only terminate on strings.
This constraint on the electric field $E$ and charge operator $\mathscr{j}$ is equivalent to the gauge-redundancy of their conjugate compact potentials $A$ and $\mathscr{a}$, \cref{eq:1-form-gauge-transform}: the Gauss law says charge cannot be created without its attached flux, and gauge invariant operators are precisely the ones that do not violate this constraint.

There is an additional gauge redundancy associated to the 1-form string field,
\begin{equation}
    \mathscr{a}_{\tilde\ell} \to \mathscr{a}_{\tilde\ell} - \sum_{\tilde t \in \boundary \tilde\ell} \lambda_{\tilde t},
    \label{eq:gauge-of-gauge}
\end{equation}
where $\lambda$ is an arbitrary compact 0-form taking values on the dual diamond vertices $\tilde{t}$.
This arises from the fact that the operators in \cref{eq:gauge-higgs-mapping} can only create closed-string excitations. 
In other words, it corresponds to the constraint
\begin{equation}
    \sum_{\tilde{\ell}\in\boundary^\dagger \tilde{t}}
\,\,\mathscr{j}_{\tilde{\ell}} = 0,
    \label{eq:string_gauss_law}
\end{equation}
where the sum is over links in the coboundary of $\tilde{t}$, i.e. all oriented dual links $\tilde{\ell}$ which end at dual vertex $\tilde{t}$. 
This says that the number of strings entering a vertex is equal to the number of strings leaving a vertex, or in other words, the divergence of $\mathscr{j}$ is zero.

\subsection{Quantum String Dynamics and Gauge Mean Field Theory}

We have demonstrated, via \cref{eq:gauge-higgs-mapping}, how to map spin operators to gauge-invariant operators of a 2-form gauge-Higgs model.
In principle this map turns any spin Hamiltonian into a frustrated (``odd'') 2-form gauge-Higgs Hamiltonian, for which one must then add an $E^2$ term and take the strong coupling limit to recover the physical Hilbert space. 
Such a map will only be particularly useful near the classically degenerate point, otherwise it will map somewhere deep in a confined or Higgsed phase of the gauge theory.
Let us therefore consider perturbing the classical spin vorticity Hamiltonian, \cref{eq:H_vort_wp}, with the various possible symmetry-allowed nearest-neighbor interactions~\cite{rauFrustratedQuantumRareEarth2019} (of which \cref{eq:H_QSV} is only one possibility).
Consider a spin Hamiltonian of a more generic form 
than Eq.~\eqref{eq:H_QSV}~\footnote{This Hamiltonian, without the $J_{yz}$ interaction (contained in the $J_{z\pm}$ term) is appropriate to describe the rare-earth pyrochlore compounds with dipolar-octupolar doublets~\cite{rauFrustratedQuantumRareEarth2019}. For pseudo-spin-1/2 and non-Kramers doublets, additional bond-dependent phase factors would have to be included~\cite{rauFrustratedQuantumRareEarth2019}.
    }
\begin{align}
    H = H_{\text{SV}} 
    &+ J_{\pm\pm} \sum_{\langle ij \rangle} (S_i^+ S_j^+ + S_i^- S_j^-)\nonumber \\
    &- J_{\pm} \sum_{\langle ij \rangle} (S_i^+ S_j^- + S_i^- S_j^+) \nonumber \\
    &+ J_{z\pm} \sum_{\langle ij \rangle} [(S_i^+ S_j^z + S_i^z S_j^-) + \mathrm{h.c.}].
    \label{eq:H-perturbed}
\end{align}
The map \cref{eq:gauge-higgs-mapping} converts this Hamiltonian into a 2-form gauge-Higgs model, i.e. a collection of gauge-invariant operators which insert or remove fluxes and create or destroy strings. 
We now discuss how these terms act to create and transport (``hop'') the string excitations. 

First note that, by the 2-form Gauss law~\cref{eq:1-form-gauss-spins}, the unperturbed spin vorticity Hamiltonian maps to
\begin{equation}
    H_{\text{SV}} \to J \sum_{\tilde \ell} {\mathscr{j}_{\tilde \ell}}^2,
\end{equation}
which is an exact rewriting of \cref{eq:H_vort_wp} by identifying the string field $\mathscr{j}$ on the dual links with the vorticity $\omega$ on the direct plaquettes.
This nicely demonstrates that the classical spin vorticity is indeed a model of closed strings on the dual diamond lattice, subject to the closed-string condition 
\cref{eq:string_gauss_law}.
Note that since $J$ is, by assumption, the largest energy scale in the Hamiltonian, the (unfrustrated) 1-form string field is deep in its confined phase.
This is equivalent to the statement that the strings cost energy proportional to their length, i.e. they carry a positive string tension, as discussed in Section~\ref{sec:string_excitations}.

We already considered the $S_i^{\pm} S_j^{\pm}$ term in \cref{eq:H_QSV}. 
Acting on a ground state, this generates a single string along the edges of two dual plaquettes with energy $8J$, and upon a second application, it generates the membrane exchange term discussed in \cref{sec:membrane_exchange}, as illustrated in \cref{fig:membrane_exchange}, effectively by creating a string, wrapping it around a closed surface, and re-annihilating it.
This process perturbatively generates the membrane exchange, \cref{eq:H-mem-exch}, or equivalently the magnetic field term, $-K \sum_{\vtiny} \cos\left[B(\vtiny)\right]$ in \cref{eq:H-2-form}. 

\begin{figure}[t!]
    \centering
    \vspace{2ex}
    \begin{overpic}[width=.7\columnwidth]{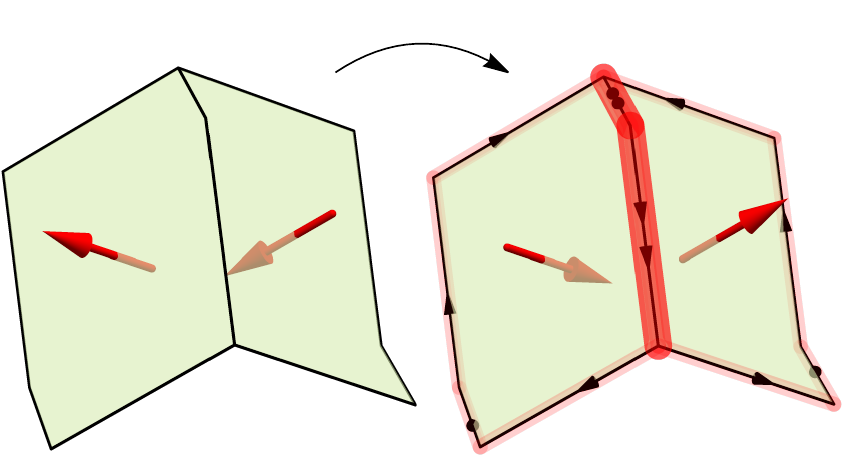}
        \put(-10,50){(a)}
        \put(-10,-20){(b)}
        \put( 43, 54){$S_i^\pm S_j^{\mp}$}
    \end{overpic}
    \\[5ex]
    \begin{overpic}[width=.7\columnwidth]{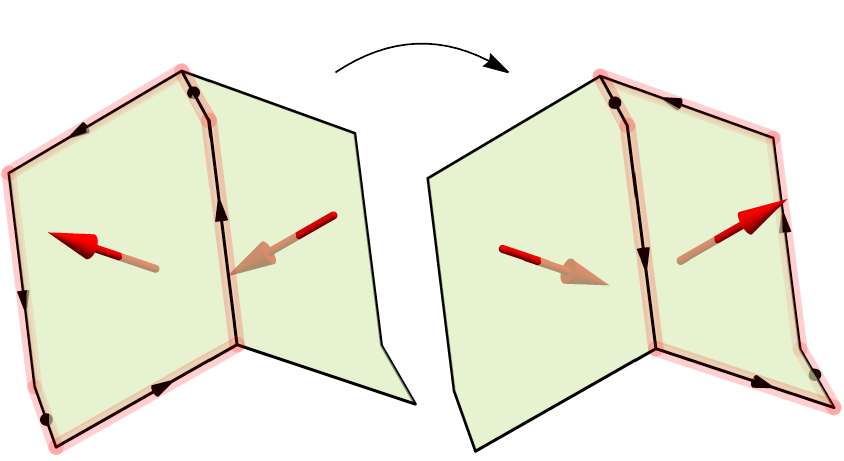}
        \put( 43, 54){$S_i^\pm S_j^{\mp}$}
    \end{overpic}
    \\[1ex]
    \caption{
    (a) Acting on a ground state, the $S_i^{\pm} S_j^{\mp}$ term flips two spins, creating a pair of strings on adjacent plaquettes with opposite circulation. This is a highly energetic state with energy $16J$, since the two strings have the same charge on their overlap on two dual edges, illustrated with two arrowheads and a thicker red line. 
    (b) Acting on a configuration with a single minimal string, this term hops the string by one plaquette, generating the string kinetic energy. 
    }
    \label{fig:flip_flop}
\end{figure}

Next, consider the $S^{\pm} S^{\mp}$ term. Acting on a ground state with no string excitations, this creates a pair of minimal strings on neighboring plaquettes with opposite circulation, illustrated in \cref{fig:flip_flop}(a). 
This state has classical energy $16J$, due to the extra energy of two strings overlapping. 
For this reason, application of this term alone does not generate tunneling between classical ground states at low orders in perturbation theory. 
On the other hand, acting on a state with a single minimal string, this term serves to hop the string to a neighboring plaquette, as illustrated in \cref{fig:flip_flop}(b). 
Thus this term generates a ``kinetic energy'' for the strings. 
Note, however, defining such a kinetic energy of strings is not immediately straightforward since strings are extended objects which have internal configurational (i.e. ``vibrational'') modes. 
It may be possible to define a band dispersion for the shortest strings, treated as particles hopping on the direct pyrochlore lattice sites (centers of dual lattice plaquettes), plus broadening corrections from longer string configurations. 

The last term in \cref{eq:H-perturbed}, $S_i^z S_j^{\pm}$, may be interpreted as a sort of transverse field, albeit weighted by the surrounding spin configuration, i.e. it can be written using operators $S_i^{\pm} (\sum_{j@i} S_j^z)$, where the second sum is over all neighbors of site $i$. 
An effective, albeit random, transverse field term may also be generated in non-Kramers systems via quenched crystalline disorder splitting the ionic ground state doublet~\cite{savaryDisorderInducedQuantumSpin2017,bentonInstabilitiesU1Quantum2018}.
Such transverse field terms serve simply to create and destroy minimal strings by flipping individual spins. 
Further terms in \cref{eq:H-perturbed} could include next-neighbor spin-spin interactions or three-spin or higher interactions, all of which map to gauge-invariant operators in the gauge-Higgs language. 
Beyond the minimal strings, any operator that flips spins can act to create or destroy strings, to make strings longer or shorter, to deform a string, or to merge or split strings, by flipping spins on plaquettes touching the string.

After having mapped a spin model of the form of \cref{eq:H-perturbed} to a 2-form gauge-Higgs model via \cref{eq:gauge-higgs-mapping}, one may, in principle, attempt a mean field decoupling to obtain a mean field ground state phase diagram for the spin Hamiltonian, an approach dubbed gauge mean field theory~\cite{savaryCoulombicQuantumLiquids2012,leeGenericQuantumSpin2012,savarySpinLiquidRegimes2013,savaryQuantumCoherenceQuantum2021}. 
In the equivalent mapping utilized for QSI~\cite{savaryCoulombicQuantumLiquids2012}, each $S_{\ell}^{\pm}$ (treated as 1-form operators) decomposes into three operators---one on the link and two on the endpoints---such that after substitution into the quadratic spin Hamiltonian, one obtains terms with up to six operators~\cite{leeGenericQuantumSpin2012}. 
In the mapping given by \cref{eq:gauge-higgs-mapping}, on the other hand, each $S_{\tilde p}^{\pm}$ (treated as 2-form operators) decomposes into seven operators---one on the dual plaquette, $\exp(\pm i A_{\tilde{p}})$, and six on its boundary links $\exp(\pm i \mathscr{a}_{\tilde{\ell}})$, such that one obtains terms containing up to fourteen operators for the bilinear spin-spin interactions, making the mean field decoupling considerably more complex than for QSI.
Notwithstanding the complexity of implementing such a program, we have demonstrated how to map, via \cref{eq:gauge-higgs-mapping}, a spin model to a 2-form U(1) gauge-Higgs model, with the putative 2-form U(1) QSL phase corresponding to the deconfined phase of this gauge theory.
Neighboring ordered phases may then be interpreted either as confined phases of the gauge field, or Higgs phases where the string field condenses~\cite{reyHiggsMechanismKalbRamond1989,hayataStringConfinementTwoform2019}.

\subsection{Magnetic Sector: Instantons and Stability}
\label{sec:instanton}

Beyond the perturbative or mean-field level that we have looked at thus far, there are non-perturbative objects in the spectrum of the 2-form U(1) gauge theory, and thus of 2-form U(1) QSLs---the vortices of the gauge potential~$A$. 
These are the ``magnetic'' excitations, topological defects in the gauge field, directly analogous to the famous Dirac magnetic monopoles in 1-form quantum electrodynamics (QED).
Consider a compact $\pe$-form field (``$\mathrm{e}$'' for ``electric''), e.g. a compact scalar ($\pe=0$) or compact vector potential $(\pe=1)$. 
Compactness allows the field to wind non-trivially around a $(\pe+2)$-dimensional surface, creating a vortex. 
In a Lagrangian description, these vortices are $\pmag$-dimensional defects (m for magnetic), where $\cramped{\pmag = D-(\pe+2)}$ in $D$ spacetime dimensions, and are referred to as magnetic excitations~\cite{teitelboimMonopolesHigherRank1986}.  
In standard 1-form quantum electrodynamics in $\cramped{D=3+1}$ dimensions, $\cramped{\pmag=4-(1+2)=1}$, and these vortices are the worldlines of the Dirac monopoles, whose analogue in QSI are called visons~\cite{hermelePyrochlorePhotonsU12004,gingrasQuantumSpinIce2014}. 
In a Hamiltonian description, they are disorder operators which act to disorder the magnetic field configurations~\cite{fradkinDisorderOperatorsTheir2017} (see \cref{apx:2-form-gauge-theory}).

For the spin vorticity model 2-form U(1) QSL, described by a ($\cramped{\pe=2}$)-form U(1) lattice gauge theory in $\cramped{D=3+1}$ spacetime dimensions, we have $\pmag=0$, meaning that these non-perturbative defects are instantons, i.e. instantaneous changes in $B(\vtiny)$.\footnote{In a 4D Euclidean lattice gauge theory, e.g. if one were to Trotterize the Hamiltonian \cref{eq:H-2-form}, the compact 3-form magnetic field $B$ has integer winding number around each 4-dimensional cell, which may be viewed as an integer flux sourced by a 0-dimensional instanton sitting inside the cell, i.e. a ``magnetic charge'' localized in both space and time.}
Their appearance in 2-form U(1) gauge theory in $\cramped{D=3+1}$ dimensions is directly analogous to the case of QED in $\cramped{D=2+1}$ dimensions, called QED$_3$, where Polyakov famously demonstrated that these instantons destabilize the deconfined phase, gapping the ground state~\cite{polyakovQuarkConfinementTopology1977}. 
These instantons may therefore destabilize the gapless 2-form U(1) QSL, preventing the system from having a gapless deconfined phase and instead driving the system to a relatively trivial ground state (such as a valence bond crystal~\cite{lhuillierIntroductionQuantumSpin2011}).

\subsubsection{Higher-Form Symmetry Perspective}

Another perspective on this issue arises from the modern understanding of higher-form symmetries, where the Polyakov mechanism may be viewed as a generalization of the Mermin-Wagner theorem for spontaneously-broken $\pe$-form symmetries~\cite{lakeHigherformSymmetriesSpontaneous2018,mcgreevyGeneralizedSymmetriesCondensed2023,gaiottoGeneralizedGlobalSymmetries2015}.
The pure 2-form U(1) gauge theory, whose lattice Hamiltonian is given by \cref{eq:H-2-form}, has a 2-form U(1) symmetry given by shifting the 2-form gauge potential $A$ by a closed 2-form.\footnote{
    On the lattice, a closed 2-form $\lambda$ satisfies $\sum_{\tilde{p}\in\boundary \vtiny} \lambda_{\tilde p} = 0$.
    }
A subset of these transformations are the un-physical gauge transformations of \cref{eq:1-form-gauge}. 
The remaining ones generate a physical 2-form U(1) symmetry. 
The deconfined phase of the gauge theory is the spontaneously broken phase for this symmetry, and the gapless photon is the corresponding Goldstone mode.
In this language, the Polyakov mechanism may be reformulated as a generalization of the Mermin-Wagner theorem (originally formulated for $D=2$ and $\pe = 0$) which applies whenever $\pmag = 0$~\cite{lakeHigherformSymmetriesSpontaneous2018}.

An alternative description of the mechanism is that there is no symmetry that conserves the magnetic field $B$ in \cref{eq:H-2-form}.
Consider QSI, which realizes a 1-form U(1) gauge theory. 
In the 1-form U(1) gauge theory the absence of magnetic monopoles corresponds to the conservation of magnetic flux, i.e. a 1-form symmetry called magnetic symmetry~\cite{gaiottoGeneralizedGlobalSymmetries2015,mcgreevyGeneralizedSymmetriesCondensed2023}. 
This symmetry is explicitly broken by the presence of monopoles---visons in QSI---but is restored at energies below the monopole gap. 
Thus the dynamical vison gap means that QSI has an emergent magnetic 1-form symmetry, which is not itself a symmetry of the microscopic spin Hamiltonian but is necessary for the stability of the Coulomb phase~\cite{savaryQuantumSpinLiquids2016}. 
This is characteristic of higher-form symmetries: they are robust against moderate explicit breaking due to the gap of the defects which break them~\cite{mcgreevyGeneralizedSymmetriesCondensed2023,paceEmergentGeneralizedSymmetries2024}.
Ordinary (0-form) symmetries are not stable to such explicit breaking. 
In (3+1)-dimensional 2-form U(1) gauge theory the magnetic field $B$ is a scalar, thus the magnetic symmetry is a 0-form symmetry, not a higher-form symmetry. 
It is explicitly broken by the presence of instantons, i.e. the 0-form magnetic field $B$ is not conserved, destabilizing the Coulomb phase.
Because the magnetic symmetry is not a symmetry of the microscopic spin model, it is not expected to be emergent at low energy---instantons are not forbidden, and generically spoil the quantum spin liquid.

\subsubsection{Frustration and Stability}

The discussion above is relevant for continuum and lattice \emph{unfrustrated} 2-form U(1) gauge theory in 3+1 dimensions.
Unfrustrated lattice 2-form U(1) gauge theory was studied in Refs.~\cite{orlandInstantonsDisorderAntisymmetric1982,pearsonPhaseStructureAntisymmetric1982} on the cubic (rather than diamond) lattice, where it was found to lack a deconfined phase. 
Indeed, we note that in developing the theory of symmetric rank-2 gauge theories, Pretko briefly considered 2-form gauge theories but dismissed them as unstable for this very reason~\cite{pretkoSubdimensionalParticleStructure2017}. 
However, we are not aware of any systematic studies of either the effects of frustrated (``odd'') gauge fields or finite local Hilbert space dimension on the magnetic instantons.
Frustration is generally expected to help stabilize the deconfined phase since it forbids a trivial $E=0$ ground state.
In particular, note that in frustrated 1-form U(1) lattice gauge theory (describing QSI), the deconfined phase persists all the way to infinite coupling~\cite{hermelePyrochlorePhotonsU12004} where the unfrustrated theory is deep in the confined phase. 
Therefore, the fact that unfrustrated 2-form U(1) gauge theory in $\cramped{D=3+1}$ is susceptible to instanton effects may not immediately preclude the possibility of realizing a stable 2-form U(1) QSL from quantum spin vorticity models. 
An ultimate determination of the stability of a 2-form U(1) QSL in quantum spin vorticity models, or the observability of physics associated to such a phase, will require detailed theoretical and numerical studies in order to assess the effects of frustration, symmetries, and other mechanisms in the proliferation of instantons and stability of the gapless 2-form U(1) QSL.
Furthermore, it is worth noting that even if the QSL is unstable and the ground state is ordered, partially quantum-coherent effects and deconfinement may occur at finite temperature, as appears to occurs in (2+1)-dimensional QSI~\cite{henryOrderDisorderQuantumCoulomb2014} and $\mathrm{QED}_3$~\cite{grignaniConfinementdeconfinementTransitionThreedimensional1996} which are both affected by instantons.
Indeed, the spin vorticity QSL discussed in this section will undoubtedly control the structure of the surrounding phase diagram for quantum pyrochlores with further-neighbor interactions, regardless of whether it is a stable (quantum ground state) phase itself.

\begin{table*}
    \includegraphics[width=\textwidth]{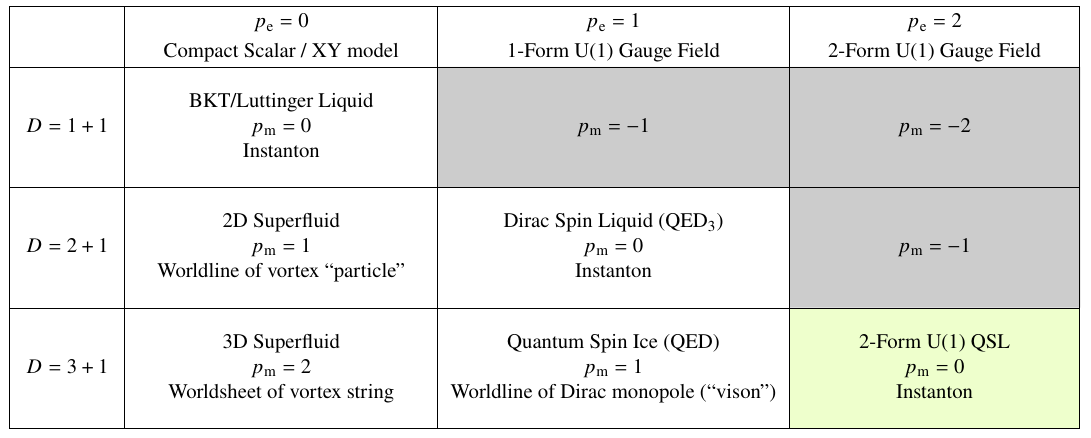}
    \caption{Here we organize the various gapless U(1) phases, classified by their low-energy description in terms of compact $\pe$-form fields in $D$ spacetime dimensions. The case $\pe=0$ is a compact scalar, $\cramped{\pe=1}$ is a U(1) 1-form gauge field, and $\cramped{\pe=2}$ is the U(1) 2-form gauge field appearing in \cref{eq:H-2-form}, which describes the low-energy limit of the putative U(1) 2-form quantum spin liquid phase. A compact scalar ($\cramped{\pe=0}$) is equivalent on a lattice to an XY model. For each case, we give the dimension $\cramped{\pmag=D-(\pe+2)}$ of the magnetic topological defects of the compact field. On the diagonal, $\cramped{\pmag=0}$, and the defect is an instanton. Below the diagonal, $\cramped{\pmag=1}$, and the defect is the worldline of the ``magnetic'' particle---a vortex in 2+1 dimensions, and the Dirac monopole in 3+1 dimensions. At the bottom left, $\pmag=2$, and the defect is the worldsheet of a vortex string, such as those that appear in superfluids and superconductors in 3+1 dimensions. 
    }
    \label{tab:vison_table}
\end{table*}

\subsection{How 2-Form U(1) Quantum Spin Liquids Fit In}

Taking a broader perspective than the quantum spin vorticity model discussed so far, let us place 2-form U(1) QSLs in the wider context of gapless U(1) liquid phases. 
\cref{tab:vison_table} organizes $\pe$-form U(1) field theories in $D$ spacetime dimensions. 
Each cell of the table gives the corresponding gapless liquid phase, the dimension $\pmag$ of the magnetic defect, and a description of this defect.
The key takeaway is that the diagonal cells correspond to the lower critical dimension of the corresponding $\pe$-form U(1) fields, which occurs when $\pmag = 0$ so that the magnetic vortex defect is an instanton, and a generalized Mermin-Wagner theorem applies~\cite{lakeHigherformSymmetriesSpontaneous2018}.
For dimensions higher than this, we obtain stable gapless phases which spontaneously break the global U(1) $\pe$-form symmetry~\cite{mcgreevyGeneralizedSymmetriesCondensed2023}---a superfluid for $\pe=0$ and deconfined phases for $\pe>0$. 
In the stable cases the magnetic defect is gapped, or equivalently magnetic symmetry is emergent at low energy~\cite{paceEmergentGeneralizedSymmetries2024}. 
The putative 2-form U(1) QSL in three spatial dimensions we discussed in this section sits in the lower right, highlighted green, on the diagonal, at its lower critical dimension.
However, the diagonal cases are in many ways the most interesting cases in the table despite, or perhaps due to, their inherent instability, so it is worth looking at these more carefully.

Starting with $\cramped{D=1+1}$, we have the compact scalar field, or the two-dimensional XY model. 
Here the instantons are the vortices of the XY model.
Despite their presence, a stable gapless phase survives via the Berezinskii-Kosterlitz-Thouless (BKT) mechanism~\cite{kosterlitzOrderingMetastabilityPhase1973,joseRenormalizationVorticesSymmetrybreaking1977}. 
This BKT mechanism arises due to the low dimensionality, the vortices effectively forming a gas interacting by a logarithmic Coulomb potential, yielding a transition to a stable gapless quantum spin liquid where the magnetic instantons become confined.

In $\cramped{D=2+1}$ the 1-form U(1) gauge theory, i.e. quantum electrodynamics or QED$_3$, is permanently confined by magnetic instanton proliferation (the Polyakov mechanism)~\cite{polyakovQuarkConfinementTopology1977}, which gaps the photon and yields a relatively trivial ground state.
This effect can be overcome, however, by adding to the model a sufficient number of gapless electric charges in the form of Dirac fermions~\cite{kleinertKosterlitzThoulessdeconfinementMechanism2+1dimensional2003,hermeleStabilityU1Spin2004}.
QED$_3$ is important as a theory describing deconfined quantum critical points~\cite{groverQuantumPhaseTransition2013,luQuantumPhaseTransitions2014,janssenCriticalBehaviorQED3GrossNeveu2017}, and it emerges as the low-energy effective theory of the U(1) Dirac spin liquid in two-dimensional frustrated quantum spin models~\cite{hermeleStabilityU1Spin2004,hermeleAlgebraicSpinLiquid2005,songUnifyingDescriptionCompeting2019,songSpinonBandTopology2020,wietekQuantumElectrodynamics2+12024} with proposed experimental realizations~\cite{kiesePinchpointsHalfmoonsStars2023}. 
In particular, Refs.~\cite{songUnifyingDescriptionCompeting2019,songSpinonBandTopology2020} were able to show that the magnetic instanton operators, which correspond to order parameters for various valence-bond ground states, are forbidden due to crystalline symmetries of some lattices.

Given that in $\cramped{D=1+1}$ and $\cramped{D=2+1}$ the effects of instantons can be overcome to stabilize a gapless spin liquid, either via the BKT mechanism or by adding massless electrically charged matter, it would seem plausible that some form of gapless 2-form U(1) QSL is stable in $\cramped{D=3+1}$. 
Even if the spin vorticity model introduced in this paper turns out not to support a gapless ground state, it serves as an entry point and minimal model to search for mechanisms that may be able to stabilize a 2-form U(1) QSL.
For example, we recall that the fermionic Luttinger liquid phase in $\cramped{D=1+1}$ (closely related to the BKT phase~\cite{giamarchiLuttingerLiquids2003}) may be viewed as a one-dimensional analog of the Dirac spin liquid in $\cramped{D=2+1}$~\cite{songUnifyingDescriptionCompeting2019}. 
In a similar vein, continuing down the diagonal of \cref{tab:vison_table}, we propose that a promising investigative direction would be to identify a (3+1)-dimensional analog of the Dirac spin liquid.
This would seemingly require a model with \emph{fermionic strings} described by an appropriate continuum string field theory~\cite{erbinStringFieldTheory2021} generalizing QED$_3$, presumably lying well beyond the parton approach used for the Dirac spin liquid~\cite{wenQuantumFieldTheory2004,songUnifyingDescriptionCompeting2019,songSpinonBandTopology2020}. 
While the possibility of such a QSL is at the present time somewhat hypothetical, its pursuit is well-motivated by the position of a 2-form U(1) QSLs as the final unexplored case in \cref{tab:vison_table}.
One of the primary reasons why the Dirac QSL is important is that it serves as a parent phase for many competing orders, which arise from different patterns of electric or magnetic charge condensation~\cite{hermeleAlgebraicSpinLiquid2005,songUnifyingDescriptionCompeting2019,songSpinonBandTopology2020}. 
If a 2-form U(1) QSL were to exist in $\cramped{D=3+1}$ serving an analogous role, it should have deconfined 2-form U(1) gauge fields.
In this vein, recent work has discovered a novel two-dimensional analog of the one-dimensional spin-Peierls instability (dimerization of a gapless quantum spin chain) by coupling the Dirac QSL to phonons~\cite{seifertSpinPeierlsInstabilityU12024}.
Again motivated by \cref{tab:vison_table}, an analogous three-dimensional instability would likely be obtained from a three-dimensional analog of the Dirac QSL.

In the context of its position as the final unexplored case in \cref{tab:vison_table}, it would be surprising if the detailed study of 2-form U(1) QSLs did not prove to be a highly fruitful endeavor given the importance that all the other cases have had for modeling a range of condensed matter phenomena. 
There are various other potentially fruitful avenues generalizing the 2-form U(1) spin liquids discussed thus far. 
One would be to search for spin models realizing fractonic 2-form U(1) spin liquids~\cite{shenoyKnfractonicMaxwellTheory2020,yanRank2U1Spin2020}.
Another direction would be to explore the possibility of spin liquids realizing non-Abelian generalizations of 2-form gauge theories, based on ``2-groups''~\cite{reyNonperturbativeProposalNonabelian2010,baezInvitationHigherGauge2011,lipsteinLatticeGerbeTheory2014}.
A third possibility is a QSL realizing a hypothetical higher-form KT phase proposed in Ref.~\cite{lakeHigherformSymmetriesSpontaneous2018}.
It will also be of interest to investigate whether long-range interactions can provide a workaround to the generalized Mermin-Wagner theorem~\cite{lakeHigherformSymmetriesSpontaneous2018,brunoAbsenceSpontaneousMagnetic2001,halperinHohenbergMerminWagner2019}.
In sum, the study of quantum spin liquids described by deconfined 2-form gauge fields offers many exciting avenues for future theoretical developments.


\section{Experimental Research Avenues }
\label{sec:experimental}

At this juncture, it is natural to ask what experimental settings might provide an opportunity for exploring the physics described in this paper.
To do so, we are guided by the perspectives gained from many years of experimental study of classical spin ice physics and its quantum extensions~\cite{gingrasQuantumSpinIce2014,udagawaSpinIce2021}.
To this end, we identify three main points of interest which we  address in order: spin vorticity models, 2-form classical spin liquids, and 2-form U(1) quantum spin liquids.

We begin with the spin vorticity model, defined abstractly by \cref{eq:H_vort_wp} to minimize an appropriate lattice notion of curl. 
Although we have in this paper focused on a version of the spin vorticity model defined on the pyrochlore lattice, realized in many rare earth magnetic compounds~\cite{gardnerMagneticPyrochloreOxides2010,hallasExperimentalInsightsGroundState2018,rauFrustratedQuantumRareEarth2019}, generalizations to other lattice geometries are straightforward.
In particular, it is worthwhile to take a moment to discuss the possibility of experimentally realizing versions of the vorticity model in two spatial dimensions. 
As we discuss in \cref{apx:2d-vorticity-model}, these models are exactly dual to ice models, which can be traced to the duality of divergence and curl in two dimensions~\cite{henleyCoulombPhaseFrustrated2010}. 
A vortex in two dimensions is a point object, and thus the excitations of 2D vorticity models are dual to the point charges of an ice model, which may be viewed as a simple but elegant classical realization of particle-vortex duality~\cite{seibergDualityWeb2+12016}. 
Given the broad interest in so-called artificial spin ice~\cite{nisoliColloquiumArtificialSpin2013}---2D synthetically engineered mesoscopic systems which satisfy a local zero-divergence constraint and are amenable to direct experimental control and study---it would be worthwhile to pursue the construction and exploration of artificial spin vorticity models, as they are naturally dual to artificial spin ices and thus provide an experimental platform to study particle-vortex duality.

Next, let us consider the prospects of experimentally realizing classical 2-form spin liquids in three spatial dimensions, starting with pyrochlore compounds~\cite{gardnerMagneticPyrochloreOxides2010,hallasExperimentalInsightsGroundState2018,rauFrustratedQuantumRareEarth2019}.
In the model considered in this paper, the spin liquid characterized by vanishing vorticity is engendered by a precise set of fine-tuned first, second and third (type $3b$) nearest-neighbor Ising exchange couplings in Eq.~\eqref{eq:H_vort_spin}. 
The necessity to have fine-tuned interactions up to third nearest-neighbor interactions that are of very similar magnitude and competing sign is much more stringent than the corresponding situation for spin ice where the spin liquid results solely from  a nearest-neighbor Hamiltonian~\cite{henleyCoulombPhaseFrustrated2010,castelnovoSpinIceFractionalization2012,bramwellSpinIceState2001}.
While natural fine-tuning is seemingly unlikely, the geometric nature of the couplings across a hexagon could potentially be achieved by exchange pathways mediated by a central atom on the $\mathcal{B}$ pyrochlore sites at the center of each hexagon, or in systems with a magnetic $\mathcal{B}$ pyrochlore lattice such as the pyrochlore iridates~\cite{jacobsenStrongQuantumFluctuations2020,pearceMagneticMonopoleDensity2022},
ruthenates~\cite{gardnerMagneticOrderCrystal2009,kuLowTemperatureMagnetic2018,huebschMagneticStructuresElectronic2022,leeLinearScalingRelationship2023} and molybdates~\cite{dunsigerMuonSpinRelaxation1996,gardnerMagneticPyrochloreOxides2010}.
It may also be simpler to realize a spin vorticity model in other lattice geometries, such as the octochlore lattice of corner-sharing octahedra which is realized in the antiperovskite structure~\cite{szaboFragmentedSpinIce2022}.
This lattice only requires fine tuning of first and second neighbors, since the plaquettes are squares rather than hexagons.

Even if a given material proved to not have its Ising interactions exactly satisfying the ratios defined by Eq.~\eqref{eq:H_vort_spin}, for a material with its couplings ``sufficiently close'' to those in Eq.~\eqref{eq:H_vort_spin} such that the spin liquid manifold is still nearly degenerate, a finite-temperature cooperative paramagnetic region should exist.
Furthermore, for such a system nearby the spin vorticity model, the strong frustration coming from a nearly-degenerate spin liquid manifold would likely result in the system falling out of equilibrium at low temperatures, which may prevent observation of long range order, as is known to occur in the classical spin ice compounds where it results in remarkable non-equilibrium dynamics~\cite{jaubertSignatureMagneticMonopole2009,matsuhiraDynamics2021,samarakoonAnomalousMagneticNoise2022,hallenDynamicalFractalAnomalous2022}. Thus novel physics stemming from proximity to a spin vorticity 2-form spin liquid may remain observable over a rather broad range of parameters at nonzero temperature.

Finally, let us address the possibilities for 2-form U(1) quantum spin liquids. 
Here solid state systems, presumably requiring Ising interactions fine-tuned to a degree smaller than the scale of the perturbative quantum exchange terms such as in \cref{eq:H_QSV}, would seem unlikely. 
On the other hand, the fact that the membrane exchange term in \cref{eq:H-mem-exch} comes about at second order in the quantum perturbations, rather than at third order as in QSI~\cite{hermelePyrochlorePhotonsU12004}, means that the window for quantum spin liquidity could in principle be wider than that in QSI. 
Perhaps a more promising route for experimentally exploring 2-form spin liquid described in this paper with effective Ising degrees of freedom are trapped ion systems where the $J_{ij}$ Ising  interactions can be engineered~\cite{molmerMultiparticleEntanglementHot1999,kimEntanglementTunableSpinSpin2009,monroeProgrammableQuantumSimulations2021,zhangQuantumSlushState2024}.
It is interesting to note that trapped ions also afford to generate quantum perturbations to the Ising pairwise interactions in the form of an effective transverse field~\cite{monroeProgrammableQuantumSimulations2021}.
Moreover, recent progress has been made that also allows to generate (quantum) transverse exchange couplings of the form $S_i^\pm S_j^\mp$~\cite{kotibhaskarProgrammableXYtypeCouplings2024}, with the possibility of creating other anisotropic quantum terms, such as the second term in \cref{eq:H_QSV}, a topic of current research.
However, the system sizes that are currently available with such a platform are not sufficiently large to encode the $J_{ij}$ interactions of the (embedded) three-dimensional nature of Eq.~\eqref{eq:H_vort_spin} and the interacting vorticity field $\omega_p$ living on the hexagons of Fig.~\ref{fig:hexagon}.

Admittedly, at this time, we are not aware of any systems for which the microscopic classical and quantum models of \cref{eq:H_vort_spin} and  \cref{eq:H_QSV}, respectively, constitute a reasonably close physical description. 
Discovery of real materials harboring the physics describe herein or the development of synthetic platforms to explore it will likely have to wait.  
However, one notes that Ref.~\cite{hermelePyrochlorePhotonsU12004} investigating weak quantum fluctuations in frustrated pyrochlore Ising antiferromagnets was also not motivated by any specific microscopic real Ising material of interest at the time of its publication. 
It is only in subsequent works that the prospects of realizing the type of quantum spin liquid kindred to that theoretically identified in Ref.~\cite{hermelePyrochlorePhotonsU12004}, in either solid state magnetic systems~\cite{molavianDynamicallyInducedFrustration2007,onodaQuantumMeltingSpin2010,rossQuantumExcitationsQuantum2011} or in Rydberg atom experiments~\cite{glaetzleQuantumSpiniceDimer2014}, were put forward.
One may hope that the models and ideas discussed in this paper will find a realization in one experimental context or another in a timely fashion.

\section{Discussion and Conclusion}
\label{sec:discussion}

In this paper, we have introduced the spin vorticity model on the pyrochlore lattice, \cref{eq:H_vort_spin}, which is constructed to enforce a zero-curl condition on every plaquette at zero temperature, generalizing the zero-divergence condition of spin ice. 
The interaction matrix of this bilinear spin model contains two topological flat bands in its spectrum, and the self-consistent Gaussian approximation predicts the presence of a classical spin liquid ground state characterized by inverted pinch points, shown in \cref{fig:vort_bands_SCGA}. 
We confirmed the existence of this classical spin liquid numerically via classical Monte Carlo simulation, as indicated by an extensive ground state entropy and sharp pinch point features shown in \cref{fig:vorticity_monte_carlo}.
Curiously, we have found that this spin liquid coexists with a weak breaking of inversion symmetry, reflected by a singularity in the specific heat and the presence of Bragg peaks in the structure factor below the critical temperature, which contain less than $1\%$ of the total spectral weight at zero temperature.  
The ground state topological order of this classical spin liquid can be understood as a membrane condensate: each spin is fractionalized into a flux-membrane on a plaquette of the dual diamond lattice, attached to a string excitation on its boundary, as shown in \cref{fig:diamond_dual_membranes}(c). 
In the spin liquid ground states, the spins, treated as fluxes, organize themselves to form closed surfaces---membranes---on which all spins are oriented in the same direction. 
Flipping such a closed membrane costs zero energy.
The fractionalized gapped excitations are 1-dimensional strings, created by flipping spins forming an open membrane.
The gauge theory description of this phase is given by 2-form electrodynamics, thus establishing it as a novel classical 2-form spin liquid.

We have further illustrated how quantum perturbations of the classical spin vorticity model map at second order in degenerate perturbation theory to a \emph{membrane exchange} model, generalizing the ring exchange model of quantum spin ice.
The effective membrane-exchange model, \cref{eq:H-mem-exch}, maps to a 2-form U(1) lattice gauge theory. 
To our knowledge, this is the first known spin liquid described by an emergent 2-form U(1) gauge theory---previously studied U(1) spin liquids are described by emergent 1-form gauge fields.
We demonstrated that a Rohksar-Kivelson point exists for this model, where the ground state is an exact massive superposition of all classical ground states, thus establishing an exactly soluble point exhibiting a ground state rightly deserving to be called a 2-form U(1) quantum spin liquid (QSL).
Moreover, we showed how to map a general quantum spin Hamiltonian near the spin vorticity model to a 2-form U(1) gauge-Higgs model by treating the string excitations as a 1-form U(1) Higgs field coupled to the 2-form U(1) gauge field, \cref{eq:gauge-higgs-mapping}.

The spin vorticity model therefore serves as a minimal model which could potentially host a 2-form U(1) QSL. 
However, we discussed how 2-form U(1) QSLs are susceptible to the effects of instantons, which act to destabilize the gapless deconfined phase.
Detailed analytical and numerical studies will be required to assess what mechanisms may suppress the instantons and the range of stability of this phase.
While we have presented here a 2-form U(1) lattice gauge theory description for the quantum spin vorticity model involving only bosonic fields, perhaps the most interesting theoretical outlook of our work is that it possibly suggests the existence of a fermionic analog which parallels the Dirac spin liquid in two spatial dimensions. 
The door is thus open to explore and categorize 2-form U(1) QSLs, to assess their stability, and to search for potential material or synthetic platforms for their experimental study.
Our work lays the groundwork to launch the study of a novel hypothetical phase of matter---2-form U(1) spin liquids---both in the spin vorticity model presented in this work, as well as other incarnations that remain to be discovered.
\\[1ex]
\textit{Note added---In the process of finalizing the proofs of this manuscript, a manuscript that discusses a number of results closely related to those presented in this paper appeared~\cite{rougemailleAmperePhaseFrustrated2025}.}

\begin{acknowledgments}
We thank Alex Hickey, Daniel Lozano-G\'{o}mez, Hannes Bernien, Peter Holdsworth, Pranay Patil, Rajibul Islam, Crystal Senko and Han Yan for useful discussions, and the first two for comments on an early version of the manuscript. We acknowledge the use of computational resources provided by Digital Research Alliance of Canada. 
This research was funded by the NSERC of Canada and the Canada Research Chair Program~(MJPG, Tier I). 
This work was in part supported by the Deutsche Forschungsgemeinschaft  under the cluster of excellence ct.qmat (EXC-2147, project number 390858490).

\end{acknowledgments}

\appendix

\section{Conventions and Interaction Matrices}

In these appendices we provide supplementary calculations, explain methodology, and provide a few introductory remarks for interested readers on 2-form gauge theory. 
Note that these appendices are structured to be easily readable in a sequential fashion, such that they do not all appear in the same order they are referenced in the main text.

This first appendix lays out our lattice and Fourier transform conventions, and provides the general definition of the interaction matrix for spin models on the pyrochlore lattice.

\subsection{Lattice Conventions}
\label{apx:conventions}

The pyrochlore lattice can be described as a face-centered cubic (FCC) lattice decorated with tetrahedra, i.e. with a four-site basis. In Cartesian coordinates, we take the FCC primitive lattice vectors to be given by 
\begin{equation}
    \bm{a}_1 = \frac{a_0}{2}(\uvec{x} + \uvec{y}), \quad
    \bm{a}_2 = \frac{a_0}{2}(\uvec{x} + \uvec{z}), \quad    
    \bm{a}_3 = \frac{a_0}{2}(\uvec{y} + \uvec{z}),
\end{equation}
where $a_0$ is the cubic conventional cell edge length. We then define the pyrochlore sublattice positions within each FCC unit cell as
\begin{equation}
    \bm{c}_1 = \bm{0},    \quad
    \bm{c}_2 = \bm{a}_1/2,  \quad
    \bm{c}_3 = \bm{a}_2/2,  \quad
    \bm{c}_4 = \bm{a}_3/2,  \quad
    \label{eq:sm_sublat_positions}
\end{equation}
which form the corners of a tetrahedron. The pyrochlore lattice positions $\bm{r}_i$ are then given by combinations
\begin{equation}
    \bm{r}_i = \sum_{k=1}^3 n_k \bm{a}_k + \bm{c}_\mu\quad (n_k \in \integers_L),
    \label{eq:sm_r_i}
\end{equation}
where the three integers $n_k \in \{0,1,\cdots,L-1\}$ specify the FCC unit cell and $\mu\in \{1,2,3,4\}$ indexes the sublattice. The index $i\in \{1,\cdots 4L^3\}$ specifies a single pyrochlore site, for example we can take
\begin{equation*}
    i = (\mu-1)L^3 + n_1 L^2 + n_2 L + n_3 + 1,
\end{equation*}
though this choice of indexing is irrelevant for our purposes.

There are two tetrahedra per FCC cell, one centered at
\begin{equation}
    \bm{\delta} = \frac{1}{4}\sum_{\mu} \bm{c}_\mu,
\end{equation}
and the other centered at $-\bm{\delta}$. We refer to the former as the $A$ tetrahedron and the latter as the $B$ tetrahedron. The local 3-fold easy axes connect the centers of neighboring tetrahedra, and we take our quantization axes $\hat{\bm{z}}_\mu$ to consistently point from the $A$ tetrahedron to the $B$ tetrahedron, i.e from the center of the $A$ tetrahedra to each of its corners, given by
\begin{equation}
    \uvec{z}_\mu = \frac{\bm{c}_\mu - \bm{\delta}}{\vert \bm{c}_\mu - \bm{\delta}\vert}.
    \label{eq:sm_quant_axis}
\end{equation}

\subsection{Fourier Transforms and Reciprocal Space}

\begin{figure}
    \centering
    \includegraphics[width=.7\columnwidth]{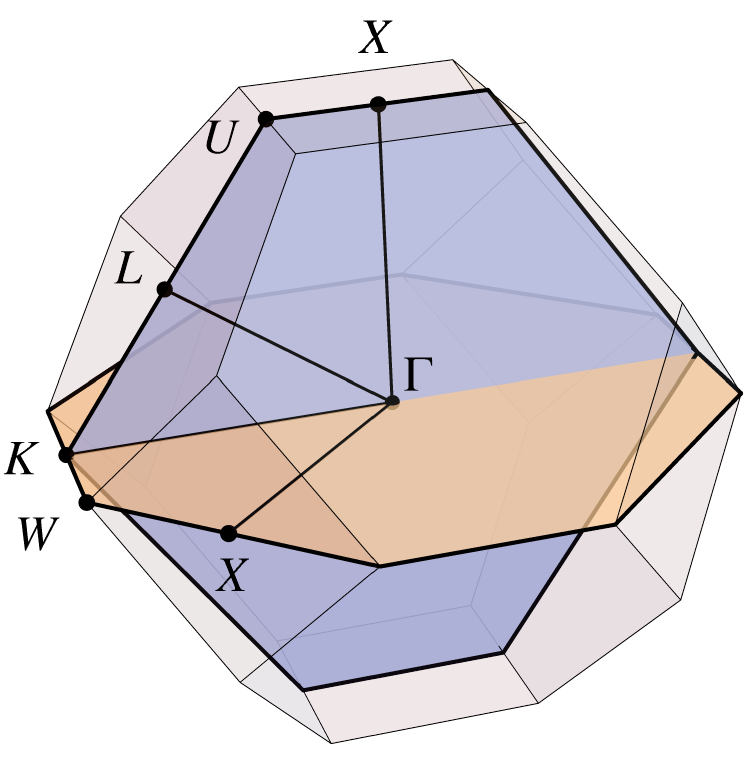}
    \caption{A three-dimensional depiction of the first Brillouin zone of the FCC lattice with high symmetry points, lines, and planes indicated, in particular the $(hk0)$ plane (orange) and $(hhl)$ plane (blue). Using $\{hkl\}$ to denote the set of symmetry-equivalent points to $(hkl)$, we have $X=\{001\}$,
    $U=\{\tfrac{1}{4}\tfrac{1}{4}1\}$, $L=\{\tfrac{1}{2}\tfrac{1}{2}\tfrac{1}{2}\}$, $K=\{\tfrac{3}{4}\tfrac{3}{4}0\}$, $W=\{\tfrac{1}{2}10\}$, and $\Gamma=(000)$.
    }
    \label{fig:SM_fbz3d}
\end{figure}

We assume throughout an $L \times L \times L$ FCC lattice with periodic boundaries, i.e. $n_k \in \integers\text{ mod } L$, containing $4L^3$ pyrochlore sites. 
We define the Fourier transformed spin variables
\begin{equation}
    \sigma_\mu^z(\bq) \equiv \frac{1}{\sqrt{L^3}} \sum_{i \in \mu} \sigma_i^z e^{-i\bq\cdot\bm{r}_i},
    \label{eq:sm_s_mu_q}
\end{equation}
where the sum is over all pyrochlore sites $i$ in sublattice $\mu$. For matrix quantities $\mathcal{M}_{ij}$,  we define the Fourier transform as
\begin{equation}
    \mathcal{M}_{\mu\nu}(\bq,\bq') \equiv \frac{1}{L^3}\sum_{i \in \mu} \sum_{j \in \nu} \mathcal{M}_{ij} \,e^{-i\bq\cdot \bm{r}_i}e^{i\bq\cdot\bm{r}_j},
    \label{eq:FT}
\end{equation}
where $\mu,\nu\in\{1,2,3,4\}$ index the four sublattices.
If $\mathcal{M}_{ij}$ is translationally-invariant then it is $4\times 4$ block diagonal in the Fourier basis,
\begin{align}
    \mathcal{M}_{\mu\nu}(\bq,\bq') &= \mathcal{M}_{\mu\nu}(\bq) \delta_{\bq,\bq'}, \nonumber\\ 
    \mathcal{M}_{\mu\nu}(\bq) &= \sum_{j \in \nu} \mathcal{M}_{ij} e^{i\bq\cdot\bm{r}_j} \quad (i \in \mu),
    \label{eq:SM_FourierMatrix}
\end{align}
where the sum is over all sites $j$ in sublattice $\nu$ with an arbitrary fixed site $i$ in sublattice $\mu$.
The allowed wavevectors commensurate with periodic boundaries are given by
\begin{equation}
    \bm{q} = \sum_{k=1}^3 m_k \frac{\bm{b}_k}{L} \quad (m_k \in \integers), 
    \label{eq:sm_allowed_q}
\end{equation}
where $\bm{b}_k$ are the reciprocal lattice vectors of the FCC primitive vectors $\bm{a}_k$, satisfying ${\bm{a}_k \cdot \bm{b}_l = 2\pi \delta_{kl}}$, which point orthogonal to the planes of the FCC lattice. 
Then $\bm{b}_k/L$ are the reciprocal lattice vectors for the ``super-lattice'' primitive vectors $L\bm{a}_i$ (the repeating super cell). 
The $L^3$ wavevectors in the first Brillouin zone (FBZ) are given by taking $m_k \in \{0,\cdots, L-1\}$. 
It is convenient to use Miller indices relative to the cubic conventional unit cell (rather than the FCC primitive cell) to specify points in reciprocal space, where a wavevector is given by
\begin{equation}
    \bq = \frac{2\pi}{a_0} \left(h \hat{\bm{x}} + k \hat{\bm{y}} + l \hat{\bm{z}}\right) \equiv (hkl).
\end{equation}
\Cref{fig:SM_fbz3d} shows the FBZ of the FCC lattice, with high-symmetry points, lines, and planes indicated.

\subsection{Interaction Matrix}
\label{apx:int_mat}

For a bilinear Ising Hamiltonian of the form
\begin{equation}
    H = \frac{1}{2}\sum_{ij} \sigma_i^z \Jmat_{ij} \sigma_j^z,
    \label{eq:SM_H}
\end{equation}
where the sum is over all pyrochlore lattice sites $i,j$, we refer to $\Jmat$ as the interaction matrix. 
Note that $\Jmat_{ij} = \Jmat_{ji}$, and $\Jmat_{ii} = 0$, so the prefactor of 1/2 in \cref{eq:SM_H} removes double counting. 
It is convenient to express the interaction matrix in terms of further-neighbor adjacency matrices, defined as
\begin{equation}
    \Amat^{(n)}_{ij} 
    = 
    \begin{cases} 
        1 &\quad \text{$i,j$ are $n$'th nearest-neighbors}, \\
        0 &\quad \text{otherwise}.
    \end{cases}
\end{equation}
The interaction matrix is then a linear combination
\begin{equation}
    \Jmat_{ij} = \sum_{n} J_n \Amat_{ij}^{(n)},
\end{equation}
with some exchange energies $J_n$. Note that the index $n$ distinguishes between neighbors which are of the same (Euclidean) distance apart but are not related by symmetry, for example there are two types of third nearest-neighbors which are inequivalent by symmetry and so can have different $J_n$~\cite{conlonAbsentPinchPoints2010,heneliusRefrustrationCompetingOrders2016}.
The nearest-neighbor spin ice (NNSI) model is given by 
\begin{equation}
    J_1 = J\quad (\text{NNSI}), 
\end{equation}
with an energy scale $J>0$, while the spin vorticity model defined in \cref{eq:H_vort_spin} in the main text is given by
\begin{equation}
    J_1 = -J, \quad J_2 = \frac{J}{2}, \quad J_{3b} = -\frac{J}{2} \quad (\text{spin vorticity}),
    \label{eq:sm_vorticity_Js}
\end{equation}
and all others zer, with again $J>0$. See \cref{apx:SVM_Hamiltonian} for the derivation of \cref{eq:H_vort_spin} and see \cref{fig:hexagon} for the second and third neighbor relative positions.

Since the interaction matrix $\Jmat$ is translationally invariant, its eigenvalues form four bands in reciprocal space.  We give below the Fourier transformed adjacency matrices for the three $\Amat_{ij}^{(n)}$ which appear in the spin vorticity interaction matrix, following the sublattice conventions laid out in \cref{apx:conventions}. For nearest-neighbors, 
\begin{equation}
\Amat_{\mu\nu}^{(1)} (\bq) 
= 2 \,
\begin{pmatrix}
0 & c_{xy} & c_{xz} & c_{yz} \\
c_{xy} & 0 & c_{y\overline{z}} & c_{x\overline{z}} \\
c_{xz} & c_{y\overline{z}} & 0 & c_{x\overline{y}}\\
c_{yz} & c_{x\overline{z}} & c_{x\overline{y}} & 0
\end{pmatrix} 
\,,
\end{equation}
and for second neighbors,
\begin{equation}
\Amat_{\mu\nu}^{(2)} (\bq) 
= 4 \,
\begin{pmatrix}
0 & c_{zz} \, c_{x\overline{y}} & c_{yy} \, c_{x\overline{z}} & c_{xx} \, c_{y\overline{z}} \\
c_{zz} \, c_{x\overline{y}} & 0 & c_{xx} \, c_{yz} & c_{yy} \, c_{xz} \\
c_{yy} \, c_{x\overline{z}} & c_{xx} \, c_{yz} & 0 & c_{zz} \, c_{xy} \\
c_{xx} \, c_{y\overline{z}} & c_{yy} \, c_{xz} & c_{zz} \, c_{xy} & 0
\end{pmatrix} \,,\\
\label{eq:interaction_matrix_1_2}
\end{equation}
with $\cramped{c_{ab} \equiv \cos(\frac{q_a + q_b}{4})}$, $c_{a\bar{b}} \equiv \cos(\frac{q_a - q_b}{4})$, where $q_a, q_b$ refer to the Cartesian components of the wavevector $q_x$, $q_y$, $q_z$ in units of $a_0^{-1}$. For type-$b$ third neighbors, 
\begin{equation}
\begin{split}
\Amat_{\mu\nu}^{(3b)} (\bq) 
&= 2 \,
\begin{pmatrix}
C_{\scriptscriptstyle -\,-\,-} & 0 & 0 & 0 \\
0 & C_{\scriptscriptstyle -\,+\,+} & 0 & 0 \\
0 & 0 & C_{\scriptscriptstyle +\,+\,-} & 0 \\
0 & 0 & 0 & C_{\scriptscriptstyle +\,-\,+}
\end{pmatrix}\,,
\label{eq:interaction_matrix_3a}
\end{split}
\end{equation}
with $C_{\pm\pm\pm} =  \cos(\frac{q_{x}\pm q_{y}}{2}) + \cos(\frac{q_{y}\pm q_{z}}{2}) + \cos(\frac{q_{z}\pm q_{x}}{2})$. The Fourier transformed interaction matrix is then given by
\begin{equation}
    \Jmat_{\mu\nu}(\bq) = \sum_n J_n \Amat_{\mu\nu}^{(n)}(\bq).
    \label{eq:sm_J_A_q}
\end{equation}

\section{SCGA Analysis of the Spin Vorticity Model}

This appendix provides details of the self-consistent Gaussian approximation (SCGA) analysis of the spin vorticity model described in the main text \cref{sec:SCGA}.

\subsection{Spin Vorticity Model Hamiltonian}
\label{apx:SVM_Hamiltonian}

In \cref{eq:omega_p} of the main text, we defined the vorticity $\omega_p$ on each plaquette $p$ as 
\begin{equation}
    \omega_p = \frac{1}{2}(-1)^p(\sigma_1^z - \sigma_2^z + \sigma_3^z - \sigma_4^z + \sigma_5^z - \sigma_6^z)_p,
\end{equation}
and the spin vorticity Hamiltonian, \cref{eq:H_vort_wp}, as
\begin{equation}
    H_{\text{SV}} = E_0 + J\sum_p \omega_p^2,
    \label{eq:SM_Hvot_wp}
\end{equation}
where $E_0$ is the ground state energy. Squaring $\omega_p$ generates bilinear couplings between all spins in a hexagonal plaquette. Nearest-neighbor terms are ferromagnetic, such as $-\sigma_1^z\sigma_2^z$, second-neighbor terms are antiferromagnetic, such as $+\sigma_1^z\sigma_3^z$, and third-type-$b$-neighbors are ferromagnetic, such as $-\sigma_1^z \sigma_4^z$. Every spin also appears squared, such as $(\sigma_1^z)^2$. Therefore, we can expand \cref{eq:SM_Hvot_wp} as
\begin{equation}
    H_{\text{SV}} = E_0 + \frac{J}{4}\sum_p \left[ \sum_{i,j\in p} \sigma_i^z \left(\delta_{ij} - \Amat^{(1)}_{ij} + \Amat^{(2)}_{ij} - \Amat^{(3b)}_{ij}\right)\sigma_j^z \right].
\end{equation}
Now, we need only account for the fact that when we sum over all plaquettes $p$, every nearest-neighbor pair appears \emph{twice}---each nearest-neighbor pair (equivalently, each tetrahedron edge in the pyrochlore lattice) is a member of two plaquettes. 
Furthermore, each individual spin is a member of six plaquettes, which can be seen in \cref{fig:diamond_dual_membranes}(c) in the main text. 
With these combinatorial factors, we obtain
\begin{equation}
    H_{\text{SV}} = E_0 + \frac{J}{4} \sum_{ij} \sigma_i^z \left(6\delta_{ij} - 2 \Amat^{(1)}_{ij} + \Amat^{(2)}_{ij} - \Amat^{(3b)}_{ij}\right)\sigma_j^z ,
\end{equation}
where each of the double sums is over all pyrochlore sites $i,j$.
Setting $E_0 = -(3/2)JN_{\text{spin}}$ cancels the constant term coming from the diagonal $\delta_{ij}$ terms (using $(\sigma_i^z)^2 = 1$), so that we obtain a bilinear spin model with no on-site interaction terms,
\begin{equation}
    H_{\text{SV}} = \frac{J}{2} \sum_{ij} \sigma_i^z \left(
    - \Amat^{(1)}_{ij} + \frac{1}{2}\Amat^{(2)}_{ij} - \frac{1}{2}\Amat^{(3b)}_{ij}\right)\sigma_j^z .
\end{equation}
Comparing to \cref{eq:SM_H}, we can read off the spin vorticity interaction matrix,
\begin{equation}
    \Jmat_{ij}^{(\text{SV})} = J\left(
    - \Amat^{(1)}_{ij} + \frac{1}{2}\Amat^{(2)}_{ij} - \frac{1}{2}\Amat^{(3b)}_{ij}\right)\,.
    \label{eq:sm_J_vort}
\end{equation}

\begin{figure}[t]
    \centering
    \includegraphics[width=\columnwidth]{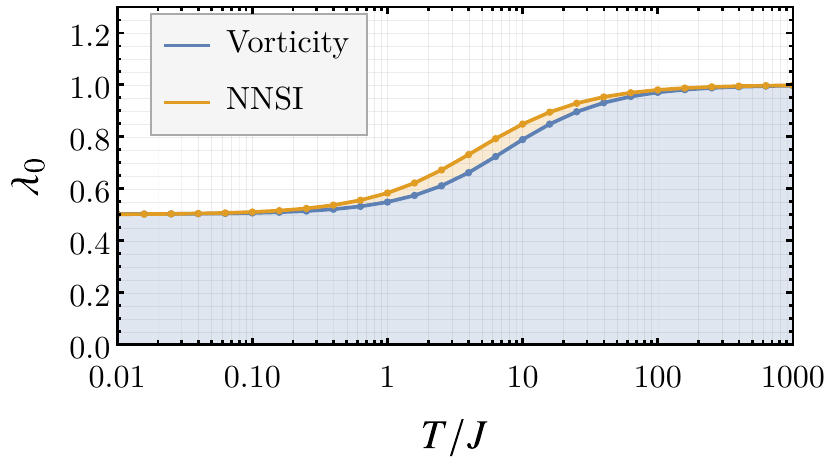}\hfill\vphantom{a}
    \caption{Evolution of the Lagrange multiplier $\lambda_0$ for both NNSI and the spin vorticity model, smoothly interpolating from 1 at high temperature to 1/2 at low temperature.}
    \label{fig:SM_lambda}
\end{figure}

\subsection{Self-Consistent Gaussian Approximation}
\label{apx:SCGA}

We utilize the self-consistent Gaussian approximation (SCGA) to compute the spin-spin correlation functions, from which we obtain the spin structure factor (\cref{eq:Sq}) shown in \cref{fig:vort_bands_SCGA} of the main text. 
Starting from a bilinear Hamiltonian of the form of \cref{eq:SM_H}, we treat the $\sigma_i^z$ as real variables and add to the Hamiltonian Lagrange multipliers $\lambda_i T (\sigma_i^z)^2/2$ to control the spin length, where the temperature $T$ is included for convenience. 
The SCGA effective Hamiltonian is then
\begin{equation}
    H_{\text{SCGA}} = \frac{1}{2}\sum_{ij} \sigma_i^z (\lambda_i T \delta_{ij} + \Jmat_{ij}) \sigma_j^z,
    \label{eq:SM_H_SCGA}
\end{equation}
where $\delta_{ij}$ is the Kronecker delta and the $\lambda_i$ will be determined self-consistently by the requirement $\langle (\sigma_i^z)^2 \rangle = 1$. The partition function can then be expressed as
\begin{equation}
    Z_{\text{SCGA}} = \int \CMcal{D}\sigma_i^z \,\exp(-\frac{1}{2}\sum_{ij} \sigma_i^z (\lambda_i\delta_{ij} + \beta \Jmat_{ij})\sigma_j^z),
\end{equation}
where $\beta = 1/T$. This is simply a multivariate Gaussian, and the correlation matrix can be readily read off,
\begin{equation}
    \CMcal{G}_{ij} \equiv \langle \sigma_i^z \sigma_j^z \rangle  =  (\Lambda + \beta \Jmat)^{-1}_{ij},
    \label{eq:SM_G}
\end{equation}
with $\Lambda_{ij} \equiv \lambda_i \delta_{ij}$. The self-consistency conditions $\langle (\sigma_i^z)^2 \rangle = 1$ are then implemented by the requirements $\CMcal{G}_{ii} = 1$. Since every site in the pyrochlore lattice is equivalent by symmetry and $\Jmat$ respects the symmetries of the lattice, the solutions must be the same on every site, $\lambda_i \equiv \lambda$, and we only need to determine the single constant $\lambda$. Therefore, we have $\Lambda = \lambda \mathbbm{1}$ in \cref{eq:SM_G}, and the self-consistency condition can then be expressed as
\begin{equation}
    \sum_i \langle (\sigma_i^z)^2 \rangle = \Tr \CMcal{G} = 4L^3, 
\end{equation}
which fixes the average total spin length squared per spin to 1. The trace is computed in the Fourier basis where $\Jmat$ is $4\times 4$ block diagonal, such that
\begin{equation}
    \Tr \CMcal{G} = \sum_{\bq \in \text{FBZ}} \Tr\CMcal{G}(\bq),
\end{equation}
where $\CMcal{G}(\bq)$ is the $4\times 4$ block of $\CMcal{G}$ at wavevector $\bq$ (see \cref{eq:SM_FourierMatrix}), 
\begin{equation}
    \CMcal{G}_{\mu\nu}(\bq)\equiv \langle \sigma_\mu^z(-\bq)\sigma_\nu^z(\bq)\rangle = (\lambda \openone_{4\times 4} + \beta \Jmat(\bq))_{\mu\nu}^{-1}.
\end{equation}
It will be useful to shift the minimum eigenvalue of $\Jmat$ to zero by defining
\begin{equation}
\tilde{\Jmat} \equiv \Jmat - \epsilon_0 \mathbbm{1}, \quad \lambda_0  \equiv \lambda + \beta \epsilon_0,
\label{eq:SM_shift}
\end{equation}
where $\epsilon_0$ is the minimum eigenvalue of $\Jmat$. This has no influence on the physics, in particular the spin-spin correlations in \cref{eq:SM_G} are unchanged upon simultaneously replacing $\lambda \to \lambda_0$ and $\Jmat \to \tilde{\Jmat}$. For NNSI, $\epsilon_0 = -2J$, and for the spin vorticity model, $\epsilon_0 = -6J$. 

In \cref{fig:SM_lambda}, we show the evolution of $\lambda_0$ for NNSI and the spin vorticity model. In both cases it interpolates monotonically from $\lambda_0 = 1$ as $T\to\infty$ to $\lambda_0 = 1/2$ as $T \to 0$. At zero temperature, the correlation matrix $\CMcal{G}$ is proportional to the projector to the flat bands. To see this, let $\bm{\psi}_n$ denote the $n$'th eigenvector of $\tilde{\Jmat}$, with corresponding eigenvalue $\tilde{\epsilon}_n \geq 0$. Then, performing a spectral decomposition of $\CMcal{G}$, we obtain
\begin{equation}
    \lim_{T\to 0} \CMcal{G} = \lim_{\beta\to \infty} \sum_{n=1}^{4L^3} \frac{\bm{\psi}_n \bm{\psi}_n^{\dagger}}{\lambda_0 + \beta \tilde{\epsilon}_n} \to \lambda_0^{-1}\sum_{\mathclap{\{n \vert \tilde{\epsilon}_n = \,0\}}} \bm{\psi}_n \bm{\psi}_n^{\dagger},
\end{equation}
where the final sum is over all eigenvectors with zero eigenvalue. 
In other words, at zero temperature $\CMcal{G}$ becomes $\lambda_0^{-1}$ times the projector to the null space of $\tilde{\Jmat}$, which by definition is the set of its smallest eigenvalues. 
In NNSI and the spin vorticity model, the null space is precisely the space spanned by the eigenvectors of the flat bands, including the band touching point at $\bq=\bm{0}$.

\begin{figure*}[t]
    \centering
    \includegraphics[width=\textwidth]{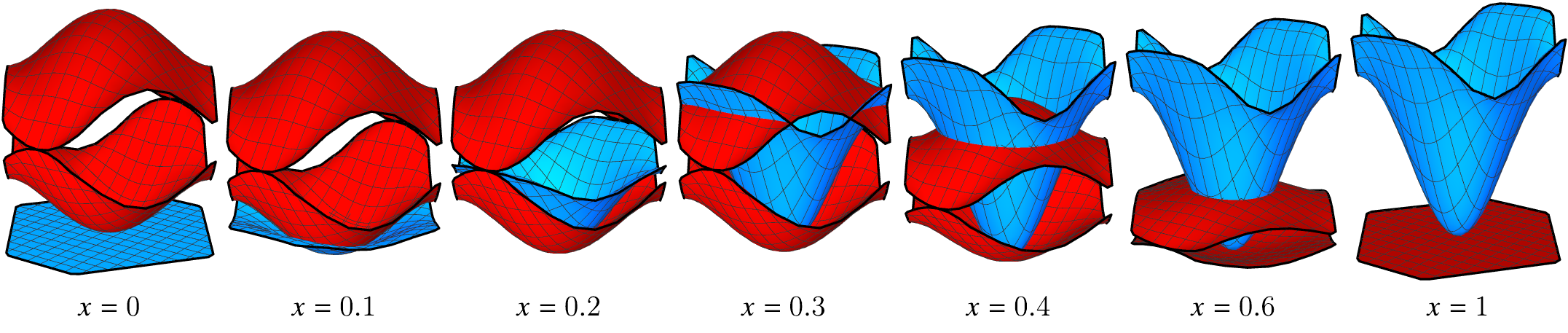}
    \caption{The bands of the interpolated interaction matrix \cref{eq:SM_J_interpolation}, where $x=0$ gives the bands of the NNSI model and $x=1$ gives the bands of the spin vorticity model. The flat bands of NNSI (blue) become dispersive in the spin vorticity model, while the dispersive bands of NNSI (red) become flat in the spin vorticity model. This demonstrates that the two models are in a sense ``reciprocal'' or ``complementary'', as reflected in the fact that one minimizes a divergence while the other minimizes a curl (see the discussions in \cref{sec:SCGA} and \cref{apx:Long_wavelength_Expansion}). The total bandwidth at $x=0$ and $x=1$ is $8J$. }
    \label{fig:sm_interpolate}
\end{figure*}

\subsection{Coarse-Grained Field Theory}
\label{apx:Long_wavelength_Expansion}

We can obtain a coarse-grained theory directly from the Hamiltonian by following the method of Conlon and Chalker~\cite{conlonAbsentPinchPoints2010}  by performing a long-wavelength expansion of the interaction matrix. 
Expanding the vorticity interaction matrix (\cref{eq:sm_J_A_q} using the parameters from \cref{eq:sm_vorticity_Js}) to quadratic order in $\bq$, we obtain (denoting $\vert \bq\vert^2 \equiv q^2$)
\begin{widetext}
\begin{equation}
    \frac{\tilde{\Jmat}^{(\text{SV})}_{\mu\nu}(\bq)}{J} \approx 
    \frac{a_0^{2}}{2}\begin{pmatrix}
     q^2 - q_x q_y - q_y q_z - q_z q_x   &   q_x q_y - q_z^2   &   q_x q_z -q_y^2   &   q_y q_z - q_x^2 \\
     q_x q_y - q_z^2   &   q^2 - q_x q_y + q_y q_z + q_z q_x   &  -q_y q_z - q_x^2  &  -q_x q_z - q_y^2 \\
     q_x q_z -q_y^2    &  -q_y q_z - q_x^2  &   q^2 + q_x q_y + q_y q_z - q_z q_x   &  -q_x q_y - q_z^2 \\
     q_y q_z - q_x^2   &  -q_y q_z - q_x^2  &  -q_x q_z - q_y^2   &   q^2 + q_x q_y - q_y q_z + q_z q_x
    \end{pmatrix}.
    \label{eq:SM_J_long_wavelength_1}
\end{equation}
\end{widetext}
We then introduce the new variables
\begin{align}
    Q(\bq) &\equiv \frac{1}{2} \sum_\mu \sigma_\mu^z(\bq), \label{eq:sm_Q}\\
    \bm{E}(\bq) &\equiv \frac{\sqrt{3}}{2} \sum_{\mu} \sigma_\mu^z(\bq) \hat{\bm{z}}_\mu, \label{eq:sm_E}
\end{align}
where $\hat{\bm{z}}_\mu$ are the local $[111]$ axes given by \cref{eq:sm_quant_axis}. Note that $Q(\bm{0})$ is the all-in-all-out (AIAO) order parameter, maximized when all $\sigma_i^z = 1$, and $\bm{E}(\bm{0})$ is effectively the bulk magnetization. These variables correspond to the following unitary change of basis~\cite{conlonAbsentPinchPoints2010}
\begin{equation}
    \begin{pmatrix}
    Q(\bq) \\
    E^x(\bq) \\
    E^y(\bq) \\
    E^z(\bq)
    \end{pmatrix}
    = 
    \frac{1}{2}\begin{pmatrix}
    1 & 1 & 1 & 1 \\
    1 & -1 & -1 & 1 \\
    1 & -1 & 1 & -1 \\
    1 & 1 & -1 & -1
    \end{pmatrix}
    \begin{pmatrix}
    \sigma^z_1(\bq)\\
    \sigma^z_2(\bq)\\
    \sigma^z_3(\bq)\\
    \sigma^z_4(\bq)
    \end{pmatrix}.
    \label{eq:SM_change_of_Basis}
\end{equation}
Transforming \cref{eq:SM_J_long_wavelength_1} to this basis, it becomes simply
\begin{equation}
    \tilde{\Jmat}^{(\text{SV})}_{\rho\sigma}(\bq) \approx 
    J a_0^2
    \begin{pmatrix}
        0 & 0 & 0 & 0 \\
        0 & q_y^2 + q_z^2 & -q_x q_y & - q_x q_z \\
        0 & -q_x q_y & q_x^2 + q_z^2 & -q_y q_z \\
        0 & -q_x q_z & -q_y q_z & q_x^2 + q_y ^2
    \end{pmatrix},
    \label{eq:JSV_inQEbasis} 
\end{equation}
where we use $\rho,\sigma$ to denote the components in the $(Q,\bm{E})$ basis given by \cref{eq:SM_change_of_Basis}. 
Note that the non-zero $3\times 3$ block in \cref{eq:JSV_inQEbasis} has matrix elements $q^2 (\delta_{ab} - q_a q_b/q^2)$, i.e. $q^2$ times the projector transverse to $\bq$. 

Plugging this into the SCGA effective Hamiltonian \cref{eq:SM_H_SCGA}, and accounting for \cref{eq:SM_shift}, we obtain the effective long-wavelength (coarse-grained) theory 
\begin{equation}
    \beta H_{\text{SCGA}}^{(\text{SV})} \!\approx \!\frac{1}{2}\int\!\! \mathrm{d}^3\bq \left[ \lambda_0 (Q^2 + \vert \bm{E}\vert^2) + \xi^2 q^2 \vert\hat{\bm{q}}\times \bm{E}\vert^2 \right]\!,
    \label{eq:SM_H_vort_SCGA_longwavelength}
\end{equation}
with $\xi^2 \equiv \beta J a_0^2$. The equivalent expression for NNSI is~\cite{conlonAbsentPinchPoints2010}
\begin{equation}
    \cramped{\beta H_{\text{SCGA}}^{(\text{SI})} \!\approx \!\frac{1}{2}\int\!\! \mathrm{d}^3\bq \left[ 8\beta J Q^2 + \lambda_0(Q^2+ \vert \bm{E} \vert^2) + \xi^2 q^2 \vert \hat{\bq}\cdot\bm{E}\vert^2 \right]\!.}
    \label{eq:SM_H_NNSI_SCGA_longwavelength}
\end{equation}
There are two key differences between the two coarse-grained theories. The most important is in the last term of each equation with coefficient $\xi^2 q^2$. 
For NNSI, the last term in \cref{eq:SM_H_NNSI_SCGA_longwavelength} tells us that the \emph{single} band that disperses away from the three-fold band touching point at $\bq = \bm{0}$ (c.f. \cref{fig:vort_bands_SCGA}(a)) corresponds to the longitudinal component of the coarse-grained field, $\hat{\bq}\cdot\bm{E}(\bq)$. 
For the spin vorticity model, the last term in \cref{eq:SM_H_vort_SCGA_longwavelength} tells us that the \emph{two} degenerate quadratically dispersing bands (c.f \cref{fig:vort_bands_SCGA}(b)) correspond to the two transverse components of the coarse-grained field, $\hat{\bq}\times \bm{E}(\bq)$. 
The nature of the flat bands is revealed by separating the field into longitudinal and transverse components,
\begin{equation}
    \vert\bm{E}\vert^2 = \vert \hat{\bq}\cdot\bm{E}\vert^2 + \vert \hat{\bq}\times\bm{E}\vert^2,
\end{equation}
and substituting back for $\vert{\bm E}\vert^2$ in \cref{eq:SM_H_vort_SCGA_longwavelength,eq:SM_H_NNSI_SCGA_longwavelength}. 
In NNSI the two transverse components lack dispersion and thus correspond to the two flat bands, while in the spin vorticity model both the longitudinal component of $\bm{E}$ as well as the scalar $Q$ lack dispersion, thus constituting the two flat bands. 

The second difference is the role of $Q$, which in NNSI corresponds to the upper band that is gapped at the Brilloun zone center with maximum energy $8J$. 
Since it is gapped in NNSI, it can be integrated out entirely and ignored in the coarse-grained theory describing the low-energy description~\cite{conlonAbsentPinchPoints2010}. 
In the spin vorticity model, by contrast, the AIAO state maximizing $Q(\bm{0})$ is a ground state, along with the three fully-polarized configurations maximizing $\bm{E}(\bm{0})$, corresponding to the fourfold degeneracy at $\bq=\bm{0}$ in the bands of the spin vorticity model interaction matrix (see \cref{fig:vort_bands_SCGA}(b)).

\subsection{Interpolation to NNSI}
\label{apx:interpolation_to_NNSI}

To visualize the ``reciprocal'' relationship between the spin vorticity model to the nearest-neighbor spin ice (NNSI) model, we can look at how their bands evolve as we interpolate between them. 
The NNSI interaction matrix is proportional to the nearest-neighbor adjacency matrix of the pyrochlore lattice,
\begin{equation}
    \Jmat_{ij}^{(\text{NNSI})} = J \Amat_{ij}^{(1)}.
\end{equation}
We thus define a continuum of models interpolating from NNSI to the spin vorticity model defined by the interaction matrices
\begin{equation}
    \tilde{\Jmat}^{\text{interp}}(x) =  (1-x) \tilde{\Jmat}^{(\text{NNSI})} + x \,\tilde{\Jmat}^{(\text{SV})},
    \label{eq:SM_J_interpolation}
\end{equation}
where for convenience we keep the minimum eigenvalue at zero according to \cref{eq:SM_shift}.
In \cref{fig:sm_interpolate}, we show the evolution of the bands as $x$ varies from $0$ (NNSI) to 1 (vorticity model). The blue bands are flat and degenerate in the NNSI model, corresponding to the extensively degenerate ground state of closed head-to-tail configurations with zero divergence (the rotational component of the Helmholtz decomposition). Conversely, the red bands are flat and degenerate in the spin vorticity model, corresponding to the irrotational (curl-free) component of the Helmholtz decomposition and the degenerate ground states of the spin vorticity model made of closed membranes. 
One may directly verify that $\tilde{\Jmat}^{(\text{NNSI})}$ and $\tilde{\Jmat}^{(\text{SV})}$ commute, and thus share a common eigenbasis. Therefore, only the eigenvalues of $\tilde{\Jmat}^{\text{interp}}(x)$ vary as a function of $x$, but \emph{not} the eigenvectors, meaning that the character of the bands as either irrotational (red) or rotational (blue) does not change as the bands deform upon varying $x$.

\section{Monte Carlo Simulations}
\label{apx:MonteCarlo}
This appendix provides details of the Monte Carlo simulations and data analysis described in the main text, \cref{subsec:MonteCarlo}.

\subsection{Sampling and Star Moves}
\label{apx:MonteCarlo_sampling}

We compute the spin-spin correlation functions and other thermodynamic observables by averaging over $N_{\text{samp}}$ sample configurations at each temperature, generated by Markov chain Monte Carlo simulations performed for an $L\times L \times L$ FCC lattice with periodic boundaries, containing $4L^3$ spins. 
Simulations for NNSI are performed using the extremely efficient algorithm introduced by Otsuka~\cite{otsukaClusterAlgorithmMonte2014} (see therein for 
details). This algorithm splits the entire spin configuration into closed and open strings, which are flipped with 50/50 probability. Since the autocorrelation times are extremely short, we take a sample after each iteration of the algorithm.

For the spin vorticity model, we utilize both single spin flips and ``star moves'' (\cref{subsec:MonteCarlo}) to thermalize the system. 
A star move flips an all-out tetrahedron to an all-in tetrahedron or vice-versa---for the spin vorticity Hamiltonian, this cluster move costs zero energy. 
For single spin flips, we choose a random spin to flip, accepting or rejecting according to the Metropolis condition, i.e. with probability $\text{min}[1,\exp(-\beta \Delta E)]$, where $\beta$ is inverse temperature and $\Delta E$ is the change in the total energy due to the flip. 
For star moves, we pick a random tetrahedron and flip it if it is all-in or all-out---this move satisfies detailed balance since the before and after configurations have the same energy and the transition probability in either direction is the same.
We take a sample configuration after each sweep of the lattice, where a single sweep means applying $4L^3$ single spin flip attempts and $2L^3$ star move attempts. 
At temperatures well below the minimum excitation energy scale of $6J$, where the vorticity spin liquid phase is well-formed, single spin flips are exponentially suppressed. 

\begin{figure*}[p]
    \centering
    \begin{overpic}[width=.914\textwidth]{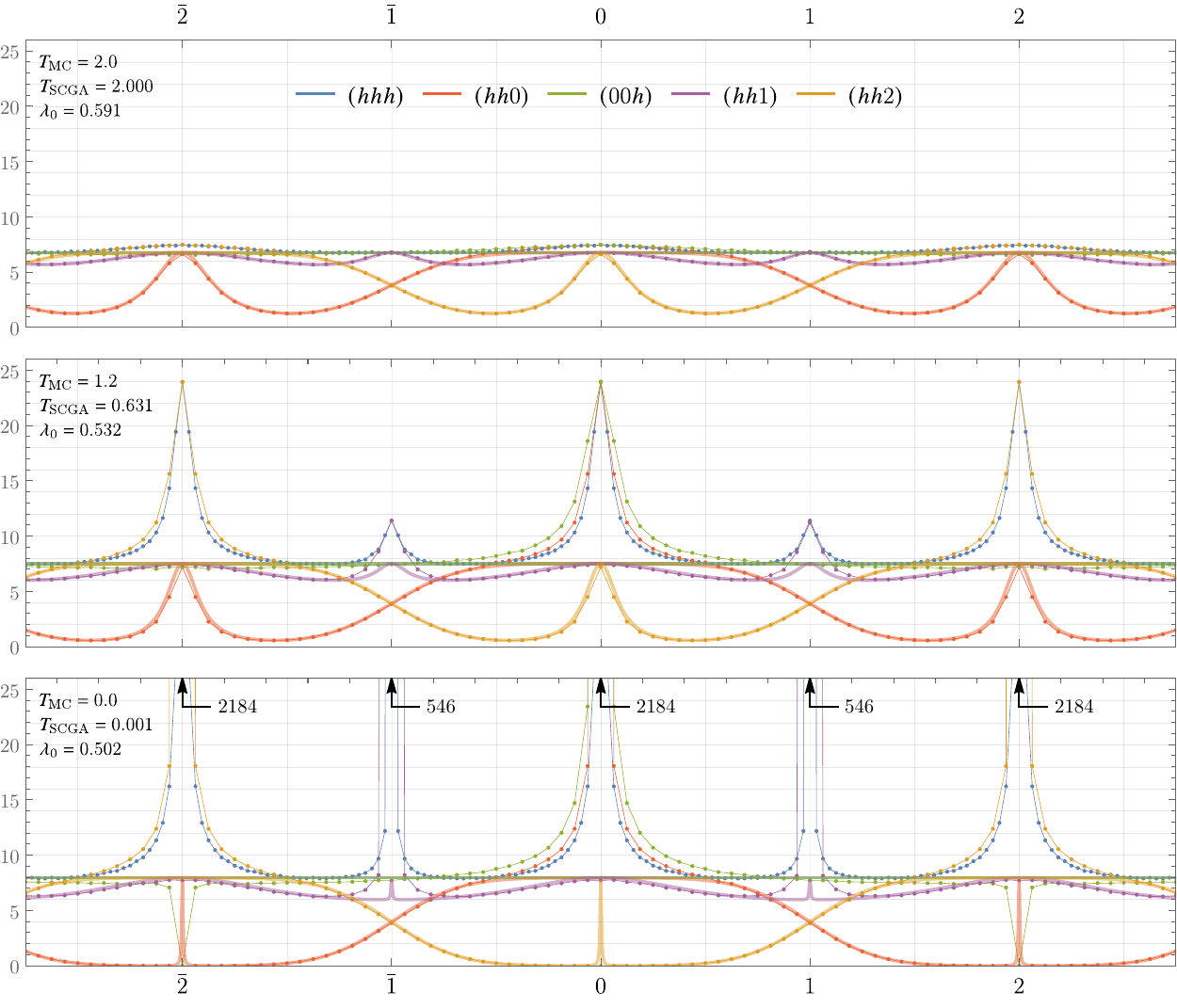}
        \put(-4,80){(a)}
        \put(-4,53){(b)}
        \put(-4,26){(c)}
        \put(-4,-3){(d)}
        \put(51,-3){(e)}
        \put(79,66){
            \includegraphics[width=.17\textwidth]{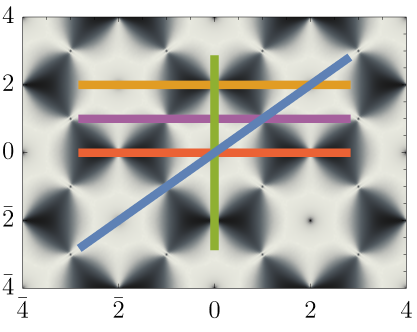}
        }
    \end{overpic}
    \\
    \raisebox{-.0\height}{
    \includegraphics[width=.49\textwidth]{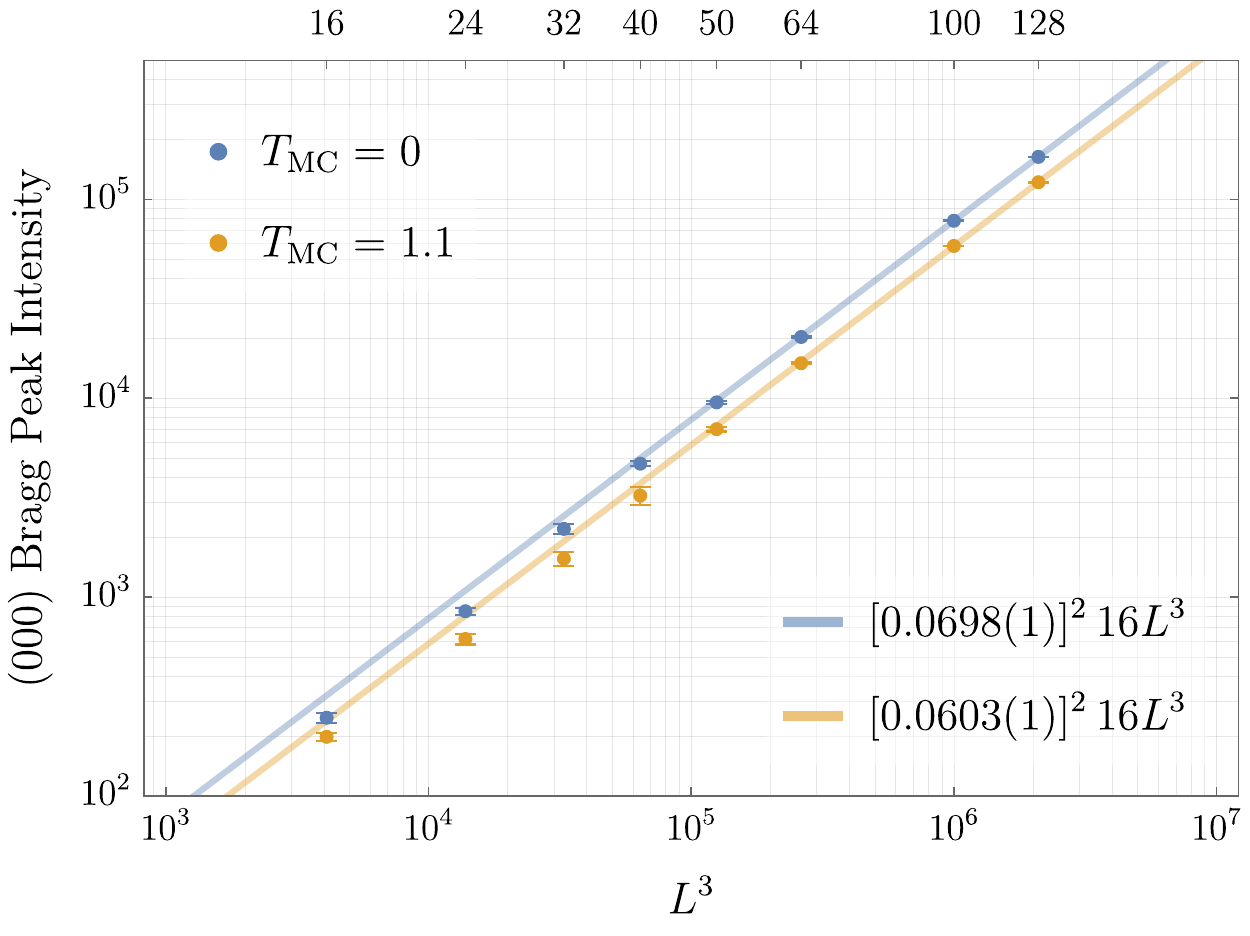}
    }
    \hfill
    \raisebox{0\height}{\includegraphics[width=.43\textwidth]{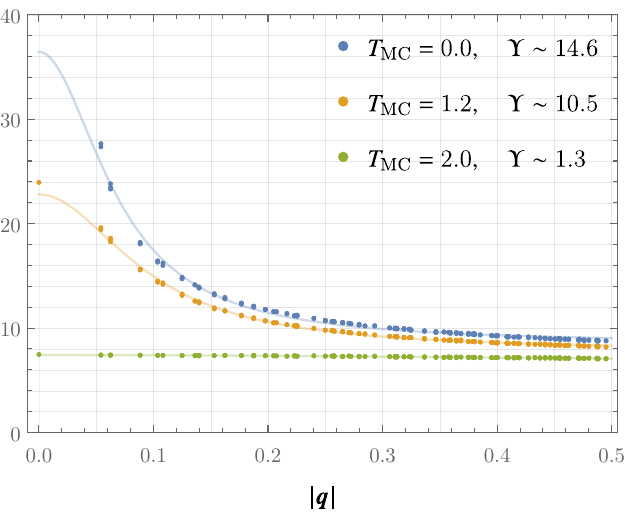}}
    \hspace{.025\textwidth}
    \caption{(a,b,c): Line cuts of the spin structure factor $S(\bq)$ for the spin vorticity model at three temperatures. Filled dots show Monte Carlo data for temperatures $T_{\text{MC}} = 2.0,1.2,0.0$, averaged over $10^5$ samples at each temperature with $L=32$. The Bragg peaks in (c) are noted with arrows indicating their intensities. Thick transparent lines show the SCGA result for temperatures $T_{\text{SCGA}} = 2.0,0.63,0.001$, chosen to quantitatively match the diffuse intensity. 
    Note that in the SCGA, $S(\bq)$ is constant along the highest-symmetry lines $(00h)$ and $(hhh)$. 
    Inset: $S(\bq)$ in the $(hhl)$ plane from which the line cuts in panels (a), (b) and (c) were obtained, with the five cut lines indicated by matching colors.
    (d) Scaling of the Bragg peak with system size, demonstrating that long-range order is only partially saturated in the thermodynamic limit. 
    (e) Lorentzian fits of the tails surrounding the $\bq=\bm{0}$ Bragg peak, in the region $\vert \bq\vert \lesssim 0.5$, at the same three temperatures as (a,b,c). }
    \label{fig:sm_linecuts}
\end{figure*}

\subsection{Line Cuts of Structure Factors}
\label{apx:LineCuts}

In \cref{fig:sm_linecuts}, we show line cuts of the Monte Carlo measurement of the spin structure factor $S(\bq)$ (\cref{eq:Sq}) appearing in  \cref{fig:vorticity_monte_carlo}(b,c,d) in the main text. 
Monte Carlo data points are shown by colored points, connected by thin solid lines as a guide to the eye. 
For comparison, we also plot the same line cuts computed in the SCGA, shown as thick transparent lines.
Note that because of the "softening" of the spin length constraint in the SCGA, observables computed at a fixed temperature $T$ in both SCGA and Monte Carlo will not necessarily agree, since the ``hard'' Ising spin length constraint in Monte Carlo means there is a finite energy gap to excitations, while in the SCGA the soft spins can fluctuate continuously. 
Thus we compare Monte Carlo data at temperature $T_{\text{MC}}$ with SCGA data at a temperature $T_{\text{SCGA}}$, the latter being chosen to give a good quantitative fit of the structure factor in the diffuse regions away from zone centers. 

The primary feature differentiating the Monte Carlo results from the SCGA is the presence of the sharp Bragg peaks surrounded by Lorentzian-like ``tails'' in the Monte Carlo data. 
These tails begin to form above $T_*$ and maintain a finite width all the way to zero temperature, despite the formation of delta-function Bragg peaks.
We emphasize that the SCGA tracks the development of the diffuse correlations of the spin liquid, away from the Bragg peaks, to a high degree of precision at all temperatures, i.e. we can always find an SCGA temperature $T_{\text{SCGA}}$ at which the SCGA quantitatively agrees with the Monte Carlo data in the diffuse region.
The second important difference between the SCGA and Monte Carlo data is the loss of pinch point intensity at the (002) position in the Monte Carlo data at the lowest temperature.
This loss of intensity is visible in the $(hh2)$ cut (yellow) at $h=0$, in the $(hh0)$ cut (red) at $h=\pm 2$, and in the $(00h)$ cut (green) at $h= \pm 2$.
We argued in \cref{sec:top_sector_ergodicity} that this loss is due to the lack of topological sector fluctuations at low temperatures since we do not have at hand an explicit cluster move that changes the topological sector in the Monte Carlo simulations (i.e. one which identifies and flips non-contractible membranes).

In the bottom right panel of \cref{fig:sm_linecuts} we also show the shape of the tails surrounding the zone-center Bragg peak as a function of $\vert\bq\vert$ in units of $2\pi/a_0$ in the long-wavelength region $\vert \bq\vert \lesssim 0.5$. The tails can be fit to Lorentzians of the form
\begin{equation}
    f(\bq) = C + \frac{A}{1 + \Upsilon^2 \vert \bq\vert^2},
\end{equation}
where $C$ accounts for the near-constant background intensity which is predicted by the SCGA. 
The quantity $\Upsilon$ may be interpreted as a correlation length, which saturates to a value on the order of $\Upsilon \sim 10^1$ at low temperature.
This lengthscale does not diverge at the transition, increasing monotonically from zero at high temperature to a finite value at $T=0$, without diverging or showing clear features at the transition temperature~$T_*$.

\subsection{Partially Saturated AIAO Order}
\label{apx:AIAO}

The pattern of Bragg peak intensities in the structure factor $S(\bq)$ (\cref{eq:Sq}) seen in \cref{fig:vorticity_monte_carlo}(d) is indicative of all-in-all-out (AIAO) ordering. To see this, first note that for a fully-ordered AIAO configuration, i.e. where every $\sigma_i^z = 1$, one can directly compute $S(\bq)$. 
Splitting the sum over pyrochlore sites $i$ into a sum over FCC and sublattice positions (\cref{eq:sm_r_i}), the double sum over FCC vectors yields a Kronecker delta restricting $\bq$ to be a reciprocal lattice vector,
\begin{equation}
    S_{\text{AIAO}}(\bq) = 
    \left(\sum_{\mu,\nu} e^{i\bq\cdot(\bm{c}_\nu - \bm{c}_{\mu})}\right)\, L^3 \delta_{\bq,\sum_k m_k \bm{b}_k}\quad (m_k \in \integers).
    \label{eq:Saiao}
\end{equation}
The intensity at each zone center can differ due to interference of the sublattice phase factors (the term in parentheses in Eq.~\eqref{eq:Saiao}),
yielding three symmetry-distinct zone-center Bragg peak intensities.
Inserting \cref{eq:sm_sublat_positions} in Eq.~\eqref{eq:Saiao}, we find
\begin{equation}
    S_{\text{AIAO}}(\bq) = 
    \begin{cases}
        16L^3 &\quad \bq=(000),\\
        4L^3 &\quad \bq=(111),\\
        0 &\quad \bq=(002).
    \end{cases}
    \label{eq:sm_AIAO_intensities}
\end{equation}
As observed in \cref{fig:vorticity_monte_carlo}(d) in the main text as well as in \cref{fig:sm_linecuts}, the zero-temperature Bragg peak intensities measured in the  Monte Carlo simulations for the spin vorticity model satisfy precisely the ratios (4:1:0) expected for AIAO order, \emph{but} with the $L=32$ peak intensities only about $0.4\%$ of those given by \cref{eq:sm_AIAO_intensities}.

To confirm that this partially saturated order is not the consequence of a finite-size effect, we have measured the scaling of the $\bq = \bm{0}$ Bragg peak with system size, shown in \cref{fig:sm_linecuts}(d), at $T=0$ using only star moves and at $T/J = 1.1$ (slightly below $T_*/J\approx1.15$) using both single spin flips and star moves. 
The peak height asymptotes to a constant slope starting around $L\sim 50$. 
This is consistent with the presence of a long lengthscale $\Upsilon$ on the order of $10$ unit cells, extracted from the width of the Lorentzian bumps beneath the Bragg peaks, c.f. \cref{fig:sm_linecuts}(e).
That is, it is necessary to reach linear system sizes of a few multiples of $\Upsilon$ before entering the scaling regime. 
From this data we can extract the thermodynamic limit of the AIAO order parameter, $Q = (1/4L^3)\sum_i \sigma_i^z$, which is bounded by $\vert Q\vert \leq 1$. 
In the large $L$ limit the $\bq=\bm{0}$ structure factor should scale as 
\begin{equation}
    S(\bm{0}) = 16L^3 \langle Q^2\rangle \xrightarrow{\quad L\,\to\,\infty\quad} 16L^3 \langle Q\rangle^2.
\end{equation}
Referring to Fig.~\ref{fig:sm_linecuts}(d), we fit $S(\bm{0})$ vs.~$16 L^3$ for the three largest system sizes, $L=64,100,128$, to extract the asymptotic behavior, finding slopes of
$\approx 0.0698^2$ and $\approx 0.0603^2$ at $T=0$ and $T=1.1$, respectively.
From this analysis we deduce that the low-temperature AIAO order parameter is saturated in the thermodynamic limit at approximately $7\%$ of its fully-saturated value of unity.

\subsection{Specific Heat and Entropy}
\label{apx:specific_heat_entropy}

For the specific heat data shown in the main text \cref{fig:vorticity_monte_carlo}(a), $N_{\text{samp}}= 10^5$ samples were averaged at each temperature for a system with $L=32$. The specific heat is measured as the variance of the energy divided by temperature squared,
\begin{equation}
    c(T) = \frac{1}{4L^3} \frac{\langle E^2 \rangle - \langle E\rangle^2}{T^2}.
    \label{eq:sm_specificheat}
\end{equation}
Uncertainties in the statistical averages are estimated via Jackknife resampling. 

The entropy per spin, $s(T)$, is computed by integrating $c/T$ as
\begin{equation}
    s(T) = s(\infty) + 
    \int_\infty^{100J} \mathrm{d}T' \frac{c(T')}{T'} + 
    \int_{100J}^{T} \mathrm{d}T' \frac{c(T')}{T'},
    \label{eq:apx_entropy_integrals}
\end{equation}
where $s(\infty) = \ln(2)$. The first integral in \cref{eq:apx_entropy_integrals} contains the entropy released between $T/J=\infty$ and $T/J=100$, the highest temperature simulated in Monte Carlo, which is estimated as follows. 
Over this range, the specific heat is approximated by its higher-temperature series form,
\begin{equation}
    c(T)\approx a T^{-2} + b T^{-3},   
\end{equation}
where the leading term goes as $T^{-2}$ since the variance of the energy in \cref{eq:sm_specificheat} is constant at high temperature. 
We therefore fit $c(T) T^2$ versus $T^{-1}$ between $T/J=50$ and $T/J=100$ to a linear function $a + bx$ (where $x=T^{-1}$) to extract the values of $a$ and $b$, from which we obtain
\begin{align}
    &\int_\infty^{100J} \mathrm{d}T' \frac{c(T')}{T'} \approx -\frac{150a + b}{3\times 10^6} \\
    &\qquad= \begin{cases}
        -(2.59\pm .02)\times 10^{-4} &\quad \text{vorticity}, \\
        -(1.47\pm.02)\times 10^{-4} &\quad \text{NNSI}.
    \end{cases}
\end{align}
The remaining entropy release below $T/J=100$ (the second integral in \cref{eq:apx_entropy_integrals}) is computed by interpolating the Monte Carlo $c(T)$ data and integrating numerically. 

The uncertainties in the integrated quantities are computed by resampling the data: each data point whose average and standard deviation is measured in Monte Carlo is treated as a random Gaussian variable to generate $N_{\text{resamp}}$ new $c(T)$ data sets, for each of which $s(T)$ is computed by numerical integration. 
The resulting values are averaged and their standard deviation is taken as the uncertainty, indicated by the error bars in \cref{fig:vorticity_monte_carlo}(b) in the main text. 
We find that the averages and uncertainties have converged well for $N_{\text{resamp}}\sim 10^2$, and the reported values are given for $N_{\text{resamp}} = 10^3$.

\subsection{Pinch Points and Topological Sector Fluctuations}
\label{apx:Topological_Sectors}

As noted in the main text, the intensity of the structure factor $S(\bq)$ for the spin vorticity model at the $\bq=(002)$ pinch point location is zero at low temperatures in our Monte Carlo simulations. 
We argued in \cref{sec:top_sector_ergodicity}  that this is because our star move algorithm is not fully ergodic since it is not capable of changing the topological sector of a spin  configuration. 
The topological sectors are indexed by $\bm{E}(\bm{0})$ (\cref{eq:sm_E}), which is proportional to the bulk magnetization. 
To see this, first define the net physical magnetic moment of a single tetrahedron, 
\begin{equation}
    \bm{E}_t \equiv \frac{\sqrt{3}}{2}\sum_{i\in t} \sigma_i^z \hat{\bm{z}}_i,
\end{equation}
in terms of which the bulk moment is
\begin{equation}
    \bm{E}(\bm{0}) = \frac{1}{2}\sum_{t} \bm{E}_t,
\end{equation}
where the factor of 1/2 accounts for double counting of spins after summing over all tetrahedra. 
Consider first a zero-magnetization all-in-all-out configuration, e.g. every $\sigma_i^z = +1$, which is a ground state of the spin vorticity model and has $\bm{E}_t = \bm{0}$ on every tetrahedron. 
The minimal membrane move is the star move: flipping an all-out tetrahedron to an all-in tetrahedron or vice versa. This clearly does not affect $\bm{E}(\bm{0})$ since $\bm{E}_t = \bm{0}$ for this tetrahedron before and after. 
Next, note that any contractible (non-winding) closed membrane flip is equivalent to a sequence of star moves on every tetrahedron inside the membrane, so these also cannot change $\bm{E}(\bm{0})$.  
Now consider a membrane which winds around the periodic boundaries, for example a flat plane orthogonal to $\hat{\bm{z}}_1\parallel[111]$ (c.f. \cref{eq:sm_quant_axis}) cutting one of the triangular sublattices of the pyrochlore lattice (\cref{fig:large_membranes_strings}(b) in the main text shows a portion of such a flat membrane).
For each tetrahedron on either side of this membrane, the moment becomes $\bm{E}_t = \sqrt{3}\hat{\bm{z}}_1/2$, so the net moment becomes $\bm{E}(\bm{0}) = N_{\Delta}\sqrt{3} \hat{\bm{z}}_1$, where $N_{\Delta}$ is the number of plaquettes in the membrane (i.e. the number of flipped spins). We conclude that the value of $\bm{E}(\bm{0})$ depends only on the topological sector.

The intensity of the pinch point at $(002)$, where there is no AIAO Bragg peak, is proportional to $\langle \vert E^z(\bm{0})\vert^2\rangle$. To see this, first note that the spin structure factor, \cref{eq:Sq} in the main text, can be expressed as
\begin{equation}
    S(\bq) = 4 \langle \vert Q(\bq) \vert^2 \rangle,
\end{equation}
with $Q(\bq)$ defined in \cref{eq:sm_Q}. Now note that from \cref{eq:sm_s_mu_q},
\begin{equation}
    \sigma_\mu(\bq + \bm{k}) = \sigma_\mu^z(\bq) e^{-i\bm{k}\cdot\bm{c}_\mu},
\end{equation}
so that we can expand \cref{eq:sm_Q},
\begin{align}
    Q((002) + \bm{k}) &= \frac{1}{2} \sum_\mu \sigma_\mu^z(\bm{k}) e^{-i(002)\cdot\bm{c}_\mu} \\
    &= \frac{1}{2}\sum_\mu \sigma_\mu^z(\bm{k}) (-\sqrt{3} \hat{\bm{z}}\cdot \hat{\bm{z}}_\mu) \label{eq:sqrt_three}\\
    &= -\hat{\bm{z}}\cdot\bm{E}(\bm{k}),
\end{align}
where $\hat{\bm{z}}$ is the global Cartesian $z$-axis unit vector along [001] and $\hat{\bm{z}}_\mu$ are the local $[111]$ axes (see \cref{eq:sm_quant_axis} and \cref{eq:sm_E}).
In \cref{eq:sqrt_three} we have used that $\hat{\bm{z}}\cdot\hat{\bm{z}}_\mu = -1/\sqrt{3}$ for sublattices 1 and 2 and $+1/\sqrt{3}$ for sublattices 3 and 4, using the conventions in \cref{apx:conventions}, matching the alternating sign of the phase factor with wavevector (002).
We thus conclude that
\begin{equation}
    S((002) + \bm{k}) = 4 \langle \vert E^z(\bm{k})\vert^2 \rangle,
\end{equation}
where $\cramped{E^z(\bm{k}) = \hat{\bm{z}}\cdot\bm{E}(\bm{k})}$, and, in particular, 
\begin{equation}
    S((002)) = 4 \langle \vert E^z(\bm{0})\vert^2\rangle.
\end{equation}
Thus the pinch point intensity of $S(\bq)$ at the $(002)$ zone center is proportional to the fluctuation of $E^z(\bm{0})$, which measures the fluctuations between topological sectors. 
Its extinction in \cref{fig:vorticity_monte_carlo}(d) in the main text is thus indicative of a lack of topological sector  fluctuations at low temperature when employing only local single spin flip and star move updates.

\begin{figure}[t]
   \centering
    \begin{overpic}[width=.85\columnwidth]{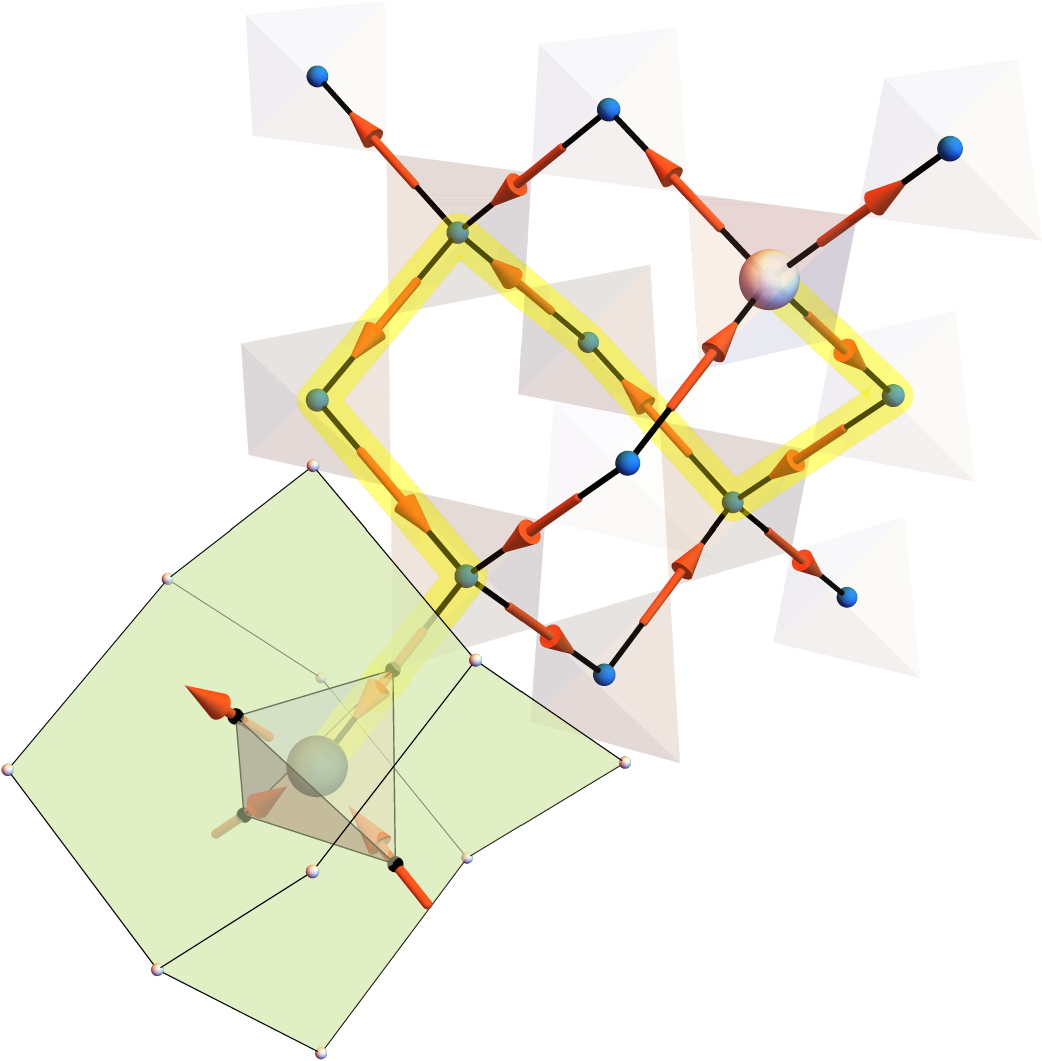}
        \put(0, 90){(a)}
        \put(0,-10){(b)}
    \end{overpic}
    \includegraphics[width=0.9\columnwidth]{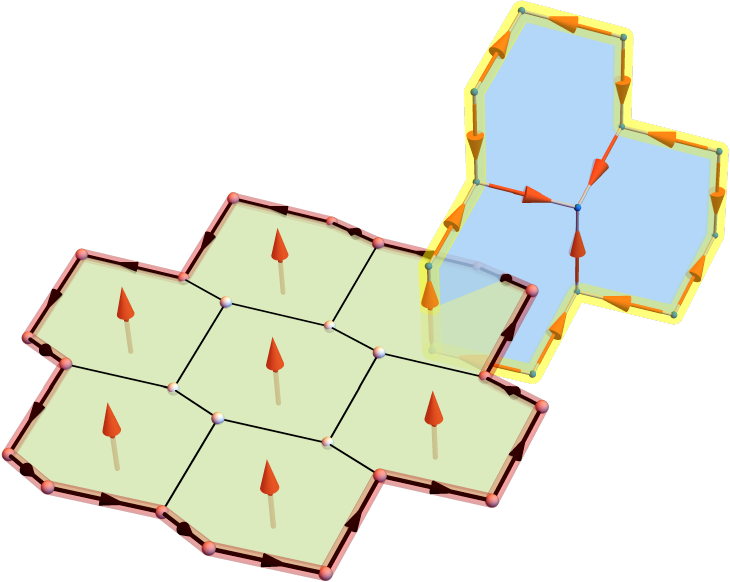}
    \\[3ex]
    \caption{(a) In spin ice, charges (black and white spheres) are detected by integrating the amount of flux through a closed surface in the dual lattice (green surface surrounding the black charge). A string (yellow) attached to the charges is not uniquely defined, but the intersection number of the string in the direct lattice and the closed surface in the dual lattice counts the amount of charge enclosed, \cref{eq:sm_div_theorem}, independent of the string position. (b) In the spin vorticity model, the excitations are charged strings (red), and these are detected by integrating the circulation of the spins around a loop in the direct lattice (yellow). This loop links the string excitations and pierces the flux membrane (green). The signed intersection of the yellow loop with the membrane counts the number of string excitations which pierce the blue surface, \cref{eq:sm_curl_theorem}.
    }
    \label{fig:gauss_ampere}
\end{figure}

\section{Topological Strings and Membranes on the Lattice}
\label{apx:string_membrane_relation}

The strings of spin ice and the membranes of the spin vorticity model play complementary roles in the two models. 
Let us first begin by reviewing the case of spin ice, the constraint $Q_t = 0$ enforces that the electric strings cannot have open ends, while its violation is sourced by a charge at each open end of a string. 
In Maxwell theory, the presence of the charge is detected by integrating the Gauss law $\nabla\cdot\bm{E} = \rho$ (with $\rho$ the charge density) over a volume $V$ which, by Gauss's theorem, is equivalent to 
\begin{equation}
    \iiint_V (\nabla\cdot\bm{E})\,\mathrm{d}V \equiv \oiint_{\boundary V} \bm{E}\cdot\mathrm{d}\bm{S} = Q_{\text{enc}},
    \label{eq:gauss_integral}
\end{equation}
where $Q_{\text{enc}}$ is the charge enclosed by the surface $\boundary V$ which bounds $V$, and the surface is oriented ``out''. Thus a positive charge within the volume corresponds to a net flux of electric field lines out through its surface.

In the pyrochlore crystalline system, these surfaces necessarily should live in the dual lattice, such that they surround the vertices of the direct lattice and are pierced by the links of the direct lattice. 
This is depicted in \cref{fig:gauss_ampere}(a), which shows a closed surface in the dual lattice (green) surrounding a point charge (black) at the end of an electric string (yellow) in the direct lattice. 
The lattice equivalent of \cref{eq:gauss_integral} is (using the notation introduced in \cref{eq:inversion_sign})
\begin{equation}
    \sum_{\tilde{p}\in\boundary{\tilde{V}}} S^z_{\tilde p} = \sum_{t \in \tilde{V}} Q_t,
    \label{eq:sm_div_theorem}
\end{equation}
where $\tilde V$ is a volume in the dual lattice. Note that the position of the electric string is not uniquely defined, but the signed intersection number of the electric string and the closed surface counts the number of charges enclosed by the surface, where the signed intersection of a $k$-dimensional objects and a $(d-k)$-dimensional object ($d$ being the dimension of space) is $+1$ if they intersect with the same orientation and $-1$ if they intersect with opposite orientation. 

Leaving now spin ice and moving to the spin vorticity model, the constraint $\omega_p=0$ enforces that the spins arrange into closed electric membranes. 
Referring to \cref{fig:gauss_ampere}(b), violations of this constraint appear as closed charged strings in the dual lattice, which are detected by computing the circulation of the spins around a closed loop of spins (yellow) which links the string (red). 
These strings may be viewed as behaving analogously to a loop of current, sourcing the field by an Ampere's law, 
\begin{equation}
    \nabla\times\bm{E} = \bm{\mathscr{j}},    
\end{equation}
which reads in integral form
\begin{equation}
    \oint_{\boundary \Sigma} \bm{E}\cdot\mathrm{d}\bm{\ell} = \iint_\Sigma \bm{\mathscr{j}}\cdot\mathrm{d}\bm{S},
    \label{eq:ampere_integral}
\end{equation}
where $\Sigma$ is any surface. 
This equation tells us that the circulation of the field around any closed loop counts the amount of current flowing through a surface $\Sigma$ which has that loop as its boundary. 
To now proceed to write the lattice version of Ampere's law \cref{eq:ampere_integral}, we utilize~\cref{eq:omega_p_boundary} and sum the vorticity over all plaquettes forming a surface $\Sigma$ in the direct lattice to obtain
\begin{equation}
    \sum_{\ell \in \boundary \Sigma} S_{\ell}^z = \sum_{p \in \Sigma} \omega_p ,
    \label{eq:sm_curl_theorem}
\end{equation}
and identify $\omega_p = \mathscr{j}_{\tilde \ell}$.\footnote{Note this may be viewed as summing \cref{eq:omega_p_boundary} over all plaquettes in $\Sigma$, and all $S_{\ell}^z$ on interior edges cancel pairwise.
}
\Cref{eq:sm_curl_theorem} then states that the circulation of the spins around the bounding edge of the surface, $\boundary \Sigma$, is equal to the number of strings piercing the surface $\Sigma$, i.e. linking $\boundary \Sigma$.
This is depicted in \cref{fig:gauss_ampere}(b), which shows a single string (red) attached to a membrane (green) in the dual diamond lattice. 
It is linked by a closed contour (yellow) in the direct lattice, which is on the boundary, $\boundary \Sigma$, of a surface $\Sigma$ (blue). 
The circulation of $S^z$ around the yellow contour counts the number of strings piercing the blue surface $\Sigma$, which is the signed intersection number of the surface with the string.

Utilizing these microscopic lattice definitions, we can define topological winding numbers for both spin ice and the spin vorticity model ground states. In spin ice, where ground state configurations have zero divergence, the topological sector is determined by the number of electric strings piercing a topologically non-trivial surface $\Sigma$ in the dual lattice, 
\begin{equation}
    w[\Sigma]:= \sum_{\tilde{p} \in \Sigma} S^z_{\tilde p},
\end{equation}
which depends only on the homology class of $\Sigma$~\cite{hatcherAlgebraicTopology2002}, i.e. which of the three holes of the torus $\Sigma$ winds around, and is invariant under the flipping of a contractible loop of head-to-tail spins. The number $w[\Sigma]$ is zero if $\Sigma$ is contractible, since it is equivalent by \cref{eq:sm_div_theorem} to the total divergence inside the volume bounded by $\Sigma$. 
In the vorticity model, the topological sector of a ground state is defined by the winding number around non-contractible loops $C$ in the direct lattice, 
\begin{equation}
    w[C] := \sum_{\ell \in C} S_{\ell}^z,
\end{equation}
which depends only on the homology class of $C$, i.e. which hole it winds around, and is invariant under the flipping of contractible membranes. The number $w[C]$ is zero if $C$ is contractible, since by \cref{eq:sm_curl_theorem} it is equal to the total vorticity on a set of plaquettes forming a surface bounded by $C$.

\begin{figure}[t]
    \centering
    \begin{overpic}[width=0.65\columnwidth]{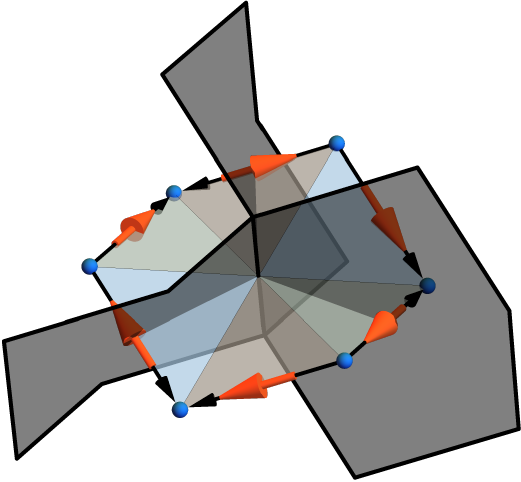}
        \put(-15,80){(a)}
        \put(-15,-25){(b)}
    \end{overpic}
    \\[4ex]
    \begin{overpic}[width=0.65\columnwidth]{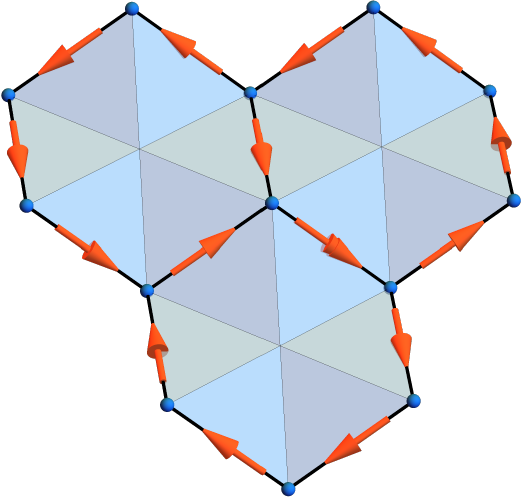}
        \put(16,65){$\omega_p = +2$}
        \put(43,27){$\omega_p = -3$}
        \put(63,65){$\omega_p = +3$}
    \end{overpic}
    \caption{(a) Spin configurations can be mapped to hard-plaquette coverings of the dual diamond lattice by identifying the two Ising states (indicated with red arrows) relative to a reference orientation (black arrowheads) with the presence or absence of a hard hexagon (dark gray). (b) In order to turn the ground state constraint $\omega_p = 0$ into a local constraint on edge-touching of hard hexagons, one should choose a reference orientation where every hexagon is fully polarized with $\vert\omega_p\vert = 3$. However, there is no such spin orientation due to the way that the direct-lattice hexagons touch at their shared corners. Here we show a configuration with two fully-polarized hexagons, which forbids the third (top left) hexagon from being fully polarized.}
    \label{fig:hard_hexagons}
\end{figure}

\section{Dimers versus Hard-Plaquette Coverings}
\label{apx:dimers-hard-hexagons}

It is worth considering how to think of the spin vorticity model in terms of dimer models, which play an important role in the understanding of liquid phases based on resonating valence bonds as well as in spin ice~\cite{moessnerQuantumDimerModels2011}.
This is done by mapping the two spin states with respect to a chosen reference configuration to the presence or absence of a dimer. 
In NNSI, we take the reference configuration to be an AIAO state, with the dimers covering the bonds of the diamond lattice. 
Then the two-in/two-out constraint $Q_t = 0$ translates to the constraint that exactly two dimers touch at each vertex (diamond site), such that the ground states of NNSI are in one-to-one correspondence with closed-string coverings of the diamond lattice with dimers~\cite{gingrasQuantumSpinIce2014,moessnerQuantumDimerModels2011}. 
It turns out that no such mapping exists for the spin vorticity model, which we now illustrate. 

The dimer language to express the constraint $\omega_p = 0$ in the spin vorticity model is more complicated than its usage in the NNSI case.
Let us take the AIAO configuration as our reference state, so that $\sigma_i^z = +1$ ($-1)$ corresponds to the presence (absence) of a dimer on each edge.
Separating the six edges direct diamond edges around the hexagonal plaquette in \cref{fig:hexagon} into the even and odd subsets, the constraint $\omega_p=0$ translates to covering an equal number of even and odd edges by a dimer. 
Expressed in this way, the constraint appears non-local and ad-hoc (i.e. lacking physical justification, such as a hard-core constraint), and one may wonder whether there is a more natural description. 

In spin ice, the dimer-touching constraint corresponds to the local zero-charge constraint $Q_t = 0$ at each diamond vertex, so it is natural to search for a similar touching constraint equivalent to $\omega_p = 0$, but located on the dual diamond lattice edges, where the string excitations live. 
Thus we consider mapping the Ising states $\sigma_i^z = \pm 1$ to the presence or absence of a \emph{hard plaquette} on the corresponding dual plaquette, rather than a dimer on the corresponding direct edge. 
This is illustrated in \cref{fig:hard_hexagons}(a), with six spins forming a $\vert\omega_p\vert = 3$ configuration which map, relative to all-in-all-out configuration (indicated by small black arrowheads), to the presence of three hard hexagons (dark gray). 

Switching from dimers to hard hexagons does not solve the issue of the ad-hoc constraint by itself, instead we need to choose a different basis in which the constraint $\omega_p = 0$ translates to the constraint that three hard hexagons touch at each dual edge. 
For a single plaquette $p$, this is achieved precisely when we choose the reference configuration to be one of the two maximally-circulating $\omega_p = \pm 3$ configurations, relative to which an $\omega_p = 0$ configuration corresponds to choosing any three of the six $\sigma_i^z$ on the hexagon to be $-1$. 
Then the constraint says that exactly three hard hexagons touch at the dual lattice edge passing through plaquette $p$. 
This is exactly analogous to spin ice, where the reference configuration is one of the two all-in all-out configurations and the constraint $Q_t = 0$ says that any two of the $\sigma_i^z$ on a tetrahedron are chosen to be $-1$. 
However, unlike the all-in-all-out configuration for which every tetrahedron satisfies $\vert Q_t \vert = 2$, there is no spin configuration for which every plaquette satisfies $\vert \omega_p \vert = 3$. 
This is illustrated in \cref{fig:hard_hexagons}(b), which shows that two adjacent hexagons can be chosen to have opposite maximal circulation, but a third hexagon sharing a spin with each of them cannot be assigned a maximal circulation configuration. 
Thus there is no hard-hexagon mapping where the constraint $\omega_p = 0$ becomes a simple local constraint that three hard-hexagons touch at each dual lattice edge, such that the system is at half filling in the ground state.

\begin{figure*}[ht]
    \centering
    \begin{overpic}[width=\textwidth]{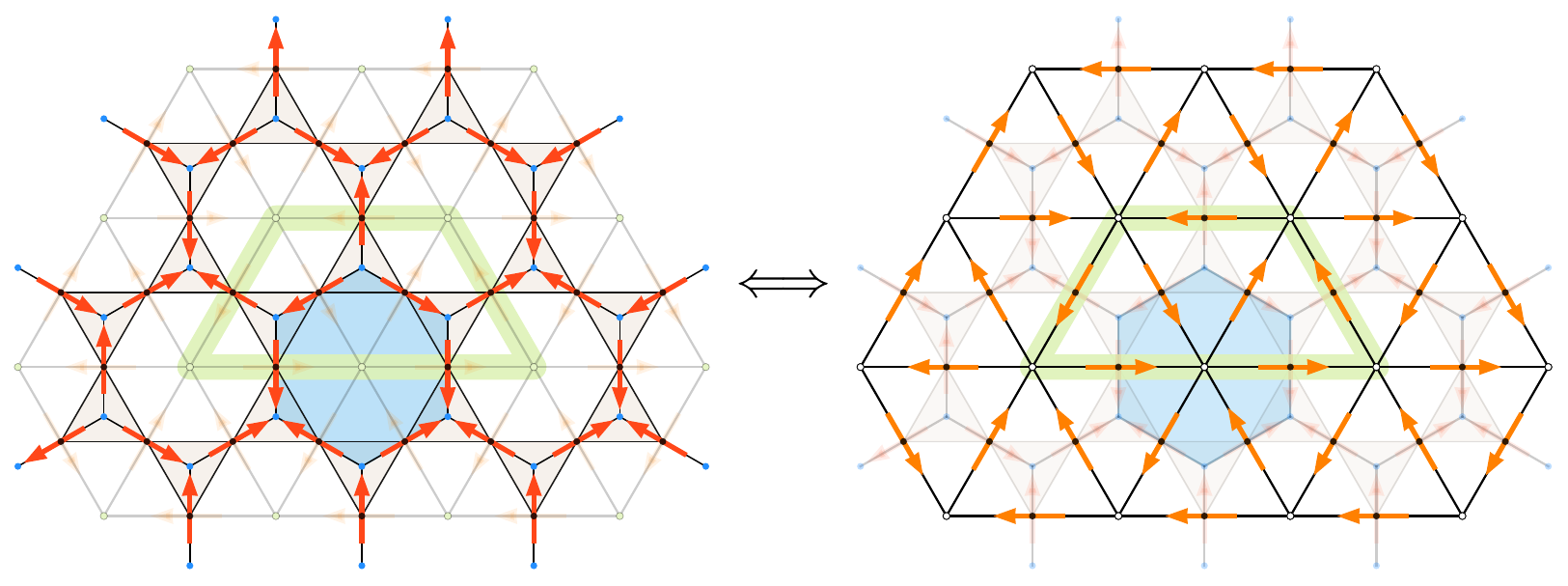}
        \put(22,-2){(a)}
        \put(75.8,-2){(b)}
    \end{overpic}
    \\[2ex]
    \caption{(a) A portion of the kagome lattice with Ising spins represented by red arrows oriented along the parent honeycomb lattice edges. Comparing to \cref{fig:pyro_diamond}(a), the kagome lattice plays the role of the pyrochlore lattice while the bipartite honeycomb lattice plays the role of the bipartite diamond lattice: the kagome is the line graph of the honeycomb, just as the pyrochlore is the line graph of the diamond. Defining a kagome version of the vorticity model whose Hamiltonian is given by \cref{eq:H_vort_wp} with vorticity defined on each hexagonal plaquette (one highlighted blue) according to \cref{eq:omega_p}, ground states have zero circulation around each plaquette. The spin configuration shown here satisfies the ground state constraint on the five shown hexagons. The dual lattice of the honeycomb is a triangular lattice, shown transparent in the background. The analog of a closed membrane in two dimensions in a closed curve. The green string highlighted on the dual triangular lattice intersects five spins, all of which point ``out'', analogous to the closed membranes in \cref{fig:small_membranes}(b,c). Flipping these spins does not change the circulation around any plaquette and thus does not change the energy, analogous to the membrane flips in the 3D spin vorticity model. Thus in a 2D spin vorticity model, the analog of the two-dimensional membranes are one-dimensional strings. (b) This model can be mapped to an ice model on the dual triangular lattice by simply rotating each spin vector by ninety degrees in the plane so that they point along the edges of the triangular lattice. The zero-vorticity constraint then becomes a zero divergence constraint in the dual model, i.e. a 3-in-3-out ice rule, which can be visually seen to be satisfied at the five vertices in the interior, each with six surrounding spins. The green string in the vorticity model in (a) becomes a string of head-to-tail spins in the ice model in (b), and flipping this string does not change the divergence at any triangular lattice vertex. This is because curl and divergence are dual operations in two dimensions. Thus two-dimensional spin vorticity models are dual to two-dimensional spin ice models. }
    \label{fig:kagome_vorticity_duality}
\end{figure*}

\section{2D Vorticity Models: Duality with Ice Models}
\label{apx:2d-vorticity-model}

One may wonder why the spin vorticity model we defined in \cref{eq:H_vort_wp} and discussed throughout this section should be defined in three spatial dimensions. One can indeed define a spin vorticity model in two dimensions (2D), such as the 2D kagom\'e lattice.
It is straightforward to see that the resulting model is in fact dual to an ice model, due to the duality in two dimensions between curl and divergence, and thus does not result in a 2-form spin liquid, but rather another 1-form spin liquid. 

To see this, consider an Ising model on the kagome lattice, shown in \cref{fig:kagome_vorticity_duality}(a) with arrows (red) indicating the directions of the local spins. 
For each plaquette $p$ of the parent honeycomb lattice, one of which is colored blue, we can define the vorticity $\omega_p$ identical to \cref{eq:omega_p}, and a spin vorticity model with Hamiltonian given by \cref{eq:H_vort_wp} which includes first, second, and third neighbor interactions whose ground states minimize the circulation around every plaquette. 
On the three-dimensional pyrochlore lattice, as discussed in \cref{sec:2-form-classical}, we could find closed surfaces in the dual diamond lattice on which all spins point the same direction, which can be flipped without changing the vorticity anywhere. 
On the two-dimensional kagome lattice, the dual of the parent honeycomb lattice is a triangular lattice, shown transparent in the background of \cref{fig:kagome_vorticity_duality}(a).
We trace a one-dimensional contour (green) intersecting five spins, all of which point ``out''.
Flipping these spins does not change the circulation around any plaquette, thus costing zero energy, and are thus the analog of the membrane flips discussed in the main text.
However, in two spatial dimensions this is a 1-dimensional string, not a 2-dimensional surface.

Indeed, these strings are in fact the strings of a spin ice model defined on the dual lattice.
Each spin configuration in \cref{fig:kagome_vorticity_duality}(a) may be mapped to a configuration of spins sitting on the links of a triangular lattice, by simply rotating every red arrow by ninety degrees in the plane, shown in \cref{fig:kagome_vorticity_duality}(b). 
Each direct plaquette with zero vorticity becomes a vertex with zero divergence in the dual system. This can be seen in \cref{fig:kagome_vorticity_duality}, where the blue plaquette has zero circulation in (a), and the corresponding vertex satisfies a 3-in-3-out ice rule in (b).\footnote{Note that conversely, the nearest-neighbor kagome ice model~\cite{willsModelLocalizedHighly2002} is dual to a vorticity model with spins on the edges of the triangular lattice by rotating all spins ninety degrees. 
    }
This is a natural consequence of the fact that in two dimensions, the curl and divergence operators are dual~\cite{henleyCoulombPhaseFrustrated2010}. 
Given a vector field $\bm{E} = E_x \hat{\bm{x}} + E_y \hat{\bm{y}}$, define its dual field $\bm{E}' = -E_y \hat{\bm{x}} + E_x \hat{\bm{y}}$, then $\nabla\cdot\bm{E} = \nabla\times\bm{E}'$ and $\nabla\times \bm{E} = -\nabla\cdot\bm{E}'$, where the two-dimensional curl is a scalar defined by $\nabla\times \bm{E} = \partial_x E_y-\partial_y E_x$ (i.e. the $\hat{\bm{z}}$-component of the 3D curl for a system confined to the $x$-$y$ plane).\footnote{This curl-divergence duality was utilized in Ref.~\cite{nisoliGaugefreeDualityPure2021} to study 2D ice models, and is also utilized to define height model representations of 2D lattice Coulomb phases~\cite{huseCoulombLiquidDimer2003,henleyCoulombPhaseFrustrated2010}.
}
We conclude that in two dimensions, there is a duality between spin vorticity models, which enforce a zero-curl constraint, and ice models, which enforce a zero-divergence constraint---both result in traditional Coulomb phases~\cite{henleyCoulombPhaseFrustrated2010}.
In the spin ice model the excitations are charges, while in the spin vorticity model they are vortices, so this may be viewed as a simple manifestation of the phenomenon of particle-vortex duality~\cite{seibergDualityWeb2+12016}.

\section{\textit{p}-Form Gauge Theory}
\label{apx:2-form-gauge-theory}

In this Appendix, we concisely summarize for the interested reader the basics of continuum higher-form gauge theory to contextualize the lattice-based discussion in the main text.
The core mathematical tools are differential forms, where a $p$-form is a rank-$p$ antisymmetric tensor field.\footnote{A detailed account can be found in differential geometry textbooks, for example, see Chapters 5 and 7 of Nakahara~\cite{nakaharaGeometryTopologyPhysics2003}. 
}
These are the natural types of fields which can be unambiguously integrated over a $p$-dimensional surface. 
We first summarize the core mathematical machinery required for the formulation of $p$-form gauge theory. 

\subsection{Differential Forms}

\emph{Differential Forms}---Consider a $d$-dimensional space $M$, and a set of $d$ basis 1-forms, denoted $\mathrm{d}x^\mu$, $\mu = 1, \ldots d$. 
One may think of these as a set of $d$ linearly independent vector fields. 
A 0-form is just a scalar function.
We can expand a 1-form, a 2-form, etc. in this basis in terms of their components as
\begin{equation}
    \alpha = \alpha_\mu \,\mathrm{d}x^\mu, 
    \quad 
    \beta = \frac{1}{2!}\,\beta_{\mu\nu}\, \cramped{\mathrm{d}x^\mu\wedge \mathrm{d}x^\nu}, 
    \quad 
    \text{etc.}
    \label{eq:diff_forms}
\end{equation}
with implied summation over repeated indices. 
The components of a $p$-form (i.e. $\alpha_\mu$, $\beta_{\mu\nu}$) are a collection of $p!$ functions, completely antisymmetric with respect to exchanging indices. 

\emph{Wedge product}---The wedge product, $\wedge$ in \cref{eq:diff_forms}, is the anti-symmetrized tensor product, meaning that 
\begin{equation}
    \diff x^\mu\wedge \diff x^\nu= -\diff x^\nu\wedge \diff x^\mu,
\end{equation}
The basis elements $\cramped{\diff x^{1}\wedge\cdots\wedge\diff x^{p}}$ play the role of the infinitesimal integration measures on $p$-dimensional surfaces, and the antisymmetry encodes a reversal of sign under a reversal of orientation. 
The wedge product of a $p$-form and a $q$-form is a $(p+q)$-form with components given by
\begin{equation}
    \alpha\wedge\beta = \frac{1}{p! q!} \alpha_{\mu_1 \cdots\mu_p} \beta_{\mu_{p+1}\cdots \mu_{p+q}} \diff x^{\mu_1} \wedge \cdots \wedge \diff x^{\mu_{p+q}}.
\end{equation}

\textit{Exterior Derivative}---The exterior derivative generalizes the notion of gradient to $p$-forms, sending a $p$-form $\omega$ to a $(p+1)$-form denoted $\diff_p \omega$, whose components are given by gradients of the components of $\omega$,
\begin{equation}
    \mathrm{d}_p\omega = \frac{1}{p!} \partial_{\mu_1} \omega_{\mu_2\cdots\mu_{p+1}} \, \cramped{\mathrm{d}x^{\mu_1}\wedge \cdots \wedge \mathrm{d}x^{\mu_{p+1}}},
    \label{eq:exterior_derivative}
\end{equation}
where $\partial_\mu=\partial/\partial x^\mu$. 
Note that, because partial derivatives commute, i.e. $\partial_\mu \partial_\nu$ is symmetric, it follows that $\diff_{p+1}\diff_{p} = 0$ for any $p$, since the contraction of a symmetric and anti-symmetric object is zero.
The subscript on the derivative is generally dropped when the rank of the form it acts on is clear, and we simply say that $\diff^2 = 0$.

\textit{Stoke's Theorem}---Utilizing the exterior derivative, we can state Stoke's theorem,
\begin{equation}
    \int_{U} \diff \omega = \int_{\boundary U} \omega,
    \label{eq:stokes}
\end{equation}
where $\omega$ is a $p$-form, $U$ is a $(p+1)$-dimensional surface, and $\boundary U$ is its $p$-dimensional boundary. 
Stoke's theorem should be viewed as a higher-dimensional generalization of the fundamental theorem of calculus, $\smash{\cramped{\int_a^b \diff f = f(b) - f(a)}}$, and indeed may be viewed as the defining property of the exterior derivative.
In three dimensions, comparison with standard vector calculus identities shows that
\begin{equation}
    \diff_0 \simeq \mathrm{grad} , \quad
    \diff_1 \simeq \mathrm{curl} , \quad
    \diff_2 \simeq \mathrm{div} \quad (d=3).
\end{equation}
Note that the identity $\diff^2=0$ encodes the well-known identities $\mathrm{curl}\,\mathrm{grad} = 0$ and $\mathrm{div}\,\mathrm{curl} = 0$. 
Indeed, vector calculus is simply a special case of exterior calculus in three-dimensional Euclidean space.

\textit{Hodge Duals and Inner Product}---We assume our manifold is equipped with a metric which defines angles, distances, and volumes. Then every $p$-form $\omega$ corresponds to a unique $(d-p)$-form $*\omega$, called its Hodge dual. 
Their components are related by
\begin{equation}
    (*\omega)_{\mu_1 \cdots \mu_{d-p}} = \frac{\sqrt{\vert g \vert}}{p!} 
    \omega^{\nu_1 \cdots \nu_p}
    \,
    \epsilon_{\nu_1 \cdots \nu_p \mu_1 \cdots \mu_{d-p}},
    \label{eq:hodge_star}
\end{equation}
where $g$ is the metric determinant, $\epsilon$ is the fully-antisymmetric Levi-Civita symbol, and upper indices are raised with the metric.
This allows us to define a notion of inner product of $p$-forms, 
\begin{equation}
    \langle \alpha,\beta \rangle := \int_M \alpha\wedge * \beta ,
    \label{eq:inner_product}
\end{equation}
where the integral is over all of $M$. 
The key idea is that $\alpha\wedge *\beta$ is a $d$-form, proportional to the volume form, which can sensibly be integrated over all of $M$. 

\textit{Coexterior Derivative}---With respect to the inner product, the exterior derivative has an adjoint operator called the coexterior derivative,
\begin{equation}
    \langle \diff \alpha,\beta \rangle = \langle \alpha, \diff^\dagger \beta \rangle,
\end{equation}
which lowers the degree of forms. 
It may be viewed as a generalization of divergence, i.e. it computes how much a $p$-form ``diverges'' from $(p-1)$-dimensional surfaces.

\subsection{Lagrangian Formulation}

We start with the spacetime action formulation of $p$-form electrodynamics. The core idea is that a $p$-form gauge field encodes the parallel transport of closed $(p-1)$-dimensional charged objects, i.e. the Aharanov-Bohm phase that such an object accumulates when tracing out a closed $p$-dimensional worldsheet~\cite{henneauxPFormElectrodynamics1986}.
Let $A$ be the $p$-form vector potential, and $F=\diff A$ be the corresponding field strength. Then the action of the theory is given by
\begin{equation}
    S = -\frac{K}{2}\int_M F \wedge *F.
\end{equation}
where $M$ is all of spacetime and $K=1/g^2$ where $g$ is the coupling constant. 
This $K$ can is analogous to the $K$ in \cref{eq:H-2-form}.
The field strength may be split into time-like and space-like parts to define the electric field $p$-form $E$ and magnetic field $(p+1)$-form $B$, $F = \diff t \wedge E + B$.
This theory has a global $p$-form symmetry given by shifting 
\begin{equation}
    A \to A + \Lambda \quad \text{with} \quad \diff \Lambda = 0.
    \label{eq:sm_shift_symmetry}
\end{equation}
A subset of the shifts are gauge transformations, i.e. those with $\Lambda = \diff\lambda$ for an arbitrary $(p-1)$ form $\lambda$, which are unphysical redundancies in the path integration. 
The deconfined Coulomb phase spontaneously breaks the physical part of the shift symmetry, also called electric symmetry, and the gapless photon is the corresponding Goldstone mode~\cite{gaiottoGeneralizedGlobalSymmetries2015,mcgreevyGeneralizedSymmetriesCondensed2023}.
A gauge-invariant order parameter indicating the symmetry breaking is given by large Wilson loops,
\begin{equation}
    W[C] = \exp(i \int_C A),
    \label{eq:sm_wilson_loop}
\end{equation}
where $C$ is a closed $p$-dimensional surface (i.e. a string when $p=1$ or a membrane when $p=2$), which satisfy a scale as the surface area of $C$ (volume of $C$) in the deconfined (confined) phase in the absence of charged matter.

The dynamics of the theory are governed by the Bianchi identity $\diff F = 0$ (following from $\diff^2 = 0$) and the equations of motion, $\smash{\diff^\dagger} F = 0$ (coming from varying with respect to $A$). If the gauge field is compact then it is allowed to have vortex singularities which violate the Bianchi identity, and if it is coupled to electric charges the equations read~\cite{henneauxPFormElectrodynamics1986}
\begin{align}
    \diff F &= j_\text{m}, \label{eq:bianchi}\\
    \diff^\dagger F &= j_{\text{e}}, \label{eq:eom}
\end{align}
where $j_{\text{m}}$ and $j_{\text{e}}$ are magnetic and electric currents, respectively. 
Splitting these two equations into timelike and spacelike components, one recovers the equivalent of the four Maxwell equations.

\subsection{Hamiltonian Formulation}

We now consider the Hamiltonian formulation in $d$ space dimensions. Fields now have only spatial components. 
The $p$-form vector potential and electric field are canonically conjugate, schematically $[\hat{A},\hat{E}] = i$ where hats denote operators, and the $(p+1)$-form magnetic field is $\hat{B} = \diff\hat{A}$.
The vector potential generates creation and annihilation of the electric field, meaning that the Wilson operator
\begin{equation}
    \hat{W}[C] = \exp(i \int_C \hat{A}),
    \label{eq:sm_wilson_loop_operator}
\end{equation}
creates unit electric flux along the closed $p$-dimensional surface $C$. 
If $C$ is the boundary of a surface, $C = \boundary \Sigma$, then this operator equivalently inserts magnetic flux through $\Sigma$ according to Stoke's theorem.\footnote{One can think of this intuitively according to Ampere's law: imagine that $C$ is a wire, then the uniform electric field inside the wire provides the electromotive force to drive a current, which generates a magnetic flux through the loop.
}

In the Hamiltonian formulation gauge invariance says that a state $\ket{A}$ of definite vector potential is gauge-equivalent to a state $\ket{A + \diff \lambda}$, i.e. these are the same physical state.
The electric field generates such transformations through the operators 
\begin{equation}
   \hat{G}_\lambda = \exp(i \int_{N}\hat{E} \wedge *\diff\lambda ) = \exp(i\int_{N} \diff^\dagger \hat{E} \wedge * \lambda),
   \label{eq:sm-gauge-operator}
\end{equation}
where $N$ is all of space (not spacetime) and $\lambda$ is an arbitrary $(p-1)$-form.
This operator must act as the identity on physical states, which therefore enforces the Gauss law operator identity on the physical Hilbert space,
\begin{equation}
    \diff^\dagger \hat{E} = 0,
\end{equation}
i.e. the electric field satisfies a generalized zero-divergence condition.

The electric shift symmetry of $A$, \cref{eq:sm_shift_symmetry}, is implemented by the operators (called Gukov-Witten operators~\cite{gaiottoGeneralizedGlobalSymmetries2015}) 
\begin{equation}
    \hat{G}_\alpha[\Omega] = \exp(i \alpha \int_\Omega *\hat{E}) = \exp(i \alpha \int_N *\hat{E}\wedge \delta_{\Omega}),
\end{equation}
where $\Omega$ is a $(d-p)$-dimensional surface and $\alpha$ is a real number parameterizing the shift. 
The conserved charge generating the symmetry is $\int_{\Omega}*\hat{E}$, the total electric flux through $\Omega$. 
In the second integral, $\delta_{\Omega}$ is a $1$-form Poincar\'e dual to $\Omega$~\cite{bottDifferentialFormsAlgebraic1982,lakeHigherformSymmetriesSpontaneous2018,thorngrenHiggsCondensatesAre2023}.\footnote{For our purposes, the Poincar\'{e} dual of a closed $k$-dimensional surface $\Sigma$ in a $d$-dimensional space $N$ is a $(d-k)$-form which acts like a generalized delta function, so that 
    \[
        \int_\Sigma \alpha = \int_N \alpha\wedge \delta_{\Sigma}
    \]
    for any $k$-form $\alpha$. Unlike a delta function, however, it is smooth, not singular.
    \label{fn:poincare_dual}
    }
This operator shifts $\ket{A}$ to $\ket{A + \alpha \delta_{\Omega}}$.
In particular, if $\Omega$ is contractible then $\delta_{\Omega}$ is an exact form and this is a gauge transformation. 
If $\Omega$ is non-contractible then this generates the physical electric $p$-form symmetry.
The electric symmetry thus corresponds to the conservation of electric flux through closed surfaces.

\subsection{Compactness and Magnetic Defects}
\label{apx:EM-duality}

If the gauge field $A$ is compact then it can form vortices, which are the magnetic defects of the theory. 
The role of these magnetic defects can be exposed by performing a duality transformation on the path integral~\cite{quevedoPhasesAntisymmetricTensor1997,peskinMandelstamtHooftDuality1978,seibergDualityWeb2+12016}.\footnote{
    Such Abelian dualities are closely related to the Kramers-Wannier duality of the 2D Ising model~\cite{druhlAlgebraicFormulationDuality1982,savitDualityFieldTheory1980,savitDualityTransformationsGeneral1982}.
    }
We work in the Lagrangian formulation of $\pe$-form U(1) gauge theory in Euclidean time signature with action
\begin{equation}
    S[A] = \frac{K}{2}\int_M \diff A \wedge * \diff  A,
    \label{eq:sm_action_dual_original}
\end{equation}
where $M$ is all of spacetime, with partition function (imaginary time path integral) $Z = \int \mathcal{D}\theta \exp(-S[A])$.
We can couple this to a background gauge field $a$ for the electric symmetry by replacing $\diff A \to \diff A - a$. 
In order to preserve the physics of the original model, we demand that $a$ be pure gauge, i.e., it must satisfy $\diff a = 0$. 
This is achieved by inserting a delta function in the partition function, which can be represented by adding an auxiliary compact $\pmag$-form field $\tilde{A}$ ($\pmag=D-(\pe+2)$) governed by the action
\begin{equation}
    S[\theta,A,\phi] = \int_M \frac{K}{2}(\diff A - a)\wedge *(\diff A - a) + \frac{i}{2\pi} \tilde{A}\wedge\diff a,
    \label{eq:sm_action_dual_augmented}
\end{equation}
which is gauge invariant under $A \to A + \Lambda$ and $a \to a + \diff \Lambda$.
Integrating over $\tilde{A}$ enforces the delta constraint $\diff a = 0$, such that $a$ can be removed by a gauge transformation to re-obtain \cref{eq:sm_action_dual_original}, thus these two theories are equivalent. 

The field $\tilde{A}$ is the dual (magnetic) vector potential, and generates the 't Hooft loop, 
\begin{equation}
    T[w] = \exp(i \int_w \tilde{A}) = \exp(i \int_M \tilde{A}\wedge\delta_w),
\end{equation}
where $w$ is a $\pmag$-dimensional surface and $\delta_w$ is Poincar\'e dual to $w$ (\cref{fn:poincare_dual}).
These are the magnetic versions of the electric Wilson operators, \cref{eq:sm_wilson_loop}.
Their action can be seen by inserting them into the path integral, which is equivalent to changing the action to 
\begin{equation}
    S \xrightarrow{T[w]} \int_M \frac{K}{2}(\diff A - a)^2 + \frac{i}{2\pi}\tilde{A}\wedge(\diff a - 2\pi \delta_w).
\end{equation}
Integrating $\tilde{A}$ now generates a delta function in the path integral enforcing that $\diff a = 2\pi \delta_w$, i.e. it inserts a $2\pi$ flux of $a$ along $w$. 
In the $a = 0$ gauge this then leaves a vortex singularity in $A$ which winds around $w$, essentially meaning that $\diff^2 A \neq 0$. 
In other words, given a $(\pe+1)$-sphere $S$ linking $w$, we have that $\int_{S} (\diff A - a) = 2\pi$. 
These are the generalization of Dirac strings, i.e. worldlines of the magnetic defects~\cite{teitelboimMonopolesHigherRank1986}.
Their operator equivalents (i.e. when $w$ lies in a single time slice) act as creation and annihilation operators for $\pmag$-dimensional magnetic ``field lines'', just as \cref{eq:sm_wilson_loop_operator} creates $\pe$-dimensional electric field lines.
Notice that since the Wilson loop measures the amount of magnetic flux through the loop, its value is changed by threading an 't Hooft loop through it.
Therefore the 't Hooft loops act as disorder operators~\cite{fradkinDisorderOperatorsTheir2017} for the Wilson loop order parameter \cref{eq:sm_wilson_loop}. 

To obtain the dual theory, we start from \cref{eq:sm_action_dual_augmented}. We can fix a gauge with $\diff A = 0$, then perform the Gaussian functional integration over $a$, obtaining the dual action
\begin{equation}
    S_{\text{dual}} = \frac{K'}{2} \int_M \diff \tilde{A} \wedge *\diff\tilde{A},
\end{equation}
where $K' = 1/4\pi^2 K$, which is a compact $\pmag$-form theory, thus establishing the duality. 
The dual action describes the same theory but from the perspective of the magnetic variables, where $\tilde{A}$ is the dual magnetic vector potential.
Note that this transformation reveals that the theory has $\pmag$-form symmetry (in the absence of magnetic monopoles), appropriately called magnetic symmetry~\cite{gaiottoGeneralizedGlobalSymmetries2015}, corresponding to conservation of magnetic flux.
The presence of magnetic charges explicitly breaks the magnetic symmetry, just as the presence of electric charges breaks the electric symmetry. 
If those charges are gapped, however, the symmetry can still be emergent at low energy~\cite{gaiottoGeneralizedGlobalSymmetries2015,mcgreevyGeneralizedSymmetriesCondensed2023,paceEmergentGeneralizedSymmetries2024}.

\bibliography{refs_vorticity}

\end{document}